\documentclass[aps,prd,nofootinbib,showpacs,preprintnumbers,longbibliography,amssymb,11pt]{revtex4-1} % not twocolumn
\pdfoutput=1

\normalsize
\usepackage{booktabs}
\usepackage{multirow}
\usepackage{makecell}
\usepackage{graphicx} % Required for inserting images
\usepackage{amsmath}
\usepackage{physics}
\usepackage{hyperref}
\usepackage[table]{xcolor}

\usepackage{siunitx}
\usepackage{amssymb}
\usepackage{slashed}
\usepackage{dsfont}
\usepackage{subcaption}
\usepackage{cleveref}
\usepackage{soul}
\usepackage{comment}

\usepackage{hyperref}

\begin{document}

\preprint{MITP-26-055}

\title{New Axion-Induced Effects in Mechanical Quantum Sensors}
\author{Louis Hamaide}
\email{Corresponding author: hamaidel@uni-mainz.de}
%\affiliation{PRISMA$^{++}$ Cluster of Excellence \& Mainz Institute for Theoretical Physics, Johannes Gutenberg University, 55099 Mainz, Germany}
\affiliation{Institut f\"ur Physik, Johannes Gutenberg-Universit\"at Mainz, Staudingerweg 7, 55128 Mainz, Germany}
\author{Felix Yu}
\email{yu001@uni-mainz.de}
\affiliation{PRISMA$^{++}$ Cluster of Excellence \& Mainz Institute for
  Theoretical Physics, Johannes Gutenberg-University, 55009 Mainz,
  Germany}
\date{\today}

\begin{abstract}
Axions and axion-like particles are well-motivated dark matter candidates which generically couple to CP-odd fermion spin and gauge field Lagrangian terms. 
We study how such axion fermion couplings and the axion electric dipole moment coupling lead to observable effects in paramagnetic or ferromagnetic materials used in mechanical quantum sensors, which are less studied in this context. 
We show how these couplings create oscillating torques and forces on these sensors, and show how these may be detected either by tuning the sensor's mechanical frequency to the axion frequency, or by borrowing methods used in other contexts. 
In particular, we discuss heterodyning the axion-induced signal from these couplings to obtain naturally broadband searches at low frequencies, extending the typically restricted parameter space of these searches by orders of magnitude. We also show how these couplings lead to oscillating shifts in the sensor's mechanical frequency, a naturally broadband and complimentary detection method which further extends the accessible axion parameter space. We show that by combining these tools with quantum limited noise floors of and macroscopically large sensor masses, we can perform broadband searches sensitive to several axion couplings for all axion masses below $m_a\lesssim 10^{-10}$~eV. We discuss the requirements for current and future mechanical quantum sensors to optimize their sensitivity to these couplings, and obtain sensitivity projections which surpass current sensitivity bounds and very competitive with future experimental proposals.
\end{abstract}

\maketitle

\tableofcontents

%%%%%%%%%%%%%%%%%%%%%%%%%%%%%%%%%%%%%%%%%%%%%%%%%%
\section{Introduction and motivation}
%%%%%%%%%%%%%%%%%%%%%%%%%%%%%%%%%%%%%%%%%%%%%%%%%%

The search to elucidate the nature of dark matter, which is believed to constitute the majority of the matter in the universe is one of the most pressing questions in modern physics. The detection of dark matter is a decades-long global effort, with a 
particular emphasis on searches for `Weakly Interacting Massive Particle' (WIMP) dark matter particles with masses roughly in the GeV-to-TeV range~\cite{PDG, Cirelli:2024ssz}. 
In recent years the absence of a positive signal from interactions of WIMP candidates in direct, indirect and collider-based searches has led to a shift in focus towards light (sub-GeV) and ultralight (sub-eV) dark matter models~\cite{Battaglieri:2017aum, Mitridate:2022tnv, Jaeckel:2022kwg, Jaeckel:2010ni}. Among these, searches for ultralight dark matter have become the center of increasingly broad theoretical and experimental efforts efforts using new technologies~\cite{Antypas:2022asj}. Ultralight dark matter fields are necessarily bosonic and have macroscopically high occupation numbers given their ultralow mass and temperature, such that they present as a coherent classical oscillating field as seen on Earth~\cite{Jaeckel:2022kwg, Hui:2021tkt}.

Among the most motivated candidates in the ultralight dark matter mass range are spin-1 vectors~\cite{HOLDOM1986196, Fayet:1980ad, FAYET1990743, Fayet:2025brx, Fayet:1980rr, Abel:2008ai, Fabbrichesi:2020wbt, Goodsell:2009xc, Nelson:2011sf, Arias:2012az, Graham:2015rva, Agrawal:2018vin, Bastero-Gil:2018uel, Dror:2018pdh, Co:2021rhi, Caputo:2021eaa, Adshead:2023qiw, Cyncynates:2023zwj, Cyncynates:2024yxm}, spin-0 scalars~\cite{Dimopoulos:1996kp, Arkani-Hamed:1999lsd, Burgess:2010sy, Cicoli:2011yy, TAYLOR1988450, Damour_2010, Damour_1994, Piazza:2010ye, Banerjee:2018xmn}, and the quantum chromodynamics (QCD) axion~\cite{Peccei:1977ur, PhysRevLett.38.1440, Weinberg:1977ma, Wilczek:1977pj, SHIFMAN1980493, Dine:1981rt, Dine:1982ah, Svrcek:2006yi, Shifman:1979if, DiLuzio:2020wdo, Choi:2020rgn, Arvanitaki:2009fg, Abbott:1982af, Preskill:1982cy, Kim:1979if, Zhitnitsky:1980tq}, a pseudoscalar particle introduced to suppress CP violation in the strong sector~\cite{Peccei:1977ur, PhysRevLett.38.1440, Weinberg:1977ma, Yu:2023gdq}, %Ai:2024cnp}, 
ensuring consistency with neutron electric dipole moment constraints~\cite{Abel:2020pzs}.
\footnote{We remark the ongoing discussion about the possible absence of CP violation in the strong interactions~\cite{Ai:2020ptm, Ai:2024vfa, Ai:2024cnp}.  In such cases, axion-like particles are serve as promising signatures of beyond the Standard Model extensions involving moduli spaces.} The QCD axion is expected to couple linearly to the electromagnetic and strong sectors via $F\tilde{F}$-type operators, as well as directly to fermions via a derivatively coupled axion. Commonly studied axion models include so-called KFSZ and DFSZ models~\cite{Zhitnitsky:1980tq, Dine:1981rt, Shifman:1979if, Kim:1979if}, which predict axions whose mass-coupling parameter space given by $(m_a, 1/f_a)$ where $m_a$ is the axion mass and $f_a$ is the axion decay constant, is typically narrowly restricted around the relation $m_a^2 f_a^2 = \chi \sim \Lambda_{\text{QCD}}^4$ where $\chi$ is the topological susceptibility of QCD, roughly given by the strong-coupling scale $\Lambda_\text{QCD}$. Further, axion-like particles (ALPs) generalize this to the case of a generic scale $\Lambda^4$, which motivates searches in the broader mass-coupling parameter space, such as those in string-inspired scenarios~\cite{Arvanitaki:2009fg, Stott:2017hvl} or other non-vanilla axion models~\cite{DiLuzio:2021pxd, Kivel:2022emq, Elahi:2023vhu}. 
We shall refer to all such scenarios as axions for simplicity. Both axion-like particles and QCD axions can notably be an ultralight particle with sufficiently high symmetry breaking scales such that the symmetry is broken pre-inflation. These axions are expected to be produced sufficiently to form the majority of the dark matter in the universe, either via the misalignment mechanism for pre-inflationary axions~\cite{Preskill:1982cy, Marsh:2015xka} or via the decay of a network of strings for post-inflationary axions~\cite{Battye:1994au, Hiramatsu:2012gg, Kawasaki:2014sqa}. Further, astrophysical structure formation bounds have suggested a lower bound on the axion of approximately $m_a\sim\mathcal{O}(10^{-20}~\text{eV})$~\cite{Eberhardt:2025caq}. 

Searches for signatures of axions via their couplings to the Standard Model (SM) have been the subject of an increasing number of searches and experimental proposals recently, with a notable focus on electromagnetic couplings and electron or nucleon spin couplings. Among these, astrophysical signatures on stellar evolution using supernovae, neutron stars, and red giants, have been successful in constraining axion mass-coupling parameter space~\cite{Hook:2017psm, Capozzi:2020cbu, Buschmann:2021juv, Balkin:2022qer, Kumamoto:2024wjd, Gomez-Banon:2024oux}, and the axion-gluon coupling has also been constrained from a big bang nucleosynthesis study~\cite{Blum:2014vsa}. Meanwhile Earth-based searches have also proven to be complementary to astrophysical searches and can probe new coupling-mass parameter space beyond the reach of astrophysical signatures~\cite{Chang:2017ruk, Abel:2017rtm, Roussy:2020ily, Arvanitaki:2021wjk, Zhang:2022ewz, JEDI:2022hxa, Schulthess:2022pbp, Banerjee:2023bjc, Zhang:2023lem, Alda:2024xxa, Fan:2024pxs, Madge:2024aot, Gue:2025nxq}. These Earth-based detection schemes often exploit the near monochromaticity (relatively small spectral bandwidth) of axion dark matter at all masses due to their small virial velocity near Earth~\cite{Hui:2021tkt,Amaral:2024tjg}, leading to natural enhancement in searches for axion dark matter which use resonant detectors.

Among resonant detectors of axion dark matter, fixed wall microwave electromagnetic cavity resonators stand out as having demonstrated particularly sensitive measurements of the axion mass-photon coupling parameter space down to the QCD axion band~\cite{CAPP:2020utb, ADMX:2025vom, QUAX:2025wtb}. This success has led to a variety of complementary setups whose superposed sensitivity has produced strong constraints on the QCD axion band in the region $m_a\sim 10^{-6}$ to $10^{-5}$~eV which is motivated by natural $\mathcal{O}(1)$ misalignment angles that lead to axion relic densities in agreement with observations~\cite{Marsh:2015xka}. The electromagnetic fields in these sensors are typically modeled as coherent, semi-classical harmonic oscillators whose noise is quantum-limited down to the `Standard Quantum Limit' (SQL)~\cite{ADMX:2018ogs, Alesini:2019ajt, Beurthey:2020yuq, BREAD:2021tpx, HAYSTAC:2024jch, ADMX:2025vom}. Such resonators generally require a trade-off, however, between resonantly-enhanced sensitivity and the correspondingly narrow axion mass coverage, limiting the axion masses they can practically cover. 

To improve on this limitation, electromagnetic cavities, particularly superconducting radio frequency cavities, have recently been studied as broadband resonant devices to search for dark matter signals with potentially much lower frequencies than the cavity's natural frequency. This is achieved through \textit{heterodyning} the axion signal~\cite{Berlin:2020vrk, Li:2025pyi} by driving the oscillations of the electromagnetic field at the natural frequency, such that the product of the electromagnetic and axion oscillations leads to oscillating axion signals at sum and beat frequencies located on either side of the natural frequency, allowing for detection with increasingly high quality factors at low axion masses. 
Although sensitivity to larger axion masses requires operating at lower quality factors, heterodyne detection is naturally broadband, allowing any axion whose mass lies within the resonant bandwidth to be detected at the on-resonance SQL.
%Despite the cost of lower quality factors for larger axion masses, crucially this approach is naturally broadband since heterodyning allows for detection at the SQL on-resonance of any axion whose mass is small enough to fit in the resonant peak. 
This technique has been projected to have sensitivity surpassing current astrophysical and Earth-based bounds by several orders of magnitude.

Despite proposed resonant cavity concepts, including resonant circuits~\cite{Silva-Feaver:2016qhh, DMRadio:2022pkf}, being sensitive to new axion parameter space, this approach has been limited to sensors which use electromagnetic fields to which axions are coupled via the electromagnetic coupling. These setups are notably insensitive to the axion's derivative couplings to fermions or linear coupling to gluons: at the low energy effective field theory level, such couplings map to the spins of the constituent electrons, protons and neutrons present in materials. While these couplings are primarily constrained by astrophysical observations, the spin precession CASPEr experiment and proposed comagnetometer setups~\cite{Budker:2013hfa, Bloch:2019lcy} are notable exceptions, as well as the work in Refs.~\cite{Graham:2017ivz, Terrano:2019clh}.  The CASPEr experiment is expected to significantly improve the bounds on the nuclear electric dipole moment (EDM) coupling by looking for neutron spin procession in a gram-scale  nuclear magnetic resonance-active material with highly polarized neutron spins~\cite{JacksonKimball:2017elr}. In CASPEr-Electric, for example, a large external magnetic field is used to tune the neutron Larmor precession frequency, while an orthogonal internal electric field is provided by a ferroelectric material. By scanning the Larmor frequency across the axion mass range, the experiment searches for resonantly enhanced axion-induced spin torques. Similarly, the CASPEr gradient experiment seeks to use spin oscillations to search for the axion-electron coupling to nucleons, and their projected sensitivity is again expected to surpass current bounds by several orders of magnitude. In addition, the axion-nucleon coupling is expected to lead to an additional spring constant and thus a frequency shift in the spin precession rate, which can be seen in power spectrum of the magnetization via an extra axion-induced peak near the unperturbed resonant peak frequency~\cite{Garcon:2017ixh, Garcon:2019inh}.

Against this backdrop, mechanical quantum sensors are a novel type of resonator, essentially harmonic oscillators in position space, which are limited by quantum noise, typically at the SQL, and studies of their suitability for axion detection have only just begun.  %. We remark that one search for axion dark matter via the axion-fermion couplings has led to over an order of magnitude improvement in the bounds from Earth-based sensors
Existing work has focused primarily on spin-mediated, dipole-dipole fifth force searches with torsion pendula~\cite{Terrano:2015sna} or mechanical rotors~\cite{ARIADNE:2017tdd}, and there are preliminary studies of the projected sensitivity of  superconductors~\cite{Higgins:2023gwq} and magnetically levitated ferromagnets~\cite{Kalia:2024eml} with quantum-limited noises. 
Specifically, Ref.~\cite{Higgins:2023gwq} used superconductors to probe the axion-photon coupling, while Ref.~\cite{Kalia:2024eml} studied how  torque on electron spins lead to torque on ferromagnetic sensors themselves: such angular motion can be detected via the change in the surrounding magnetic field using SQUIDs. 
In addition to these phenomenological signals, we discuss for the first time 
the translational effects on other types of magnetic materials, such as paramagnets, and the effects induced by the so-called axion-EDM coupling~\cite{JacksonKimball:2017elr}. In addition, we inaugurate the study of mechanical quantum sensors to the axion coupling to protons and neutron and the axion-EDM coupling in nucleons using polarized neutron spins.

Mechanical quantum sensors are harmonic oscillators whose motion is measured using electromagnetic readout, most commonly optical fields in the near-infrared, although magnetic and electric readout schemes are also possible (see, {\it e.g.},~\cite{Aspelmeyer:2013lha, Beckey:2023shi, Winstone:2023whl} for reviews of optically read out systems). These sensors are typically solid-state objects whose constituent components are treated collectively as a single mechanical quantum degree of freedom. Their motion may be confined by optical~\cite{Magrini:2020agy, Millen_2020, Iacoponi:2024mzu, Hamaide:2025buy}, magnetic, or electric fields~\cite{PhysRevLett.127.061804, Blum:2014vsa}, or through mechanical coupling to an external support, as in suspended pendula, clamped membranes, and cantilevers~\cite{Aspelmeyer:2013lha}. Mechanical quantum sensors span a broad range of experimental platforms and have been proposed or employed to search for physics beyond the SM, including scalar and vector dark matter~\cite{Carney:2019cio, Manley:2019vxy, Chowdhury:2025avw}, fifth forces~\cite{Manley:2024svv}, and gravitational waves~\cite{Aggarwal:2020umq, Aggarwal:2025noe}. For our purposes, it is useful to distinguish between electromagnetically trapped and mechanically coupled sensors, since their different couplings to the surrounding environment lead to substantially different thermal properties.

As shown {\it e.g.}~in Ref.~\cite{Hamaide:2025buy}, the relative absence of thermal coupling in optically trapped sensors relaxes the requirement for low temperatures. However, this can also lead to low spin polarization, notably in paramagnetic materials, whose spins are inversely proportional to the material's temperature. Therefore, in this work, we investigate optically and mechanically suspended sensors as complementary probes. The study of quantum and thermal noises of optomechanical quantum sensors in both types of systems can be accurately predicted (see {\it e.g.}~Ref.~\cite{Aspelmeyer:2013lha,Hamaide:2025buy,Geraci:2018fax}), and these systems are cooled sufficiently such that they are limited by the quantum noises of the trapping and readout fields, which can be tuned to reach the SQL noise floor. Remarkably, preliminary studies of different mechanical quantum sensors have implied that the SQL is independent of the type of trap or sensor material, such that our study is therefore intended to broadly indicate the sensitivity of mechanical quantum sensors to several axion couplings.

Beyond the conventional approach of tuning a mechanical sensor’s resonant frequency to the axion Compton frequency, the versatility of mechanical quantum sensors allows different materials and detection strategies to be tailored to axion-fermion and axion-EDM couplings, providing access to signatures that have not previously been explored. First, as we will show, axion couplings in magnetically interacting materials such as paramagnets lead to translational forces on the sensor, notably in the presence of a magnetic field gradient. This can be used to search for the axion signal at the translational mode's SQL noise floor, which can be tuned to optimize the reach of the sensor to a large range of axion signals. Furthermore, below the minimum frequency at which the SQL can be reached in Earth-based sensors, heterodyning, as proposed in Ref.~\cite{Berlin:2020vrk}, enables mechanical quantum sensors to search for low frequency axion signals by oscillating the spin polarization's gradient. 

Finally, as studied in the case of spin precession, the axion coupling to the spins of mechanical quantum sensors will itself oscillate over time, leading to an additional axion-induced spring constant which shifts the oscillation frequency of the sensor. This can be detected in the power spectrum of the sensor's motion analogously to the magnetization power spectrum as suggested in the CASPEr proposal~\cite{Garcon:2017ixh}. Crucially these effects have not been used with mechanical quantum sensors for axion detection. Our study thus intends to explore how these methods can be used in this context and suggest which mechanical sensor trapping methods and materials are most ideally suited in the near future to sense these effects using noise at the SQL while keeping thermal noises subdominant.

In this work, we begin by reviewing the axion Lagrangian in Sec.~\ref{sec:axion_physics}.  In Sec.~\ref{sec:dynamics}, we present the various dynamics, namely force and torques in both homodyne and heterodyne schemes as well as frequency shifts, that axion couplings generate with in paramagnetic and ferromagnetic mechanical quantum sensors. In Sec.~\ref{sec:Noises}, we then study the noise and thermal limitations of different mechanical quantum sensor configurations and identify the setups that maximize the signal-to-noise ratio for these effects, accounting for the required experimental resources, such as magnetic fields, suitable materials, and cooling. We subsequently estimate their sensitivities to the axion-electron, axion-proton, axion-neutron, and axion-EDM couplings.  In App.~\ref{app:EOM_precession}, we briefly discuss the coupled angular precession equations of motion and the necessary approximation to decouple them.
Natural units of $\hbar = c = 1$ are used  throughout this work.

%%%%%%%%%%%%%%%%%%%%%%%%%%%%%%%%%%%%%%%%%%%%%%%%%%
\section{Axion physics and axion dark matter as a coherent background field}\label{sec:axion_physics}
%%%%%%%%%%%%%%%%%%%%%%%%%%%%%%%%%%%%%%%%%%%%%%%%%%

In this section, we present the signatures of axions and ALPs (collectively referred to as axions) in mechanical quantum sensors. We start with the effective Lagrangian
\begin{align}
\mathcal{L} = &- \dfrac{1}{4} F_{\mu \nu} F^{\mu \nu}
+ \sum\limits_{f = e, p, n} \left( \overline{\psi}_f \, i \slashed{D} \, \psi_f 
+ m_f \overline{\psi}_f \psi_f \right) 
+ \dfrac{1}{2} \left( \partial_\mu a \right)^2 - \dfrac{1}{2} m_a^2 a^2 
\nonumber \\
& - \sum\limits_{f = e, p, n} \left( g_f \, \partial_\mu a \, \overline{\psi}_f \, \gamma^\mu \gamma^5 \psi_f \right) + i \dfrac{g_d}{2} 
\sum\limits_{f = p, n} \left( a \overline{\psi}_f \, \sigma_{\mu \nu} \gamma^5 \psi_f  F^{\mu \nu} \right)
- g_{a\gamma} a F_{\mu \nu} \tilde{F}^{\mu \nu} \,\, ,
\label{eq:axionLagrangian}
\end{align}
where we write the SM degrees of freedom well below the QCD scale, and we thus neglect the axion couplings to QCD pseudoscalars. The $g_f$ and $g_{a\gamma}$ couplings carry mass dimension GeV$^{-1}$, while $g_d$ has mass dimension GeV$^{-2}$ and is assumed equal for protons and neutrons, 
and $F_{\mu \nu}$ is the electromagnetic tensor. The $g_d$ interaction is a possible axion--electric dipole moment (EDM) coupling, following~\cite{Crewther:1979pi,Graham:2013gfa}. We note that we include only linear couplings to the axion and leave the phenomenology of quadratic couplings to future work.

This effective Lagrangian is a particular choice of basis, since we can still perform global axial transformations of the individual fermions, $\psi \to \psi' = e^{i \alpha \gamma^5 a / f_a} \psi$~\cite{Kivel:2022rzr, Yu:2023gdq} 
thus introduces axion couplings to pseudoscalar $(e, p, n)$ fermion bilinears~\cite{Berlin:2023ubt}.  We will work, however, in the basis where such pseudoscalar bilinear couplings are absent.\footnote{An example where the physical equivalence of the two terms related by the axial transformation of the fermion field is given in Ref.~\cite{Pospelov:2008jk}.}  In addition, since we work in a linear representation of the axion degrees of freedom, there is no apparent periodic shift symmetry in the axion field. Since our observables will only involve linear couplings of the axion, we will retain this formally imprecise Lagrangian description throughout this work for simplicity. Moreover, we will only focus on the axion coupling to axial-vector $(e, p, n)$ fermion bilinears and the axion coupling to the electric dipole moment of nucleons, and assume in this study that the effects of the $g_{a\gamma}$ coupling are sufficiently small and play a negligible effect on the dynamics of mechanical quantum sensors. Finally, in what follows we will only consider the axion interacting with the SM via one coupling at a time. 

In Sec.~\ref{sec:dynamics}, we will estimate the sensitivity of mechanical quantum sensors to the electron, proton and neutron couplings and the EDM coupling in Eq.~\ref{eq:axionLagrangian}, assuming the axion provides a cold dark matter density background for the Earth's motion through the galaxy. In essence, mechanical quantum sensors can detect the nonzero axion background field via some local momentum transfer, in analogy to traditional direct detection methods for particle dark matter.  In comparison to WIMP dark matter, however, the axion dark matter typically exhibits very high occupancy numbers, which we can estimate using $\rho_{\text{DM}} \simeq 0.4$~GeV/cm$^{3}$~\cite{deSalas:2020hbh} for the local DM energy density and focusing on axion masses below 1~eV. The local number occupancy within a de Broglie volume~\cite{Carney:2019cio, Hamaide:2025buy} can be estimated as
\begin{align}
n = \dfrac{\rho_{\text{DM}}}{m_a^4 v_a^3} \sim 10^{15} \left( \dfrac{1 \text{ meV}}{m_a} \right)^4 \ ,
\end{align}
The corresponding de Broglie wavelength of the axion is estimated by $\lambda \sim 1 / (m_a v_a)$, where $m_a$ and $v_a$ are the axion mass and axion dark matter velocity, which is much longer than the spatial scale of the experiments we will discuss, and where we use $v_a \sim 10^{-3}$ as the virialized dark matter velocity in the local galaxy. The high occupancy number justifies treating the axion field as a classical pseudoscalar field with a coherent phase on distances dictated by $\lambda$.  We remark that the assumption of $\rho_{\text{DM}} \simeq 0.4$~GeV/cm$^{3}$ is the expected local density of dark matter in the Standard Halo Model~\cite{deSalas:2020hbh} and serves as an effective normalization for the axion field amplitude.

As the Earth moves through the local axion scalar field, a laboratory experiment views the axion as a background energy density whose strength is determined locally by the scalar field amplitude. Given the small virial velocity seen in the laboratory frame, which is independent of the axion mass, the classical axion field is further expected to be highly coherent with coherence time $T_\text{coh,a} = 2\pi/m_av_a^2$. For integration times $T_\text{int} \lesssim T_\text{coh,a}$, the axion field is well approximated by an oscillating cosine $a(t) = a_0 \cos (m_a t)$ with fixed amplitude taken by a Rayleigh distribution whose mean is given by the  non-relativistic field amplitude $a_0=\sqrt{2\rho_\text{DM}}/m_a$, obtained from the approximation $\rho_{\text{DM}} \approx \dfrac{1}{2} m_a^2 a_0^2$. Although during each coherent period (of duration $T_\text{int}$) the amplitude of the field is independently drawn from the Rayleigh distribution, note that in this work we will make the simplifying assumption of taking the mean $a_0$, which is well-motivated when detecting the axion field over a large number of coherence times.
The coherence time also induces a bandwidth (in frequency space) of any axion-induced signal in SM matter given by $\delta\omega_a=2\pi / T_\text{coh,a}$, which defines a corresponding quality factor $Q$ of the axion when taken as an oscillator, $Q = m_a / \delta \omega_a \approx 10^6$.

%%%%%%%%%%%%%%%%%%%%%%%%%%%%%%%%%%%%%%%%%%%%%%%%%%
\section{Dynamics of mechanical quantum sensors coupled to axions}
\label{sec:dynamics}
%%%%%%%%%%%%%%%%%%%%%%%%%%%%%%%%%%%%%%%%%%%%%%%%%%

We now discuss the dynamics of magnetized mechanical quantum sensors, including torques, forces and mechanical frequency shifts, in the presence of individual axion couplings.

%%%%%%%%%%%%%%%%%%%%%%%%%%%%%%%%%%%%%%%%%%%%%%
\subsection{From microscopic to macroscopic interactions}
%%%%%%%%%%%%%%%%%%%%%%%%%%%%%%%%%%%%%%%%%%%%%%

From Eq.~\ref{eq:axionLagrangian}, we consider the following two Lagrangian interaction terms:
\begin{align}
\mathcal{L} \supset \sum\limits_{f = e, p, n} \left( g_f \left( \partial_\mu a \right) \overline{\psi}_f \, \gamma^\mu \gamma^5 \psi_f \right) 
-\frac{i}{2} g_d\,
\sum\limits_{f = p, n} \left( a \overline{\psi}_f \, \sigma_{\mu\nu} \gamma^5 \, \psi_f F^{\mu\nu} \right) \ ,
\label{eq:LaxionInts}
\end{align}
where again the first parentheses show an axion derivatively-coupled to the 
electron, proton, or neutron axial-vector current fermion bilinear and the second parentheses show an axion coupled to the electric dipole moment of the proton or neutron.  We remark that this interaction induces an oscillating rather than a static proton or neutron EDM, and is therefore consistent with stringent constraints on permanent nucleon EDMs provided that any time-independent contribution remains below existing bounds.

Our immediate goal will be to capture the effect of these interactions as inhomogeneous terms in the equation of motion for a mechanical quantum sensor, following Ehrenfest's Theorem.  First, we perform a non-relativistic expansion: note there are several approaches to constructing the non-relativistic dynamics of fermions, including the Foldy-Wouthuysen method~\cite{Foldy:1949wa} and Pauli reduction. The essential idea is that the Dirac Lagrangian, since it is manifestly Lorentz-covariant and linear in spacetime derivative, necessarily mixes particle and anti-particle degrees of freedom in the corresponding equations of motion, and thus a non-relativistic limit to discuss single particle dynamics arising from Eq.~\ref{eq:LaxionInts} should bisect the spectrum of field modes. This can be done systematically to all orders in $1 / m_f$ in the Foldy-Wouthuysen method for the Dirac Hamiltonian, but only order by order using Pauli reduction. Details of the Foldy-Wouthuysen method for axion interactions are shown in Ref.~\cite{Smith:2023htu}, and a comparison with Pauli reduction was presented in Ref.~\cite{Berlin:2023ubt}, while heuristic arguments to derive the leading non-relativistic effects are given in Refs.~\cite{Graham:2017ivz, Agrawal:2023lmw}.  

Second, the axion couplings from Eq.~\ref{eq:LaxionInts} act at the microscopic level, while the mechanical sensor responds to coherent momentum transfer. We would thus proceed with the many-body interacting Hamiltonian for our quantum sensor via the standard linearity construction, but we foresee that a complete accounting of the many-body effects between the constituent protons, neutrons, and electrons of a macroscopic quantum sensor in an axion background is unnecessarily verbose for our purposes and we will simplify the treatment considerably. Because of the coherence of the axion field and fermions in the sensor, we can linearly sum over all axion-fermion interactions in the sensor to obtain a net torque or force on the sensor.  This notably assumes that internal degrees of freedom (many-body interactions) such as phonons are relatively suppressed.  A mechanical quantum sensor can then be idealized as a macroscopic object whose free equation of motion (EOM) is simply a quantum harmonic oscillator. For a sensor with a total number of a given fermion species of  $N_f$, the center of mass position is then $\vec{x} = \dfrac{1}{N_f} \sum_i \vec{x}_i$, and its total momentum is $\vec{p} = \sum_i \vec{p}_i$, where $i$ runs over all of the constituent fermions in the mechanical sensor.  

Using the Foldy-Wouthuysen method~\cite{Foldy:1949wa, Smith:2023htu} and employing the coherent summation over constituent fermions described above, we can write the interacting Hamiltonian %for derivatively coupled axions 
for a given fermion as
\begin{align}
H_\text{int} &\simeq \dfrac{g_f}{2m_f} \left\{ \partial_t a,\,
\vec{P} \cdot \vec{\sigma} \right\} - g_f
\left(  \vec{\nabla} a \cdot \vec{\sigma} \, + 
\dfrac{1}{4m_f^2} \left( \left\{ \vec{P}\,{}^2, \, \vec{\sigma} \cdot \vec{\nabla} a \right\} - 2 ( \vec{P} \cdot \vec{\nabla} a ) (\vec{P} \cdot \vec{\sigma} )
\right) \right) \nonumber \\
&- g_d \, a \, \vec{\sigma} \cdot \vec{E} - g_d \, \dfrac{a}{2m_f} \vec{\sigma} \cdot \left( \vec{B} \times \vec{P} - \vec{P} \times \vec{B} \right) \ ,
\label{eq:axionHint}
\end{align}
where the first line corresponds to axion-fermion couplings and the second line corresponds to the axion-EDM coupling, $\vec{P} = \vec{p} - q_{\text{eff}} \vec{A}$ is the gauge-covariant 3-momentum operator of the sensor and $\vec{\sigma}$ are the Pauli matrices representing the spin operator acting on the sensor. Note Eq.~\ref{eq:axionHint} is truncated at second order in $1/m_f$. We also remark that the macroscopic sensor will have the total charge $q_{\text{eff}} = 0$ since it will be electrically neutral.  

We thus take $\vec{P}=\vec{p} = -i \vec{\nabla}$ to commute with the spatial variation of $a(\vec{x}, t)$, leading to our main expression for the interacting Hamiltonian,
\begin{align}
H_\text{int} \simeq &- g_f
\left(  \vec{\nabla} a \cdot \vec{\sigma} \right) +
\dfrac{g_f}{m_f} \left( \partial_t a \right) \left(
\vec{p} \cdot \vec{\sigma} \right) - \dfrac{g_f}{2m_f^2} 
\vec{p}\,{}^2 \left( \vec{\sigma} \cdot \vec{\nabla} a \right) + \dfrac{g_f}{2 m_f^2} (\vec{p} \cdot \vec{\nabla} a ) \left( \vec{p} \cdot \vec{\sigma} \right)
\nonumber \\
&- g_d \, a \, \vec{\sigma} \cdot \vec{E} + g_d \dfrac{a}{m_f} \vec{\sigma} \cdot \left( \vec{B} \times \vec{p}  \right) \ ,
\label{eq:axionHint2}
\end{align}
where we will ignore $\mathcal{O}(p^2/m_f^2)$ terms for the same reason as for the axion-fermion couplings.

The equation of motion of the macroscopic sensor location $\vec{x}$ is determined by the Poisson bracket $i \left[ H_{\text{tot}}, \vec{p} \right]$. As expected, we get a harmonic oscillator-type equation of motion with an inhomogeneity arising from the interacting Hamiltonian. 
In addition to the sensor position, the angular orientation $\vec{\theta}$ of the sensor, denoting the Euler angle with respect to fixed coordinate axes aligned with the velocity of Earth across the Milky Way DM halo, can change depending on the constituent fermion spin couplings to the axion background. These inhomogeneities will manifest as axion-induced torques or forces acting on the sensor position and angle, which is the focus of the remainder of this section.

A levitated sensor~\cite{Millen_2020, Hamaide:2025buy}, for example, has its center‑of‑mass motion confined in a locally harmonic potential generated by magnetic fields, electric fields or precisely tuned lasers. When the sensor moves in an axion background, the axion field contributes an oscillatory component to the effective potential at the axion Compton frequency ({\it i.e.}, on timescales much shorter than the axion coherence time). Spatial gradients of this axion‑induced potential create forces that drive the sensor’s motion. Such gradients can arise either from the axion field itself or from spatial variations in the sensor’s charge or spin polarization, allowing forces to be generated along different directions of motion.

%%%%%%%%%%%%%%%%%%%%%%%%%%%%%%%%%%%%%%%%%%%%%%%%
\subsection{Possible magnetic properties of mechanical sensors}
%%%%%%%%%%%%%%%%%%%%%%%%%%%%%%%%%%%%%%%%%%%%%%%%

We will consider mechanical quantum sensors with paramagnetic or ferromagnetic\footnote{We note that other types of electronic magnetism such as electronic ferrimagnetism may also be of interest to probe the couplings Eq.~\ref{eq:axionHint2}, likely due to lower optical absorption or ease of fabrication, notably for their improved optical properties compared to ferromagnets. We also note that diamagnetism, which couples to the orbital angular momentum of the electron, is not expected to couple to axions.} properties, such that the spins of the constituent electrons are aggregated into an overall spin vector $\sum_i P_f\,\vec{\sigma}_i = P_f N_f  \,\vec{\sigma}_f$ of the sensor, where $P_f$ is the spin polarization per particle of a given species (assuming a material's primary magnetic contribution is dominated by one fermion species $f$), $N_f$ is the number of particles of that species, and $\sigma_f$ are the Pauli matrices of a single fermion. In what follows we will drop the subscript $f$ for notational convenience, write the overall spin as $PN\vec{\sigma}$, and clarify the species we consider before obtaining equations of motion of the sensor.  Importantly, in the case of electron magnetism, ferromagnetic materials have a spin polarization per electron $P \sim 10^{-2}$ below the Curie temperature. Paramagnetic materials, on the other hand, can achieve even larger electron polarizations, but high concentrations of polarizable spins generally enhance many-body interactions, shortening magnetization relaxation times and increasing thermal magnetization noise, as discussed in Sec.~\ref{paragraph:magnetization_noise}. We leave a systematic optimization of such materials for future work and adopt nitrogen (P1) centers in diamond as a representative benchmark.
%In the future: gadolinium-doped yttrium aluminium garnet (Gd:YAG), gadolinium-doped yttrium lithium fluoride (Gd:YLF) or ytterbium 3+ doped yttrium lithium garnet (Yb$^{3+}$:YLF)?
In this work, we assume a $\mathcal{O}(10$~ppm) concentration of P1 centers while assuming relaxation times of tens of seconds~\cite{Jarmola_2012, Isaac_2017} such that the portion of polarizable electrons in the material $P\sim 10^{-5}$.

We remark that, for different materials, the direction of the magnetization vector of the sensor $\vec{M}=N_fP_f\vec{\mu}_f\propto\vec{\sigma}$, where $\mu_f$, the magnetic dipole moment of the fermion's spin, is not necessarily the same as its angular orientation $\vec{\theta}$. In a ferromagnetic material, this magnetization vector direction is exactly opposite the constituent spins of the polarized electrons, which are locked in a lattice, and hence, up to a parity transformation, the equation of motion for $\vec{\theta}$ is the aggregate of the equations of motion of the individual electron spins.
Notably, the energy required for the sensor's lattice of spins to rotate is much less than the energy required for internal rotation, where the spins rotate relative to each other.  A rotation of the magnetization via the axion coupling therefore manifests as a rotation of the ferromagnet.  In particular, the macroscopic equation of motion for the angular orientation $\vec{\theta}$ of the sensor is weighted by the moment of inertia tensor $I$, reflecting the possible tensorial response of the sensor to different directions of driving frequencies.   Ferromagnetism and electronic paramagnetism therefore form experimental scenarios with distinct strengths and weaknesses.

In the case of proton or neutron magnetism, given the unavailability of nuclear ferromagnets, we will only consider materials exhibiting paramagnetism. Although nuclear spins naturally antialign in pairs to fill every energy level~\cite{Stroberg:2019mxo}, nuclei with odd number of either protons or neutrons are expected to have one unpaired proton or neutron respectively.\footnote{Although spin couplings do occur between protons and neutrons inside a nucleus, given the different energy (shell) levels of electrically neutral neutrons compared to protons whose potential includes a Coulomb potential, they fill their energy shells approximately independently.} Unpaired nucleons can then be aligned at sufficiently low temperatures in the presence of an external magnetic field.\footnote{Although we do not consider it in this work, this `brute force' magnetization may also be improved using methods such as dynamic polarization or spin pumping.}

We therefore focus on materials that remain solid under ultrahigh-vacuum and cryogenic conditions, support large spin polarizations and long relaxation times ($\mathcal{O}(10^3~{\rm s})$ or longer), and exhibit low absorption at the near-infrared wavelengths relevant for optical trapping and readout. We assume the spin polarization is sufficiently large that the contribution from diamagnetism, which does not couple to the axion interaction considered here, can be neglected. The associated temperature and noise requirements are discussed in detail in Sec.~\ref{sec:thermal_limits}. While a systematic optimization of candidate materials is beyond the scope of this work, several materials provide useful benchmarks. Solid $^{129}$Xe~\cite{Walter:2025ldb} and fluorine-containing compounds such as lithium fluoride (LiF) and calcium fluoride (CaF$_2$)~\cite{abragam1982nuclear} satisfy many of these requirements and can contain an $\mathcal{O}(10\%)$ fraction of polarizable nucleons. Deuterated materials such as lithium deuteride (LiD) similarly provide a sizable fraction of polarizable neutrons and can therefore probe axion-neutron couplings~\cite{BALL2003101}. Finally, lead-containing ferroelectrics such as lead magnesium niobate--lead titanate (PMN-PT), as proposed for CASPEr-Electric~\cite{JacksonKimball:2017elr}, provide an $\mathcal{O}(10^{-3})$ fraction of polarizable $^{207}$Pb nuclei together with the large internal electric fields required to probe the axion-EDM coupling.

We note that the use of materials with similar thermal properties and levels of unpaired neutrons or protons are expected to lead to similar axion-induced effects.
As we will see, the phenomenology of paramagnets is distinct from that ferromagnets as the total polarization in paramagnets is proportional to the amplitude of the external magnetic field. Furthermore, the spins in paramagnets rotate internally to align along the axis of the external magnetic field regardless of the physical orientation of the sensor. We therefore treat paramagnetic and ferromagnetic sensors separately in the following sections. Paramagnets are used to probe axion-fermion and axion-EDM couplings, while ferromagnets are considered specifically for the axion-electron coupling.

Exploiting the coherence of the axion field across the sensor, we translate the microscopic interactions in $H_{\rm int}$ into macroscopic forces, torques, and frequency shifts. We examine how the dependence of each interaction on $\vec p$, $m_f$, and $\vec\sigma$ determines the sensor response, and optimize the sensor configuration separately for each interaction to maximize sensitivity to the corresponding axion coupling.  We give an overview of the various couplings and detection methods in Table~\ref{tab:summary}.

%\cellcolor{gray!30}
\begin{table*}[t!]
\begin{tabular}{|c |c | c | c | c|}
\cline{3-5} \multicolumn{2}{c|}{} & \multicolumn{2}{c|}{\quad Force/Torque sensing \quad} & \multicolumn{1}{c|}{\quad \multirow{2}{*}[-0.6em]{Frequency shift} \quad}\\
\cline{3-4} \multicolumn{2}{c|}{} & \makecell{\quad Homodyne \quad\\[-0.6em]\quad$m_a>10^{-14}$~eV \quad}  &  \makecell{\quad Heterodyne \quad\\[-0.6em]\quad$m_a<10^{-14}$~eV \quad} & \\ 
\hline \multirow{2}{*}{\makecell{\quad Electron \quad\\[-0.6em]\quad coupling $g_e$ \quad}} & \quad Ferromagnet \quad & \multicolumn{2}{c|}{\quad Secs.~\ref{paragraph:EOM_study_first_term_derivative_coupling}~\&~\ref{sec:prospects_ferro_derivative_coupling}\quad} & \quad Secs.~\ref{sec:axionpeakderivative_ferro}~\&~\ref{sec:g_f_ferro_freq_shift_prospects}
\quad\\
\cline{2-5} &\quad Paramagnet \quad& \multicolumn{2}{c|}{\multirow{2}{*}[-0.6em]{\quad Secs.~\ref{paragraph:EOM_study_first_term_derivative_coupling2}~\&~\ref{sec:prospects_derivative_coupling}\quad}} & \multirow{2}{*}[-0.6em]{\quad Secs.~\ref{sec:axionpeakderivative_diapara}~\&~\ref{sec:g_f_freq_shift_prospects}
\quad}\\
\cline{1-2} \multicolumn{2}{|c|}{\makecell{Proton coupling $g_p$ \\[-0.6em] (paramagnet)}} & \multicolumn{2}{c|}{} & \multicolumn{1}{c|}{}   \\ 
\hline \multicolumn{2}{|c|}{\makecell{EDM coupling $g_d$ \\[-0.6em] (paramagnet)}} & \multicolumn{2}{c|}{\quad Secs.~\ref{sec:force_EDMcoupling}~\&~\ref{sec:prospect_force_EDM} \quad} & \quad Secs.~\ref{sec:peak_split_EDM}~\&~\ref{sec:edm_freq_shift_prospects} \quad \\   
\hline 
\end{tabular}
\caption{Summary of each method (columns) mechanical quantum sensors can use to probe axion-fermion and axion-EDM couplings, depending on the materials and couplings (rows). Each combination of a detection method and coupling may lead to a unique scaling of the sensitivity with respect to axion mass. This is modified further for proton and EDM couplings when accounting for scaling of the spin polarization via the temperature (which is no longer constant for $m_a\gtrsim 4\times 10^{-11}$~eV and $m_a\gtrsim 4\times 10^{-10}$~eV in force/torque and frequency shift searches respectively, as presented in Sec.~\ref{sec:thermal_limits}).}
\label{tab:summary} 
\end{table*}

%%%%%%%%%%%%%%%%%%%%%%%%%%%%%%%%%%%%%%%%%%%%%%%%
\subsection{Axion-induced forces and torques from couplings to fermions}
\label{sec:derivativecoupling}
%%%%%%%%%%%%%%%%%%%%%%%%%%%%%%%%%%%%%%%%%%%%%%%%

%%%%%%%%%%%%%%%%%%%%%%%%%%%%%%%%%%%%%%%%%%%%%%%%
\subsubsection{\texorpdfstring{$-g_f \left( \vec{\nabla} a \cdot \vec{\sigma} \right)$}{gf}-induced effects -- ferromagnets}
\label{paragraph:EOM_study_first_term_derivative_coupling}
%%%%%%%%%%%%%%%%%%%%%%%%%%%%%%%%%%%%%%%%%%%%%%%%

\begin{figure}[htb!]
    \centering
\includegraphics[width=0.95\textwidth]{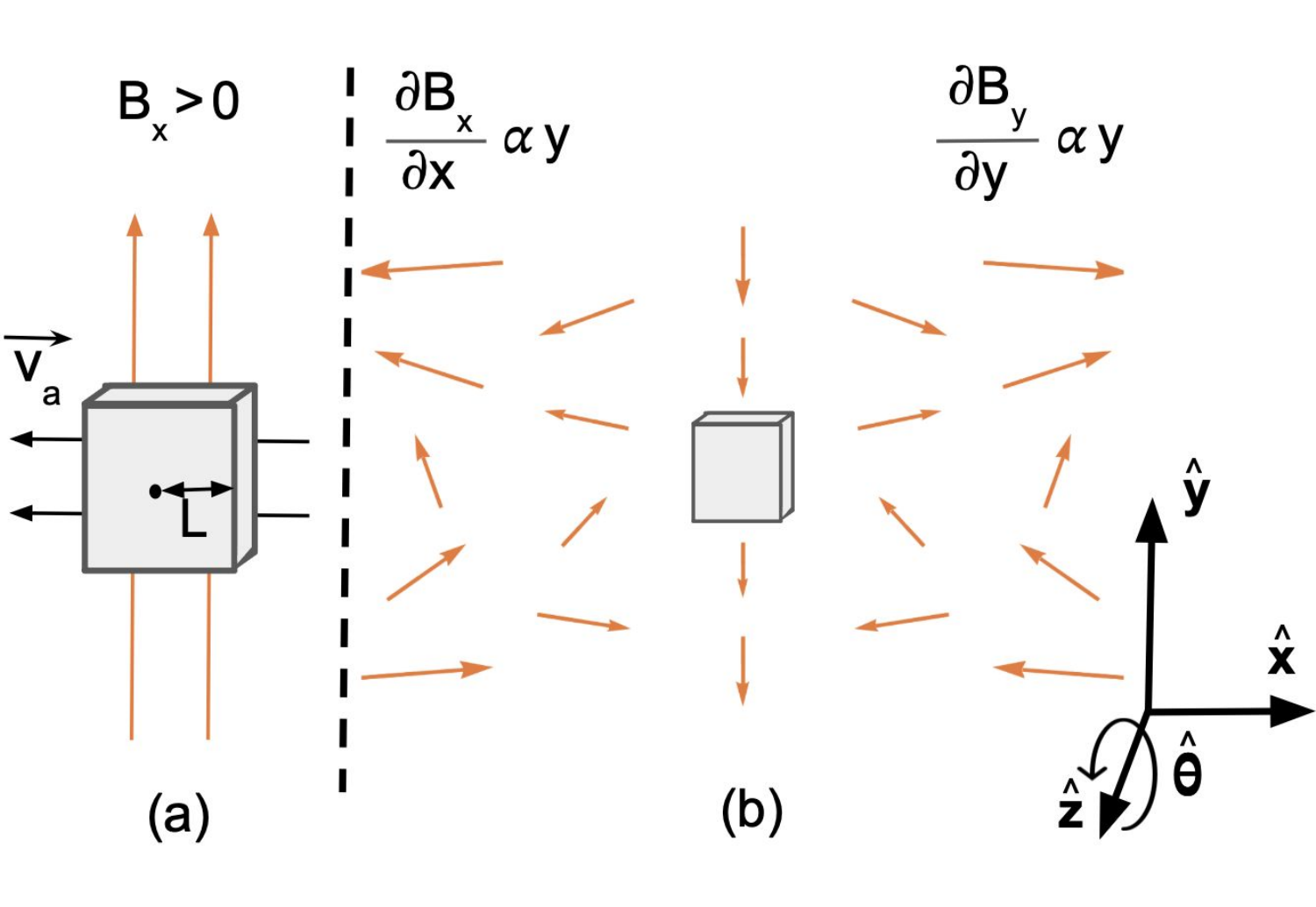}
    \caption{
    (\textbf{a}) Figure of the setup in the ferromagnetic case, where rotation angle $\theta$, which we assume is optically trapped, is expected to oscillate due to torque on the sensor to antialign with the axion wind in the $\vec{v}_a$-direction. The magnetization of the material is aligned with the external magnetic field (orange arrows). This orientation can be flipped at regular intervals by instantaneously flipping the sign of the magnetic field and simultaneously applying a rapid magnetic field burst to flip the magnetization of the sensor. Although we assume a $B$-field pointing purely in the $\mathbf{\hat{y}}$-axis in this study, a similar level of torque can be achieved for ferromagnets in the $y>0$ region of a sextupole setup shown in the following subfigure, albeit this simultaneously leads to translational forces. (\textbf{b}) Figure of the sensor placed at $(x,y)=(0,0)$ in the paramagnetic case, where we assume a sextupole magnetic field (gray arrows) and that the axion wind in the direction of $\vec{v}_a$ is aligned in the $\mathbf{\hat{x}}$-axis. When the $B$-field is oriented as in the figure, this configuration can be used for force sensing, both at the axion frequency when placed at $y\neq0$, or when heterodyning by driving the sensor to oscillate coherently along the $\mathbf{\hat{y}}$-axis between the $y<0$ and $y>0$ regions. When rotating the $B$-field by $\theta=\pi/2$ along the $x-y$ plane, this setup induces a frequency shift in the new $\mathbf{\hat{x}}$-axis.}
    \label{fig:setup}
\end{figure}

We first study the $-g_f \left( \vec{\nabla} a \cdot \vec{\sigma} \right)$ term from Eq.~\ref{eq:axionHint2} using a ferromagnetic material and focusing on the electron-axion couplings, $g_e$.  Here, the
axion field's gradient, the ``axion wind''~\cite{Graham:2013gfa}, mimics an external $B$-field.\footnote{We note that a very high axion-induced effective B-field above the material's magnetic coercivity could also cause the spins themselves to rotate internally relative to the atomic lattice, but we will assume sufficiently small axion couplings such that the torque on the material dominates.} Given the average velocity of Earth relative to the axion cloud is near the virial velocity of the axion, we take $\vec{v}_a$ to be fixed in the opposite direction to Earth's motion. We may therefore align the spin of our sensor relative to the axion velocity vector such that the polar angle $\theta_0$ between normalized polarization vector $\vec{\sigma}$ of the spins of the sensor and $\vec{v}_a$ around the $\hat{\mathbf{z}}$-axis is set to $\theta_0=\pi/2$ initially, as illustrated in Fig.~\ref{fig:setup}. This can be done by setting the harmonic oscillator potential to its minimum when the  $\vec{\sigma}$ spin vector initially lies along the $\hat{\mathbf{y}}$-axis which we set to the orientation where $\theta=0$, and by cooling the angular motion sufficiently to ensure that $\langle\theta^2\rangle\ll 1$. 

This relative angle with the axion wind causes torque in a ferromagnetic mechanical sensor around the $\mathbf{\hat{z}}$-axis due to its spins being rigidly aligned to the orientation of the lattice of spins, until the magnetization anti-aligns with the external magnetic field.  In the initial state, we note that $P_x(0)$ (the average spin of a single polarized electron projected along the $\hat{\mathbf{x}}$-axis) is then zero, while $P_y(0)>0$. In the initial state, we note that $\sigma_x(0)$ (the average spin of a single polarized electron projected along the $\hat{\mathbf{x}}$-axis) is then zero, while $\sigma_y(0)=\pm 1/2$. When the sensor rotates around the $\mathbf{\hat{z}}$-axis, the change in angular momentum is accompanied by a precessional gyroscopic rotation around the $\mathbf{\hat{x}}$-axis needed to preserve the total angular momentum. Reciprocally, torque in the angular equation of motion of rotation angle $\theta$ around the $\mathbf{\hat{z}}$-axis are partly canceled by  angular motion $\dot{\phi}$ around the $\mathbf{\hat{x}}$-axis. We can see this in the coupled rigid-body angular equations of motion,
%%%%%%%%%%%%%%%%%%%%%%%%%%%%%%%%%%%%%%%%%%%%%%%%
%%%%%%%%%%%%%%%%%%%%%%%%%%%%%%%%%%%%%%%%%%%%%%%%
%%%%%%%%%%%%%%%%%%%%%%%%%%%%%%%%%%%%%%%%%%%%%%%%
%%%%%%%%%%%%%%%%%%%%%%%%%%%%%%%%%%%%%%%%%%%%%%%%
\begin{align}
    \begin{cases}
        I_\phi\ddot{\phi} + I_\phi \gamma_\phi \dot{\phi} + I_\phi \Omega^2_\phi \phi + \dot{\theta}PN\sigma_y(t) = \tau_N^\phi(t) \\
        I_\theta\ddot{\theta} + I_\theta \gamma_\theta \dot{\theta} + I_\theta \Omega^2_\theta \theta - \dot{\phi}PN\sigma_x(t) + g_e P N m_av_a  a(t)\,\sigma_x(t) \sin \theta = \tau_N^\theta(t) \,\, ,
    \end{cases}
\end{align}
where we assume the potential's Hessian is diagonal, $(I_\phi,I_\theta)$ are the moments of inertia of the sensor around the $(\mathbf{\hat{z}},\mathbf{\hat{x}})$ axes respectively, $(\gamma_\theta,\gamma_\phi)$ and $(\Omega_\theta,\Omega_\phi)$ are the damping rates and the angular harmonic oscillator frequencies of $(\theta,\phi)$, $N$ is the number of electrons in the sensor, we used $\vec{\nabla} a = m_a \vec{v}_a a(t)$, and the projection of the polarization along the $\hat{\mathbf{x}}$-axis is given by $P\sin \theta $ in the small angle limit.  A similar discussion can be found in Ref.~\cite{Kalia:2024eml}.  The stochastic torque terms $(\tau_N^\theta(t),\tau_N^\phi(t))$ on the right hand side comes from fluctuations in the optical field reading out and possibly trapping the system, which limits the size of the detectable axion effect and will be discussed in Sec.~\ref{sec:force_noise_optimization}. We can approximate $N = m / (2 m_n)$, where $m$ is the mass of the sensor and $m_n \approx 0.94$~GeV is the approximate nucleon mass. We note that given the orientation of the magnetization, the axion torque only appears in the EOM of $\theta$, and that in the small $\phi$-angle approximation the EOM of the rotation around the $\mathbf{\hat{y}}$-axis is uncoupled to the other two such that we may ignore rotation around this axis in this work, and hence $(\theta,\phi)$ are equivalent to the spherical coordinates of Ref.~\cite{Kalia:2024eml}. 

As shown in Appendix~\ref{app:EOM_precession}, assuming a sensor whose angular motion is limited by quantum noise and the effect of axion couplings, the effect of the $g_e$ coupling in sufficiently small that the effect of the precession is below the detectable threshold and these equations of motion are approximately decoupled. Therefore we ignore the EOM of $\phi$ in the rest of this work and focus on the $\theta$ EOM approximated by
\begin{align}
\label{eq:torque_first_term}
I_\theta\ddot{\theta} + I_\theta \gamma_\theta \dot{\theta} + I_\theta \Omega^2_\theta \theta + g_e P N m_av_a  a(t)\,\sigma_x(t) \sin \theta = \tau_N^\theta(t) \ .
\end{align}
Assuming $\sigma_x$ is approximately constant in time, using $\langle\theta^2\rangle \ll 1$, this corresponds to a harmonic oscillator driven by an oscillating torque given by 
\begin{align}
    \tau_a(t) = g_e P N m_a v_a a(t)\sigma_x \ ,
    \label{eq:torquesignal}
\end{align}
whose amplitude $\tau_a(\omega_a)$ scales with the amplitude $a_0$ of the axion field. Since the maximum angle (away from $\theta=0$) of the motion is expected to be small, we use this amplitude for calculating the power spectrum of the axion-induced torque in this work. 

When the axion oscillations are much slower than that of the sensor, however, in particular looking at a frequency of motion where quantum noises is very low, as we will discuss in detail in Sec.~\ref{sec:force_noise_optimization}, an oscillating torque can also be obtained at the desired frequency by flipping the sign of the magnetization $\sigma_x$ in time such that $\sigma_x\rightarrow\sigma_x(t)$ flips sign at the frequency of the angular harmonic oscillator motion.

In this work, we assume the spins can be flipped in an external magnetic field that is larger than the material's magnetic coercivity, {\it e.g.}~using a conducting coil with an AC current like in modern electronics whose magnetic field is shaped to create a harmonic oscillator potential with sufficiently deep minima whose angle $\pm\theta_0$ with the axion wind alternates (flips) in time. 
For simplicity, we assume that the timescale of the magnetic field and magnetization flip is much shorter than the interval between flips, so that the magnetization can be treated as discontinuous jumps in time. In this limit, the magnetization is given by a Fourier series of discrete frequencies
$\Omega_n = (2n+1) \Omega_\theta$ for positive integers $n$, with amplitudes $A_n=\pi/4(2n+1)$. 

In practice, the spectral bandwidth at each frequency is not zero and is limited by the small (experimentally limited) bandwidth $\gamma_m$ which we assume is $\gamma_m\sim \gamma_\theta$. We also remark that the discrete structure of the spectrum remains valid even when the magnetization changes continuously over short time intervals (compared to the interval between flips), and we leave a precise determination of the power spectrum via an in-depth experimental analysis for future work. In this work we focus on detecting the first term in the Fourier series with frequency $\Omega_\theta$ ($n=0$), which contains the majority of the power, and such that the force in this case requires an additional factor of $4/\pi$.

Assuming sufficiently high bandwidth resolution, the oscillating axion field splits each spectral line into two peaks of equal amplitude at
\begin{align}
\Omega_\pm = \Omega_\theta \pm \omega_a \ .
\end{align}
This is the standard result obtained when \textit{heterodyning} a signal, which is a common technique in optics for detection of low frequency electromagnetic fields~\cite{BowenMilburn2016}, including those driven by axions~\cite{Berlin:2020vrk}. This contrasts the case when the axion Compton frequency already matches the frequency at which we wish to detect the signal (\textit{i.e.}~in the absence of magnetization flips), we will refer to the previous detection method as \textit{homodyne} detection.

%%%%%%%%%%%%%%%%%%%%%%%%%%%%%%%%%%%%%%%%%%%%%%%
\subsubsection{\texorpdfstring{$-g_f \left( \vec{\nabla} a \cdot \vec{\sigma} \right)$}{gf}-induced effects -- paramagnetic materials}\label{paragraph:EOM_study_first_term_derivative_coupling2}
%%%%%%%%%%%%%%%%%%%%%%%%%%%%%%%%%%%%%%%%%%%%%%%

We now turn to the case of materials with paramagnetic spins. Given the sensor's magnetization is unpolarized in the absence of a magnetic field and will follow an external static magnetic field's direction independently of the orientation of the material, a spatially homogeneous magnetic field is required to polarize the material. Although the $g_f$ coupling to the axion may lead to a torque on the spin of the sensor (\textit{e.g.}~as with CASPEr~\cite{JacksonKimball:2017elr}), this does not lead to observable effects in the motion of the sensor.
To obtain an oscillating force or torque on the sensor it is hence necessary to have an oscillating spatial (\textit{e.g.}~translational) gradient of the interaction Hamiltonian. Although the axion gradient fulfills this requirement, its effect carries another $m_a v_a$ suppression factor and can be subdominant.

In particular, the gradient of an external $B$-field can be the dominant contribution over the axion gradient if $\partial P/\partial x > P m_a v_a$, where we choose the gradient in $\mathbf{\hat{x}}$-axis without loss of generality. We note that constraints on magnetic field gradients lead us to study translational gradients only, rather than angular gradients. In order for the potential to see such a gradient however, it is crucial for the polarization along the $\mathbf{\hat{x}}$-direction, which we take to be that of the axion wind, see Fig.~\ref{fig:setup}, to change if the position of the sensor changes. 

In the case of nuclear magnetic dipole moments which are paramagnetically polarized, although low temperatures allow for higher spin polarizations if given time to align, significantly larger ($\mathcal{O}($s) or larger) relaxation times $T_1$ imply that the polarization cannot change on the time scales of the motion $\sim 1/\omega$. This implies that the \textit{perceived} potential gradient of the material is highly suppressed by an $\mathcal{O}(1/(\omega T_1)$ factor if this gradient were simply due to the gradient in the amplitude of the polarization (in 3D). However, if the direction of the magnetic field (instead of its amplitude) significantly change when the sensor moves in a specific direction, the magnetization vector of the sensor may rotate nearly instantaneously as long as the change in direction occurs at frequencies below the Larmor frequency of the material. 

In the presence of a large magnetic field, this frequency is typically quite high compared to the inverse of the relaxation time (often reaching $\mathcal{O}(100$~MHz) or more). Crucially, in such a magnetic field, since the axion gradient $\vec\nabla a$ is biased in the $\mathbf{\hat{x}}$-axis along which the axion ``wind'' is aligned, the axion effect is proportional to the spin polarization projected along the $\mathbf{\hat{x}}$-axis. Therefore, assuming the polarization projection along the $\mathbf{\hat{x}}$-axis is proportional to the position along the $\mathbf{\hat{x}}$-axis, the axion-induced potential naturally has a \textit{gradient} along the $\mathbf{\hat{x}}$-axis in which the material will feel a force. We note this does not violate the fact that static magnetic fields do no mechanical work: the mechanical work arises from the axion-induced interaction term, while the static $B$-field merely provides the spatial structure that converts coherent motion into a time‑varying effective coupling.

Assuming a magnetic field configuration that respects these constraints, we proceed to the discussion of the translational EOM of the sensor. In particular, we note that rotation in the sensor's magnetization vector leads to an equivalent effect in the rotational EOM of the sensor as for a rotating ferromagnet, where the magnetization angle is now given by the ratio $P_x/P_y$. As shown in Appendix~\ref{app:EOM_precession}, the effect of this angle in the angular EOMs is expected to be negligibly small (similarly to the ferromagnetic case) and is therefore not expected to lead to further observable effects on the sensor's motion. We thus focus on the EOM along the $\mathbf{\hat{x}}$-axis given by
\begin{align}
\label{eq:first_term_EOM_paramagnetic} m\ddot{x}+m\gamma\dot{x}+m\Omega^2x+g_fm_av_aa(t)\sigma_x(\vec{x}(t))\frac{\,\partial P(\vec{x}(t))}{\partial x}N &=F_N(t) \,\, ,
\end{align}
where $m$ is the total sensor mass, $\gamma$ and $\Omega$ are the damping rate and harmonic oscillator frequency of the (translational) motion along the $\mathbf{\hat{x}}$-axis, $f = e$, $p$ or $n$, and $F_N(t)$ is the optically induced force fluctuations on the system, analogously to $\tau_N(t)$ considered previously. Crucially here, the polarization is taken to be the projection of the polarization along the $\mathbf{\hat{x}}$-axis which oscillates freely in time if the angle of the polarization rotates while the polarization amplitude $\abs{P}$ remains (approximately) constant. 
This corresponds to a harmonic oscillator driven by an external force 
\begin{align}
\label{eq:force_paramagnetic} F_a(t)=g_fm_av_aa(t)\sigma_x(\vec{x}(t))\frac{\,\partial P(\vec{x}(t))}{\partial x}N \,\, ,
\end{align}
whose amplitude denoted $F_a(\omega_a)$ can be obtained by taking the amplitudes of the time-dependent axion field $a(t)$ and spin $\sigma_x(t)$, similarly to $\tau_a$.
We also note that the spin $\sigma_x(\vec{x})$ 
now depends on the sign of the external magnetic field component and explicitly depends on the position $\vec{x}$ of the sensor. Since the spin only takes discrete values $\pm 1/2$, we have left the spatial derivative to act only on the polarization $P$. In analogy with the ferromagnetic case, we will refer to the detection of this force as a homodyne  axion detection scheme.

From Eq.~\ref{eq:first_term_EOM_paramagnetic}, we see that if the axion oscillates too slowly compared to the material oscillation frequency at which we benefit from low noises, creating an oscillating force term is possible by implementing a more elaborate magnetic field configuration (and thus polarization). Using a configuration whose gradient along the $\mathbf{\hat{x}}$-axis changes sign when the sensor moves along an axis orthogonal to the $\mathbf{\hat{x}}$-axis. Although an extensive search for magnetic field configurations which best enhance axion detection is beyond the scope of this work, a simple 2D sextupole magnetic field configuration~\cite{wolski2011maxwell,marks1994conventional} given by 
\begin{align}
    \vec{B}(\vec{x})= B_0(xy,\frac{1}{2}(x^2-y^2),0)
\end{align}
meets such requirements. These are commonly used for control of charged particle beams~\cite{Gomberoff:2007byo, marks1994conventional, Prawanta:2019jwv, wolski2011maxwell}. We note that the magnetic field from a pair of parallel wires with opposing current is a viable alternative.

In such a magnetic field background, flipping the magnetic field around the $\mathbf{\hat{x}}$-axis such that $\vec{B}(y)\rightarrow \vec{B}(-y)$ for instance, analogously to the flipping of the magnetization of a ferromagnetic material, will generate an oscillating orthogonal motion in the $\mathbf{\hat{x}}$-axis. Another possibility, which we will consider throughout this work, is to drive motion of the sensor along the $\mathbf{\hat{y}}$-axis ({\it e.g.}~at $x=0$) such that the gradient $\partial P/\partial x$ also oscillates in time. Assuming the frequency of the harmonic oscillator along the $\mathbf{\hat{y}}$-axis is tuned to be $\Omega$ (the same as the resonant frequency in the $\mathbf{\hat{x}}$-axis), this achieves the same heterodyning effect as discussed above for ferromagnets, \textit{i.e.}~with an axion-induced force at frequencies $\Omega\pm\omega_a$. 

To obtain this oscillating behavior in the orthogonal direction we emphasize this requires the oscillator be in a \text{coherent} excited state which reproduces classical-like motion as a wavepacket instead of the sensor being in a single energy eigenstate of the harmonic oscillator along the $\mathbf{\hat{y}}$-axis. This can be done when the probability of the sensor being in any given energy eigenstate of the harmonic oscillator is given by a Poisson distribution whose mean state is determined by matching the associated variance of the sensor's oscillations with the desired classical amplitude. We note that in such states the standard deviation of the wavepacket $\sigma_x$ scales as the square root of the classical amplitude of the motion in the $\mathbf{\hat{y}}$-axis~\cite{10.1088/978-0-7503-2179-2ch5} such that for sufficiently large motion the assumption that a point-like sensor feels the magnetic field taken at the center of the wavepacket along the $\mathbf{\hat{y}}$-axis is a good approximation. We note that while in this section we remain relatively agnostic of the sensor trapping, levitation and driving mechanism, which we discuss in more detail in Sec.~\ref{sec:Noises}, we emphasize that driving can be achieved using oscillation of the trapping field, or via application of an additional homogeneous field (\textit{i.e.}~optical or magnetic) oscillating in time designed to apply an oscillatory force on the sensor in the $\mathbf{\hat{y}}$-axis.

%NOTE This is too in depth for this paper so removed: constant force may lead to offset equilibrium position that lead to non-linearities in PSD, however specially designed countering mechanical spring, or shorter integration times (among other methods), can work around this.}

%%%%%%%%%%%%%%%%%%%%%%%%%%%%%%%%%%%%%%%%%%%%%%%%
\subsubsection{\texorpdfstring{$ \dfrac{g_f}{m} \partial_t a \left( 
    \vec{p} \cdot \vec{\sigma}\right) $}{gf}-induced effects -- paramagnetic materials}
%%%%%%%%%%%%%%%%%%%%%%%%%%%%%%%%%%%%%%%%%%%%%%%%

Next, we consider the second term $\dfrac{g_f}{m} \partial_t a \left( \vec{p} \cdot \vec{\sigma} \right)$ from Eq.~\ref{eq:axionHint2} for paramagnetic materials. For an oscillator moving non-relativistically, the net momentum of the sensor can be approximated as $\vec{p}=m\vec{v}$. This term of the axion coupling leads to the following translational EOM along the $\mathbf{\hat{x}}$-axis
\begin{align}
m\ddot{x} + m \gamma\dot{x}+m\Omega^2x+ g_fm_a N \left(\frac{d}{dt}(a(t)\sigma_x(\vec{x}(t)) P(\vec{x}(t)))-\frac{\partial}{\partial x}(a(t)\vec{v}\cdot\vec{\sigma}(\vec{x}(t))P(\vec{x}(t)))\right) &= \nonumber \\ 
m\ddot{x}+m\gamma\dot{x}+m\Omega^2x+ g_fm_a N a(t) \left( m_a \sigma_x(\vec{x}(t))P(\vec{x}(t)) + \sigma_x (\vec{x}(t)) \frac{\partial P(\vec{x}(t))}{\partial t} + \right. &
\nonumber \\
\left. \sigma_x (\vec{x}(t))(\vec{v}\cdot\vec{\nabla})P(\vec{x}(t)) - \vec{v} \cdot \vec{\sigma}(\vec{x}(t))\frac{\partial P(\vec{x}(t))}{\partial x} \right) &=F_N(t) \ ,
    \label{eq:EOM_xdir}
\end{align}
where we used $d/dt=\partial/\partial t+(\vec{v}\cdot \vec{\nabla})$ in the second line. 

Assuming the same static external magnetic field as in Sec.~\ref{paragraph:EOM_study_first_term_derivative_coupling2}, we set 
$\partial_t \left( \sigma_x(\vec{x}) P(\vec{x}) \right) = 0$ at fixed $\vec{x}$, the time dependence of the polarization only arises through the spatial gradient of the polarization, \textit{i.e.}~the
$\sigma_x (\vec{x} (\vec{v}\cdot\vec{\nabla}) P(\vec{x})$ term.  We also assume that the spin is aligned along only one direction, either along the $\mathbf{\hat{x}}$-axis or the $\mathbf{\hat{y}}$-axis, and that its gradient along the $\mathbf{\hat{x}}$-axis is non-zero. Looking first at $\sigma_y\neq 0$, this simplifies the expression to 
\begin{align}
    m\ddot{x}+m\gamma\dot{x}+m\Omega^2x- g_fm_aa(t)Nv_y\sigma_y(\vec{x}(t))\frac{\partial P(\vec{x}(t))}{\partial x}=F_N(t) \ ,
\label{eq:EOMsigmay}
\end{align}
where in this case $P(\vec{x})$ represents the polarization associated with spins in the $\mathbf{\hat{y}}$-axis. Eq.~\ref{eq:EOMsigmay} hence represents a harmonic oscillator driven by an external force equal to 
\begin{align}
    F_a(t) = 
    g_f m_a a(t) N v_y \sigma_y(\vec{x}(t))\dfrac{\partial P(\vec{x}(t))}{\partial x} \ .
\label{eq:force1}
\end{align}%

Returning to Eq.~\ref{eq:EOM_xdir}, we now consider the case where $\sigma_x \neq 0$ (and $\sigma_y = \sigma_z = 0$) and assume the gradient of $P(\vec{x})$ along the $\mathbf{\hat{y}}$-axis is non-zero. Assuming $v_x\neq 0$ only leaves only terms which have an extra factor of $m_a\ll 1$. We therefore only discuss the case when $v_y\neq0$, which simplifies the expression to
\begin{align}\label{eq:third_term_EOM_ymotion}
    m\ddot{x}+m\gamma\dot{x}+m\Omega^2x- g_fm_a N a(t)v_y(t) \sigma_x(\vec{x}(t))\frac{\partial P(\vec{x}(t))}{\partial y} + \mathcal{O}(m_a^2)=F_N(t) \ .
\end{align}
We notice that the axion-induced force in both Eqs.~\ref{eq:third_term_EOM_ymotion}~\&~\ref{eq:EOMsigmay} contain a factor of $v_y$. These terms may be made to be non-zero and oscillating by driving coherent motion of the sensor along the $\mathbf{\hat{y}}$-axis, such that we could excite motion in the $\mathbf{\hat{x}}$-axis. However, the amplitude of the oscillating velocity one can expect from mechanical quantum sensor is typically restricted to be very small compared to the axion velocity ($v_a\sim10^{-3}$), {\it e.g.}~one can expect up to $v_y\lesssim 10^{-6}$ when taking $\Omega\lesssim 600$~kHz and an amplitude for the oscillating position $\langle y^2\rangle\sim 1~\mu$m. Thus we expect a suppression of the axion-induced forces in Eqs.~\ref{eq:third_term_EOM_ymotion}~\&~\ref{eq:EOMsigmay} compared to Eq.~\ref{eq:force_paramagnetic}, such that we can safely neglect the effects when studying of the phenomenology  of axions in mechanical quantum sensors. Note that due to the appearance of this factor we will also ignore these effects in Secs.~\ref{sec:axionpowerderivative}~\&~\ref{sec:peak_split_EDM} when discussing frequency shifts induced by axion couplings.

%%%%%%%%%%%%%%%%%%%%%%%%%%%%%%%%%%%%%%%%%%%%%%%
\subsubsection{Axion-induced power spectrum from couplings to fermions}
\label{sec:axionpowerderivative}
%%%%%%%%%%%%%%%%%%%%%%%%%%%%%%%%%%%%%%%%%%%%%%%
We will now estimate the axion-induced signal power spectrum in our mechanical sensor arising from the various cases shown in Secs.~\ref{paragraph:EOM_study_first_term_derivative_coupling2}~\&~\ref{paragraph:EOM_study_first_term_derivative_coupling}. By analyzing the signal power in frequency space for a time-dependent amplitude $x(t)$, where we integrate (measure) over a finite time $T_{\text{int}}$, we can write
\begin{align}
    S_{xx}^a(\omega) =
    \dfrac{1}{T_{\text{int}}}\int_{T_\text{int}} \langle x(t)x(0) \rangle e^{-i\omega t}dt  \ ,
\end{align}
We note $S_{xx}^a(\omega)$ is approximately symmetric around $\omega_a$.
The natural bandwidth of the axion power spectrum around the Compton frequency is typically well approximated by $\delta \omega_a \sim m_a v_a^2$~\cite{Hui:2021tkt,Marsh:2015xka}, but may also depend on whether the signal is being driven (heterodyned), since heterodyning mixes the axion field with the driven oscillator and introduces the damping rate $\gamma$ into the effective bandwidth. Although in the absence of heterodyning the axion signal bandwidth is well approximated by $ m_a v_a^2 \approx 10^{-6} m_a$, in the heterodyne case each peak is broadened by both the axion bandwidth and by the bandwidth of the magnetization flips $\gamma_m$ such that the resulting power spectrum is dominated by the larger of the two bandwidths. 

We also note that the driving of the sensor's motion leads to an oscillating gradient (force) in the interaction potential $-\vec{\mu}\cdot\vec{B}$, which is centered at $\Omega$ rather than $\Omega\pm\omega_a$. The expected damping rate (spectral width) of this force is approximated by the largest contribution among the damping rate of the motion in the $\mathbf{\hat{y}}$-axis (which we take to be $\gamma$, the natural coherent wavepacket width of the sensor along the $\mathbf{\hat{y}}$-axis relative to the total amplitude of the oscillation (which is approximated by $(\sigma_y/y_0)\Omega_y\sim 10^{-5}\Omega_y$)~\cite{Hamaide:2025buy}, and the width due to to the longitudinal relaxation time $1/T_1\sim 10^{-3}$~Hz of the material as it oscillates in a quadratic background magnetic field. Although a discussion of the limitation of the latter effect is for a more detailed study of experimental constraints, it is expected to lead the power spectrum of this force to overlap with the axion-induced power spectrum for $m_a\lesssim 10^{-18}$~eV, below which switching the magnetic field around the $\mathbf{\hat{x}}$-axis is likely preferable assuming a high quality square wave magnetic field profile in time may become necessary.

Throughout this work we will assume $\gamma \sim \gamma_\theta \sim 1.5\times 10^{-5}$~Hz independently of $\Omega$ or $\Omega_\theta$. Further assuming $\gamma_m \sim \gamma$, the total axion-induced torque or force power spectrum bandwidth can thus be written in general as $1 / T_\text{coh} = \max (\gamma, m_a v_a^2)$, where $T_\text{coh}$ denotes the axion \textit{signal} coherence time, which is not always the axion field coherence time $T_\text{coh,a}$. In particular, 
for $\gamma \sim 1.5\times 10^{-5}$~Hz, the signal bandwidth corresponds to $\gamma$ for all $m_a\lesssim 10^{-14}$~eV, coincident with the axion mass region in which we wish to heterodyne the axion signal, as we will justify in Sec.~\ref{sec:force_noise_optimization}. Finally, we approximate the signal power spectrum by a top-hat function given by
\begin{align}
\label{eq:signal_fermion_coupling}
    S_{\tau\tau}^a(\omega)\approx  \frac{\tau_a^2(\omega_a)}{\max(\gamma,m_av_a^2)}%= 10^{-9}~\text{N}^2\cdot\text{m}^2/\text{Hz}~\left(\frac{g}{\text{eV}^{-1}}\right)^2\left(\frac{\omega_a}{1~\text{kHz}}\right)^{-1} %\left(\frac{T_\text{int}}{1~\text{year}}\right) 
    \,\, ,
\end{align} 
in the frequency range $\omega\in[m_a-1/2T_\text{coh},m_a+1/2T_\text{coh}]$, and zero otherwise.

Moreover, we note the power spectrum can be analogously obtained for the case of paramagnetic materials experiencing an axion-induced force, which is denoted by $S_\text{FF}^a(\omega)$. Indeed, $S_\text{FF}^a(\omega)$ can be obtained similarly to $S_{\tau\tau}^a(\omega)$ by replacing $\tau_a$ with $F_a$ from Eq.~\ref{eq:force1}, giving
\begin{align}
S_{\text{FF}}^a (\omega) \approx \dfrac{F_a^2 (\omega_a)}{\text{max}(\gamma, m_a v_a^2)} \ ,
\label{eq:SFFaderivative}
\end{align}
over the same frequency range as $S_{\tau\tau}^a$, and zero otherwise. Note the bandwidth in the denominator is the same as in $S_{\tau\tau}^a(\omega)$, because, below the relevant axion-mass threshold, heterodyning introduces the sensor's damping rate $\gamma$ which dominates in the same axion mass range as for the ferromagnetic case.

%%%%%%%%%%%%%%%%%%%%%%%%%%%%%%%%%%%%%%%%%%%%%%%%%%%
\subsection{Axion-induced peak splitting from couplings to fermions}
\label{sec:axionpeakderivative}
%%%%%%%%%%%%%%%%%%%%%%%%%%%%%%%%%%%%%%%%%%%%%%%%%%%

%%%%%%%%%%%%%%%%%%%%%%%%%%%%%%%%%%%%%%%%%%%%%%%%%
\subsubsection{\texorpdfstring{$-g_f \left( \vec{\nabla} a \cdot \vec{\sigma} \right)$}{gf}-induced effects -- ferromagnets}\label{sec:axionpeakderivative_ferro}
%%%%%%%%%%%%%%%%%%%%%%%%%%%%%%%%%%%%%%%%%%%%%%%%%
We now discuss the possibility of detecting a shift in the frequency of the harmonic oscillator induced by the axion-electron coupling. Considering the torque term of Eq.~\ref{eq:torque_first_term}, if we instead re-orient our harmonic oscillator such that the preferred angle of the magnetization ({\it i.e.}~angle at the bottom of the harmonic oscillator potential) relative to the axion wind is  $\theta_0 = 0$, the EOM for motion around this new mean orientation can be approximated as 
\begin{align}\label{eq:EOM_freq_ferro_2}
I \ddot{\theta} + I \gamma_\theta \dot{\theta} + I\Omega^2_{\theta} \left( 1 + \dfrac{g_e m_a v_a a(t) N \sigma_x P}{I \Omega^2_{\theta} } \right) \theta \approx \tau_N(t) \ ,
\end{align}
where $\sigma_x$ notably no longer depends on time compared to the torque inducing ferromagnetic scenario, and we assumed $\theta$ was small such that $\sin\theta\approx\theta$. Eq.~\ref{eq:EOM_freq_ferro_2} shows that the sensor's harmonic oscillator frequency is shifted according to $\Omega_\theta^2 \rightarrow \Omega_\theta^2 + \delta \Omega_\theta^2(t)$, where
\begin{align}\label{eq:gf_ferro_peak_split}
    \delta \Omega_\theta^2 (t) = 
    \dfrac{g_e m_a v_a a(t) N \sigma_x P}{I} \ .
\end{align}
As we will discuss in Sec.~\ref{sec:Noises}, if the amplitude of the frequency shift $\delta \Omega_\theta^2$ is sufficiently large, it can be observable via the angular power spectrum of the sensor.

%%%%%%%%%%%%%%%%%%%%%%%%%%%%%%%%%%%%%%%%%%%%%%%%%
\subsubsection{\texorpdfstring{$-g_f \left( \vec{\nabla} a \cdot \vec{\sigma} \right)$}{gf}-induced effects -- paramagnetic materials}\label{sec:axionpeakderivative_diapara}
%%%%%%%%%%%%%%%%%%%%%%%%%%%%%%%%%%%%%%%%%%%%%%%%%
Next, we consider frequency shifts that could be induced in the case of a paramagnetic material.  To see such an effect, we assume the polarization gradient has a constant second derivative over a distance $\sim 1\,\mu$m where the sensor is trapped.  We note this can be achieved by rotating the $B$-field shown in Fig.~\ref{fig:setup} by an angle $\theta=\pi/2$ such that the constant second derivative would be align in the direction of the axion wind, \textit{i.e.}~in the $\mathbf{\hat{x}}$-axis. In order for the polarization along the $\mathbf{\hat{x}}$-axis to change when the sensor moves along the $\mathbf{\hat{x}}$-axis, we further assume a large external homogeneous magnetic field component $B_1=10$~T along the $\mathbf{\hat{y}}$-axis, such that the magnetic field is now described on the surface $y=0$ by
\begin{align}\label{eq:B_field_freq_shift}
    \vec{B}=(-B_0x^2/2,B_1,0)\,\, .
\end{align}
The component of the polarization of the sensor along the $\mathbf{\hat{x}}$-axis is then estimated to first order in $x$ as 
\begin{align}\label{eq:polarization_x_freq_shift}
    P\approx \tanh(\mu_fB_1/k_BT)\frac{B_0x^2}{2B_1} \,\, ,
\end{align}
where the fraction represents the angle of the polarization vector in the $x-y$ plane. Starting from Eq.~\ref{eq:first_term_EOM_paramagnetic}, we can then write $\partial P/\partial x\approx (\partial^2 P/\partial x^2)x\propto B_0x$ which yields
\begin{align}
    m\ddot{x} + m\gamma\dot{x} + m\Omega^2 \left(1 + \frac{g_f m_a v_a a(t) N \sigma_x (\vec{x}) \dfrac{\partial^2  P(\vec{x}(t))}{\partial x^2}}{m\Omega^2}\right) x &=\tau_N(t) \ .
\label{eq:freqshiftdiapara}
\end{align}
We see that the axion coupling is thus equivalent to an additional spring constant on the sensor, which in turn leads to a time-dependent axion-induced frequency shift of $\Omega^2 \rightarrow  \Omega^2 + \delta \Omega^2(t)$ where 
\begin{align}
\delta\Omega^2(t) = \dfrac{g_f}{m} m_a v_a a(t) N \sigma_x (\vec{x}) \dfrac{\partial^2 P(\vec{x})}{\partial x^2} \ .
\label{eq:gf_peak_split}
\end{align}
We also remark that near the origin, the quadratic magnetic field configuration coupled to the spins may induce a frequency shift even in the absence of the axion due to the paramagnetic interaction potential $-P N \,(e\vec{\sigma}/m_f)\cdot\vec{B}$ where $\abs{P}\propto \abs{B}$ (this is similar to how charge neutral particles are trapped magnetically or electrically~\cite{Millen_2020,Gov:1999fa}). This effect, however, would not oscillate coherently in time as an axion signal, and we postpone the discussion its effect on the detectability of axion couplings in Sec.~\ref{sec:optimization_freq_shift}.

Finally, we note that the terms in Eq.~\ref{eq:axionHint2} which are $\mathcal{O}(p^2/m_f^2)$ may also lead to observable effects. However, although these effects are expected to be able to contribute to the kinetic energy of the sensor, which can lead to frequency (and damping rate) shifts in the EOM, such equations of motion are also expected to scale as $v^2$, where $v$ is the velocity of the sensor in any direction where this effect may appear. Thus based on the relative suppression we obtained for forces from $\mathcal{O}(p/m)\sim v$ or higher terms in Eq.~\ref{eq:axionHint} which led us to ignore such terms, higher order terms are expected to be further suppressed and we therefore ignore them in this work.

%%%%%%%%%%%%%%%%%%%%%%%%%%%%%%%%%%%%%%%%%%%%%%%%%%
\subsection{Axion-induced forces from axion-EDM coupling}\label{sec:EDMcoupling}
%%%%%%%%%%%%%%%%%%%%%%%%%%%%%%%%%%%%%%%%%%%%%%%%%%

%%%%%%%%%%%%%%%%%%%%%%%%%%%%%%%%%%%%%%%%%%%%%%%%
\subsubsection{\texorpdfstring{$g_d\,a(t)\,\vec{\sigma}\cdot\vec{E}$}{gd with E-field}-induced effects}\label{sec:force_EDMcoupling}
%%%%%%%%%%%%%%%%%%%%%%%%%%%%%%%%%%%%%%%%%%%%%%%%
We now discuss the detectability of the coupling using the $\vec{\sigma} \cdot \vec{E}$ effect from Eq.~\ref{eq:axionHint2}. Compared to terms in the axion-fermion couplings studied in Sec.~\ref{sec:derivativecoupling}, this term is neither suppressed by the ultralight axion's mass nor by the velocity of the sensor harmonic oscillator. We note the $-g_d a \vec{\sigma} \cdot \vec{E}$ term of Eq.~\ref{eq:axionHint2} resembles the $-g_f \vec{\nabla} a \cdot \vec{\sigma}$ term via replacing the axion gradient with the $\vec{E}$-field, as long as we focus on nucleons. The translational EOM of the sensor along in the $\hat{\mathbf{x}}$-axis is then given by
\begin{align}
    m\ddot{x}+m\gamma\dot{x}+m\Omega^2x+g_d\, N \frac{\partial}{\partial x}\left(a(\vec{x},t)P(\vec{x})\vec{\sigma}(\vec{x})\cdot \vec{E}\right) = F_N(t) \ .
\end{align}
Similarly to the axion-fermion couplings with paramagnets, we assume the effect of a rotation of the polarization vector on the angular EOMs when moving in the $\mathbf{\hat{x}}$-axis is sufficiently small to be neglected (see App.~\ref{app:EOM_precession} for details). Unlike the axion-fermion couplings however, this direction does not need to be aligned with the axion wind. We assume instead the direction of the spin and $E$-field are aligned to maximize the effect, which we therefore choose to be in the $\hat{\mathbf{x}}$-axis without loss of generality.  By inspection when expanding the partial derivative the term becomes
\begin{align}
    g_dN\,\frac{\partial}{\partial x}(a(x,t)\,P(\vec{x})\,\sigma_x E_x) = g_d N \left( \sigma_x E_x \, P(\vec{x}) \, \frac{\partial}{\partial x}a(\vec{x},t) + a(t) \sigma_x \, E_x \, \frac{\partial}{\partial x}P(\vec{x}) + a(t) P(\vec{x}) \, \sigma_x \, \frac{\partial}{\partial x}E_x \right) \ ,
    \label{eq:partialexpansion}
\end{align}
where we included here the axion's gradient term for comparison with the other terms.  We see that there are a few different possibilities to create a force on the sensor. In the first term of the right hand side of Eq.~\ref{eq:partialexpansion}, the oscillation of the axion field contributes a gradient $m_a v_a$, which can further oscillate at frequencies near the oscillator's mechanical resonance. We now examine the last two terms and see how to improve on the amplitude of the axion gradient-induced signal with large gradients of polarizations or gradients of electric fields.

Assuming nuclear ferromagnets are not accessible for our experiment (see {\it e.g.}~\cite{Saunders_2018} for further discussion of nuclear magnetism), we focus on ``NMR active" materials ({\it i.e.}~nuclear paramagnets) with a polarization per proton $P$. The polarization per proton for such magnetism can be estimated as $P=\kappa\tanh(\mu_n B/k_BT)\sim \kappa\mu_n B/k_BT$ in the regime where $\mu_n B\lesssim k_BT$, where $\mu_n$ is the nuclear magneton and $\kappa$ is the dilution factor of the material representing the portion of polarizable nuclear spins.  At its largest, when keeping the oscillator at ultralow cryogenic temperatures in large external magnetic fields, a polarization of $P\sim \kappa \sim 10^{-3}$ is expected ({\it e.g.}~with ferroelectric materials such as those used in the CASPEr  experiment~\cite{JacksonKimball:2017elr}).

We first consider maximizing the gradient of the proton spin polarization, \textit{i.e.}~$\partial_x P(\vec{x})$ in Eq.~\ref{eq:partialexpansion}. Such a gradient can be used either for heterodyning at low frequencies or homodyning by employing a static polarization gradient and relying solely on the axion field’s intrinsic oscillations.  In the homodyning case, a static 
$B$-field with a large spatial gradient along the $\hat{\mathbf{x}}$-axis suffices, similarly to the case of the axion-fermion couplings in Sec.~\ref{paragraph:EOM_study_first_term_derivative_coupling2}.

In the heterodyning case, we may use a $B$-field which produces a polarization gradient $\partial P(\vec{x})/\partial x$ which depends on the position in an orthogonal axis to the $\mathbf{\hat{x}}$-axis\footnote{We note that a configuration which presents a varying polarization gradient in the $\mathbf{\hat{x}}$ may in principle have a similar effect, however driving motion in the $\mathbf{\hat{x}}$-axis is undesirable, particularly when consider detecting axions at the SQL on-resonance, see \textit{e.g.}~\cite{Aspelmeyer:2013lha,Beckey:2023shi} and references therein for further discussion.}, again similarly to Sec.~\ref{paragraph:EOM_study_first_term_derivative_coupling2}. Coherent motion of the sensor along the $\mathbf{\hat{y}}$-axis can then generate an effective oscillating polarization gradient and force on the sensor. This requires a nonzero mixed derivative of the polarization, $\partial^2 P(\vec{x}) / \partial x \partial y \neq 0$, where the polarized spins can be chosen to point in  $\sigma_x$. Taking the motion of the sensor to be $y_0 \cos ( \Omega t)$, where $y_0 \approx 1$~eV$^{-1}$, and recalling that the polarization is proportional to $\partial^2 B / \partial x \partial y$, we assume a sufficiently sharp microscopic gradient can be engineered that exceeds the axion gradient $m_a v_a \ll 1$~eV for now, and explicit demonstrate this in Sec.~\ref{sec:Noises}.

In both the heterodyning and homodyning detection scenarios, we may either assume a static homogeneous electric field $\vec{E} = E\,\mathbf{\hat{x}}$, like those felt by protons inside a ferroelectric material, or an external electric field (either constant or oscillating). Not only are many ferroelectric materials naturally NMR-active with relatively low absorption of light, these materials possess very large internal electric fields which are typically larger than experimentally achievable external electric fields. This applies both to $E$-fields which are static and those which oscillate at the frequency of the sensor, despite the latter naturally allowing for heterodyne detection\footnote{Indeed, even when taking a $I = 1$~kW/m$^2$ low frequency radio source (as recently demonstrated in~\cite{nano15141062}), 
not only the resulting amplitude of the electric field is roughly two orders of magnitude smaller than the internal electric fields available in a ferroelectric material, but the gradient is given by $\Omega_E E_0 \ll 1$~eV$^3$, which is further suppressed compared to the possible spatial gradients of polarizations.} In this work, we assume the same ferroelectric material as used by CASPEr-Electric~\cite{JacksonKimball:2017elr}, and a leave a more extensive search among possible materials to future work. We also note Schiff screening suppresses the internal electric field at leading order, leaving only the Schiff moment, which is proportional to the applied external field and reduced by a factor $\epsilon \sim 10^{-2}$~\cite{JacksonKimball:2017elr,Mukhamedjanov:2004ki}. Nevertheless, we expect that, although substantial, Schiff screening only moderates the large ferroelectric material electric fields and the polarization gradient can still parametrically exceed the axion gradient.

We therefore expect to achieve larger polarization gradients by using ferroelectrics and focusing on maximizing $\partial_x P(\vec{x})$, and we will ignore any contributions from electric field or axion gradients.
Using the polarization gradient term $\partial_x P(\vec{x})$ for both the heterodyne and homodyne cases, 
we estimate the axion-induced oscillating force as
\begin{align}\label{eq:EDM_force}
    F_a (t) = \dfrac{g_d}{2} N a(t)\sigma_x(t) E_x \dfrac{\partial P(\vec{x}(t))}{\partial x} \ ,
\end{align}
whose amplitude $F_a(\omega_a)$ is obtained by taking the amplitude of the oscillating axion field (and polarization in the heterodyne case). Thus, in analogy with Eq.~\ref{eq:SFFaderivative}, the signal power spectrum near the signal's peak can be estimated as 
\begin{align}
S_\text{FF}^a(\omega) =  \frac{F_a^2(\omega_a)}{\max(\gamma,m_av_a^2)} \ ,
\end{align} 
for $\omega \in [m_a-1/2T_\text{coh}, \, m_a+1/2T_\text{coh}]$ and zero otherwise.

%%%%%%%%%%%%%%%%%%%%%%%%%%%%%%%%%%%%%%%%%%%%%%%%
\subsubsection{
\texorpdfstring{$g_d \, \dfrac{a}{m_f} \, \vec{\sigma} \cdot ( \vec{B} \times \vec{p} )$}{gd with B-field}-induced effects}
%%%%%%%%%%%%%%%%%%%%%%%%%%%%%%%%%%%%%%%%%%%%%%%%
 
We briefly discuss the phenomenology of the last term in Eq.~\ref{eq:axionHint2}. For comparison with the first term, we may assume attainable external magnetic field gradients of the same size to obtain large polarization gradients. However the electric field inside a ferroelectric is expected to be larger than  readily attainable external magnetic fields. Therefore, despite the simplicity of not requiring an electric field at all to probe this term, the term is expected to be relatively suppressed compared to the first term studied in Sec.~\ref{sec:force_EDMcoupling}. Further the velocity of the sensor which appears in the non-relativistic approximation due to the presence of $\vec{p}/m_f\approx \vec{v}\ll 1$, is expected to lead to further suppression for non-relativistic oscillators. Thus, although the effect in the second term may lead to unique effects harnessable by mechanical quantum sensors (unlike non-solid state sensors like CASPEr~\cite{JacksonKimball:2017elr} where velocities are zero), these effects remain suppressed in mechanical quantum sensors.

%%%%%%%%%%%%%%%%%%%%%%%%%%%%%%%%%%%%%%%%%%%%%%%%%%
\subsection{Axion-induced peak splitting from axion-EDM coupling}\label{sec:peak_split_EDM}
%%%%%%%%%%%%%%%%%%%%%%%%%%%%%%%%%%%%%%%%%%%%%%%%%%

In analogy with Sec.~\ref{sec:axionpeakderivative}, where we showed that the axion-fermion coupling in paramagnets could lead to a frequency shift an appropriate orientation of the magnetic field, we may look at the effect of the axion-EDM coupling on the frequency of the sensor using a specially chosen paramagnetic material with polarizable protons and/or neutrons. Focusing on the $g_da(t)\vec{\sigma} \cdot \vec{E}$ term in Eq.~\ref{eq:axionHint2}, and assuming the polarization  $P(\vec{x})$ is quadratic along the $\mathbf{\hat{x}}$-axis around $x = 0$ (without loss of generality) using the same quadratic magnetic field configuration as in Sec.~\ref{sec:axionpeakderivative}, this leads to a similar effect, \textit{i.e.}~rescaled harmonic oscillator frequency as seen in the equation of motion
\begin{align}
m\ddot{x} + m\gamma\dot{x} + m\Omega^2 x + g_d a(t)\sigma_x(\vec{x}(t))N \dfrac{\partial^2 P(\vec{x}(t))}{\partial x^2} \, E \, x =F_N(t) \ ,
\end{align}
where we again assume $\partial^2 P(\vec{x})/\partial x^2$ is constant along the $\mathbf{\hat{x}}$-axis over the length of the harmonic oscillator potential. This leads to a similar frequency shift as discussed in Eq.~\ref{eq:gf_peak_split}, where  $\Omega^2\rightarrow \Omega^2 + \delta \Omega^2(t)$, with 
\begin{align}\label{eq:gd_peak_split}
\delta \Omega^2 (t) 
= \dfrac{g_d}{m} a(t) \sigma_x (\vec{x}(t)) N \frac{\partial^2 P(\vec{x}(t))}{\partial x^2} E
    \ ,
\end{align}
which may be visible in the power spectrum of the motion of the sensor, as we will discuss in Sec.~\ref{sec:Noises}.

%%%%%%%%%%%%%%%%%%%%%%%%%%%%%%%%%%%%%%%%%%%%%%%%%%
%%%%%%%%%%%%%%%%%%%%%%%%%%%%%%%%%%%%%%%%%%%%%%%%%%
 \section{Searching for axions with mechanical quantum sensors}\label{sec:Noises}
%%%%%%%%%%%%%%%%%%%%%%%%%%%%%%%%%%%%%%%%%%%%%%%%%%
%%%%%%%%%%%%%%%%%%%%%%%%%%%%%%%%%%%%%%%%%%%%%%%%%%

We now discuss the detectability prospects of axion-induced effects studied the previous section in mechanical quantum sensors.  We study both levitated and suspended mechanical sensors that can work at the SQL for a range of resonant frequencies, and which are expected to feel the effect of the axion interactions through homodyne and heterodyne force sensing as well as frequency shift sensing.

%%%%%%%%%%%%%%%%%%%%%%%%%%%%%%%%%%%%%%%%%%%%%%%%%%
\subsection{Optimizing the search strategy for forces and torques in optomechanical sensors}
%%%%%%%%%%%%%%%%%%%%%%%%%%%%%%%%%%%%%%%%%%%%%%%%%%

%%%%%%%%%%%%%%%%%%%%%%%%%%%%%%%%%%%%%%%%%%%%%%%%%%
\subsubsection{Force and torque noises in optomechanical sensors}\label{sec:force_noise_optimization}
%%%%%%%%%%%%%%%%%%%%%%%%%%%%%%%%%%%%%%%%%%%%%%%%%%
When considering the range of possible axion-induced effects discussed in Sec.~\ref{sec:dynamics}, our first practical concern must be the huge range of possible axion masses, ranging from $\mathcal{O}(10^{-20}-10^{0})$~eV that can be predicted in vanilla axion models.  In addition, axions outside the QCD axion band can also serve as possible solutions to the strong CP problem, as demonstrated in recent non-vanilla axion model constructions~\cite{Kivel:2022emq, Elahi:2023vhu}, which expands the range of axion masses and also implicitly the range of viable $g_f$ and $g_d$ couplings from Eq.~\ref{eq:axionLagrangian}. Typically, exquisite sensitivity to a given axion mass and coupling involves a corresponding tradeoff in broad coverage, which necessitates retuning the experimental setup and extending the runtime to probe new parameter space.  Importantly, we show that naturally broadband techniques such as heterodyning and frequency shifts whose sensitivity does not rely on precise matching to the SQL at the axion frequency allow us to dramatically extend the typical range expected of mechanical quantum sensors.

To estimate the detectability of the axion-induced signals described in Sec.~\ref{sec:derivativecoupling} and Sec.~\ref{sec:EDMcoupling}, 
we compare the corresponding torque and force spectra to the quantum noise spectrum of an (opto)mechanical resonator, whose position is read out via an optical field.  For our study, we will define the sensitivity reach  as the region of parameter space for which the signal-to-noise power ratio (SNR) satisfies SNR~$\gtrsim$~1, and leave a more detailed likelihood analysis for future work. Maximizing sensitivity hence corresponds to maximizing the SNR for any given coupling and axion mass. The SNR for torques and forces of a single frequency bin $\delta\omega$ are defined by
\begin{align}\label{eq:SNR_torque}
\text{SNR}_\tau(\omega_a, T_\text{coh}) &\equiv \frac{\sqrt{\int_{\delta\omega} S_{\tau\tau}^a(\omega) d\omega}}{\sqrt{\int_{\delta\omega} S_{\tau\tau}^N(\omega) d\omega}} \simeq \frac{\tau_a(\omega_a)}{\sqrt{S_{\tau\tau}^N(\omega_a)/T_\text{coh}}} \ , 
\end{align}
and 
\begin{align}\label{eq:SNR_force}
\text{SNR}_\text{F}(\omega_a, T_\text{coh}) &\equiv \frac{\sqrt{\int_{\delta\omega} S_\text{FF}^a(\omega) d\omega}}{\sqrt{\int_{\delta\omega} S_\text{FF}^N(\omega) d\omega}} \simeq \frac{\text{F}_a(\omega_a)}{\sqrt{S_{\text{FF}}^N(\omega_a)/T_\text{coh}}} \ ,
\end{align}
where $S_{\tau \tau}^N(\omega)$ and $S_\text{FF}^N (\omega)$ are the torque and force noise power spectra of our sensor, which we will assume are approximately constant (white noise) over the relatively small integration region $\delta \omega$ centered around the dark matter Compton frequency $\omega_a$. In the second equalities, we took $\delta \omega = 1/T_\text{int}$, the Fourier-limited resolution of a single integration time $T_\text{int}$, and we further assumed $\delta \omega \leq 1/T_\text{coh} = \max(\gamma, m_a v_a^2)$. Defining $N_T \equiv T_{\text{int}} / T_\text{coh} > 1$ as the number of coherence intervals within the integration time, and noting that the axion signal power is constant within $[\omega_a -1/2T_\text{coh}, \omega_a +1/2T_\text{coh}]$, the single bin SNR is independent of $N_T$ for any value $N_T > 1$.

To fully account for the effect of integration time which permit independent measurements of the SNR in many frequency bins, we begin by fixing $N_T = 50$ at all frequencies, which corresponds to about 10 months of total integration time over the entire frequency range over which we will search for axions and whose upper and lower limits we will discuss shortly. For an oscillator operated at SQL at $\omega = \omega_a$ and with sufficiently long integration time to ensure $N_T = 50$ at $\omega_a$, the resulting spectral resolution $\delta \omega \approx 1 / T_{\text{int}}$ is much narrower than the width of the axion signal bandwidth. As discussed in Refs.~\cite{Carney:2019cio,Arvanitaki:2014faa,Manley:2020mjq,Budker:2013hfa,Chaudhuri:2014dla,Hamaide:2025buy}, this leads to an enhancement of the sensitivity reach via an incoherent quadrature sum over the individual SNR contributions in each bin, producing an \textit{effective} SNR well-approximated by
\begin{align}
\text{SNR}_\tau^{\text{eff}}(\omega_a, T_\text{coh}) = \text{SNR}_\tau(\omega_a, \sqrt{N_T} T_\text{coh}) \ , \quad
\text{SNR}_\text{F}^{\text{eff}}(\omega_a, T_\text{coh}) = \text{SNR}_\text{F}(\omega_a, \sqrt{N_T} T_\text{coh}) \ .
    %T_\text{coh} \rightarrow \sqrt{N_T} T_\text{coh} \ .
\end{align}
This improves the original \textit{single bin} SNR in Eqs.~\ref{eq:SNR_torque}~and~\ref{eq:SNR_force} by a factor of $N_T^{1/4}\approx 2.7$, and this approach is equivalent to the frequentist likelihood-based treatment described \textit{e.g.}~in Ref.~\cite{Berlin:2020vrk}.

We now distinguish what type of mechanical quantum sensor may be best suited to the detection of forces and torques, in particular we focus on comparing qualities of optically trapped and mechanically trapped configurations, as this distinction will affect the achievable noise of the sensor. We also emphasize that in this work we assume a macroscopic sensor such that optical trapping can in principle be achieved using radiation pressure as was studied in the MOLeQuTE proposal~\cite{Hamaide:2025buy}. Indeed, as discussed {\it e.g.}~in Ref.~\cite{Aspelmeyer:2013lha,Hamaide:2025buy},most state-of-the-art mechanically trapped mechanical resonators ({\it e.g.}~pendula, Weber bars, membranes) can operate in two principal modes. In the first, the mechanical frequency is effectively fixed, such that the readout laser power can be tuned independently of the mechanical frequency to allow the system to reach the SQL noise floor at any desired frequency, including at the mechanical frequency which constitutes 'on-resonance' SQL detection.  
In the second mode, the (reflective) sensor is aligned with a mirror to form a detuned optical cavity, generating an optically-generated spring that provides a highly tunable mechanical frequency.  However the required laser power to obtain this effect is larger than that what is required for SQL on-resonance, and it proportional to the square of the induced mechanical frequency, scaling similarly to the off-resonance SQL noise floor. Thus on-resonance SQL cannot be achieved with this method, and at all frequencies the SQL can only be achieved off-resonance.  This implies tunability currently comes at the cost of sensitivity in these sensors.

When the SQL is reached on-resonance, the noise spectrum is
\begin{align}
    S_\text{FF}^N (\omega_a) = m \gamma\,\omega_a \ ,
    \label{eq:SNoiseon}
\end{align} 
where we assumed the mechanical frequency $\Omega$ is tuned such that $\Omega = \omega_a$. This is significantly lower than the off-resonance SQL noise given by
\begin{align}\label{eq:SNoiseoff}
    S_\text{FF}^N (\omega_a) = m \omega_a^2 \ ,
\end{align}
by a factor $Q^{-1} = \gamma / \Omega\ll 1$, where $Q$ is referred to as the sensor's mechanical quality factor. We note however that these noises can be higher than thermal noises, as discussed in Sec.~\ref{sec:thermal_limits} and shown \textit{e.g.}~in Ref.~\cite{Hamaide:2025buy}. 

Although the sensitivity in Eq.~\ref{eq:SNoiseon} is enhanced compared to Eq.~\ref{eq:SNoiseoff}, it is restricted to a narrow bandwidth of order $\gamma$.  This is in opposition to off-resonance SQL detection, where the noise is constant for all frequencies below the frequency at which the SQL is achieved. Following Ref.~\cite{Hamaide:2025buy}, we may also assume in this work that an optimal search strategy consists of tuning the sensor to be at SQL off-resonance over a discrete set of frequencies $\omega_i$ each separated by a factor of $\sqrt{2}$, such that the bounds are separated by at most a factor of $\sqrt{2}$ near the transition points. As a consequence, an off-resonance SQL search strategies will have an $\mathcal{O}(m_a)$ bandwidth across dark matter Compton frequency space, much larger (by a factor of $Q$) than on-resonance SQL searches.

We note the same considerations apply to torque detection. At off-resonance SQL, the angular noise spectrum is
\begin{align}
    S_{\tau \tau}^N (\omega) \simeq I \omega^2 \ , 
\label{eq:off-resonance-SQL-torque}
\end{align}
where $\omega$ is the angular frequency of motion, while at on-resonance SQL the noise is reduced to 
\begin{align}
    S_{\tau \tau}^N (\omega) \simeq I \gamma \Omega \ ,
\end{align}
with $\gamma$ the angular damping rate (which we will assume is similar in value to the translational damping rate). In selecting a sensor, we note it is advantageous to use one with the largest mass available, provided that the spin polarization (and thus the signals discussed in Sec.~\ref{sec:dynamics}) of the material remains fixed as the mass is increased. This helps the SNR because because the quantum noise power spectrum at SQL shown in Eqs.~\ref{eq:SNoiseoff}~\&~\ref{eq:SNoiseon} scales linearly with the sensor mass while the signal power spectrum scales as the square, such that SNR~$\propto m^{1/2}$.

More recently, optically trapped mechanical quantum sensors, including macroscopic sensors~\cite{Hamaide:2025buy,Michimura:2016urt} and dielectric nanoparticles~\cite{Magrini:2020agy,Millen_2020}, have been developed using a single optical field both to trap the sensor and to read out its position without the need for delicately detuning an optical cavity. These devices possess intrinsically high tunability of their mechanical frequency, but, as in the case of the optical spring, this tunability requires laser powers exceeding those needed to achieve SQL on-resonance. Consequently, their quantum noise floor at the axion frequency $\omega = \omega_a$ corresponds to off‑resonance SQL.
Despite this limitation, optically trapped sensors have demonstrated SQL performance at frequencies significantly closer to resonance than sensors relying on optical springs, and forthcoming technical developments~\cite{Hamaide_2026} may plausibly enable on‑resonance SQL operation. These sensors also offer other practical advantages: they achieve off‑resonance SQL over a broader frequency range $\mathcal{O}(10 \text{ Hz})$ to $\mathcal{O}(10$~MHz) and can operate at temperatures up to $T \sim 100$~K,
in contrast to mechanically trapped sensors, which typically require millikelvin cooling. For these reasons, we prioritize optically trapped sensors in regimes where off‑resonance SQL is preferred and higher temperatures do not degrade the dark matter signal.

%%%%%%%%%%%%%%%%%%%%%%%%%%%%%%%%%%%%%%%%%%%%%%%%%%
\subsubsection{Homodyne vs.~heterodyne optimization}
%%%%%%%%%%%%%%%%%%%%%%%%%%%%%%%%%%%%%%%%%%%%%%%%%%

We begin with the discussion of detecting the dark matter-induced forces or torques in a regime where the sensor can be tuned to be at the off-resonance SQL at any frequency. This is considered optimal compared to using on-resonance sensing due to practical considerations when scanning large bandwidths with high quality factor sensors. Indeed, achieving SQL on-resonance across many decades of axion Compton frequencies $\omega_a$ would require multiple sensor types optimized for different mechanical frequencies, and precise retuning of each sensor for every frequency bin of width $\gamma\ll\Omega$ over which the on-resonance noise is approximately constant. This procedure is therefore expected to be prohibitively time‑consuming and resource‑intensive with current technologies. For these reasons, in this work we adopt a simpler and broadband strategy using an optically trapped sensor operated at SQL off-resonance, while leaving open the possibility that future technical advances may render SQL on‑resonance searches more practical.
However, achieving SQL off-resonance (which scales as $\omega_a^2$) is increasingly difficult for low dark matter frequencies, where thermal noises which are often independent of frequency, or seismic noise which grows quickly at low frequencies, may dominate. At dark matter frequencies lower than the minimum frequency at which SQL off-resonance noise is achievable, we therefore use heterodyne detection of axion‑induced forces and torques using mechanical quantum sensors. In this regime, the axion signal appears as spectral peaks at frequencies $\Omega \pm \omega_a$, where $\Omega$
denotes the mechanical driving (resonant) frequency. 

If heterodyning is employed for all axions with $\omega_a \lesssim \Omega$, we may consider operating the sensor at a fixed mechanical frequency $\Omega$ and may expect to benefit from on‑resonance SQL sensitivity. To see whether this is the case, we consider optimizing SQL on-resonance noise for detecting an axion with $\omega_a \lesssim \Omega$. For this axion Compton frequency, the damping rate must satisfy $\gamma \simeq 2\omega_a$, or equivalently, $Q = \frac{\Omega}{\gamma} \simeq \frac{\Omega}{2\omega_a}$.
For smaller damping rates, the axion-induced sidebands fall outside the narrow region in which SQL on-resonance sensitivity is achieved. Thus, a tradeoff arises: reducing $\gamma$ improves sensitivity to lower‑frequency axions, but simultaneously shifts the axion sidebands away from the optimal noise region. One may then attempt to optimize sensitivity at each axion Compton frequency by tuning the sensor to operate at SQL on-resonance at $\omega_a$. Imposing $\gamma = 2\omega_a$ yields a noise spectrum
\begin{align}
    S_{\text{FF}}^N (\omega_a) = m \gamma \Omega = 2 m \omega_a^2 \ ,
\end{align}
which is nearly identical to the off-resonance SQL noise $m \omega_a^2$. Consequently, heterodyne detection does not provide an improvement over homodyne detection at frequencies where the sensor can reach off-resonance SQL, and simply employing a sensor at the off-resonance SQL for detecting low frequency axions, identically for detection of higher frequency axions, is the optimal search strategy. 

The main difference between employing these sensors for homodyne or heterodyne searches (besides driving the sensor as described in Sec.~\ref{sec:dynamics}) is that homodyne searches perform a scan (albeit a `broadband' scan where each bin is $\mathcal{O}(m_a)$), while a heterodyne search does not change the frequency at which it operates to detect axions of different frequencies, but is sensitive to all frequencies near or below the frequency at which the sensor is operating at the SQL off-resonance.
For Earth-based sensors (both levitated and suspended/clamped) we denote $\Omega=\omega_\text{SQL}^\text{min}$ the minimum frequency at which off-resonance SQL can be reached before seismic noise or thermal noise dominate. Despite the use of 4-stage seismic damping and cryogenic cooling implemented for a sensor of large mass, we will assume a setup similar to the MOLeQuTE proposal~\cite{Hamaide:2025buy} which found
\begin{align}
    \omega_{\text{SQL}}^\text{min} \simeq 30~\text{Hz} \ ,
\end{align}
as the lowest frequency at which off-resonance SQL can be maintained, \textit{i.e.}~the heterodyne frequency. We will also (conservatively) choose to use this minimum for torque sensing, although seismic torque noise is expected to be relatively suppressed compared to seismic translational forces, and we leave a more detailed study of this bound to future work.
%For axions with $\omega_a < \omega_{\mathrm{SQL}}^{\min}$, one may consider tuning the damping rate to $\gamma \simeq 2\omega_a$, which yields
%\begin{align}
%    S_{\text{FF}}^N (\omega_a) = m \gamma \Omega = m \omega_a \omega_{\text{SQL}}^\text{min} \ .
%\end{align}
%However the seismic and thermal noise dominate over SQL noise at $\omega_{\mathrm{SQL}}^{\min}$, and for $\gamma < \omega_{\mathrm{SQL}}^{\min}$ thermal noise dominates. This precludes tuning $\gamma$ in the heterodyne regime. Instead, axion signals must be searched for using a single fixed damping rate and a single fixed mechanical frequency, with all axion masses probed against the same background noise.

We also note that when looking at the axion-induced power spectrum, if we assume the minimum achievable damping rate for mechanical driving is $\gamma \sim 1.5\times 10^{-5}\,\mathrm{Hz}$ \cite{Hamaide:2025buy,Carney:2019cio,Millen_2020}, then for all axions with $\omega_a \leq \omega_{\mathrm{SQL}}^{\min}$, the signal bandwidth is given by
\begin{align}
\frac{1}{T_{\mathrm{coh}}}
= \max(\gamma,\,\omega_a v_a^2)
= \gamma \ .
\end{align}
In other words, the heterodyne detection regime coincides with the regime where the axion-induced signal bandwidth is limited by the damping rate. Combined with the identical off-resonance SQL noise power spectrum used to detect all axion masses in this regime, we may already see that for axion-electron and axion-proton couplings, for which the amplitudes of the forces and torques are independent of axion mass, the sensitivity to the couplings is expected to be independent of $\omega_a$.

%%%%%%%%%%%%%%%%%%%%%%%%%%%%%%%%%%%%%%%%%%%%%%%%%
\subsubsection{Thermal properties and limitations}\label{sec:thermal_limits}
%%%%%%%%%%%%%%%%%%%%%%%%%%%%%%%%%%%%%%%%%%%%%%%%%
We now discuss the temperature of the sensor and its effect on noise in such optomechanical detection schemes in the short term. Although it is beyond the scope of this work to carefully assess the temperature of the sensor and study all sources of thermal noise, we give a preliminary accounting of the most important factors at play to estimate the temperature and noises of the sensor when searching for signals at various axion frequencies. The temperature of the sensor is controlled by heating from absorption of the trapping optical field and by the cooling mechanism(s) available to it, which are further classified as either active or passive cooling. 

%%%%%%%%%%%%%%%%%%%%%%%%%%%%%%%%%%%%%%%%%%%
\paragraph{Passive cooling -- ferromagnetic materials\\}
%%%%%%%%%%%%%%%%%%%%%%%%%%%%%%%%%%%%%%%%%%%
In the case of optically levitated sensors,
and in the absence of gas, the sensor's primary cooling mechanism is necessarily blackbody radiation. Further, the sensor is being operated 
below its melting point which lies around $T\sim 1000$~K.
At these temperatures, the thermal noise from blackbody radiation was shown in the Ref.~\cite{Hamaide:2025buy} to be subdominant to other noises at all frequencies. In particular, seismic noise was found to dominate at low frequencies for Earth-based sensors, while the SQL noise floor dominates at high frequencies (above $\omega_{\text{SQL}}^{\text{min}}$)\footnote{This temperature limit is further expected to decrease when considering space-based sensors where seismic noise is suppressed.}. In this work, we will assume a long slim cuboid (microrod) of cross section $\mathcal{O}(1~\mu$m$^2$), 10~cm length and mass $m=3\times 10^{-6}$~g\footnote{Although a detailed study of the fabrication and magnetization of such a sensor if left to future work, we note that strong magnetization along a ferromagnet's `hard' axis can typically be achieved by sintering materials with magnetocrystalline anisotropy.}. We may consider these values as illustrative examples for now, but will justify these as the desired values for our setup in Secs.~\ref{sec:prospects_forces_torques}~\&~\ref{sec:prospects_freq_shift}. . We note this geometry is slightly more advantageous for blackbody cooling compared to Ref.~\cite{Hamaide:2025buy} due to its higher aspect ratios. We may therefore take the temperatures expected from the geometry in Ref.~\cite{Hamaide:2025buy} as conservative estimates for our case.

When operating a ferromagnetic sensor, we must consider the absorbed energy both from the magnetization flipping and from absorption of the optical field, since both heat the material which can ultimately lead to the demagnetization of the material if its temperature reaches typical Curie temperatures $T\gtrsim 500$~K. We may therefore estimate the off-resonance SQL frequency at which the sensor reaches its Curie temperature by taking into account both effects. Beginning with the magnetization flips performed with a frequency $\omega_\text{SQL}^\text{min}$, one can estimate the heating power to be $P=V\times\alpha_c\times \omega_\text{SQL}/2\pi\sim 10^{-11}$~W~$(\omega_\text{SQL}/10~\text{Hz})$ where $V=m/\rho$ is the volume of the material and $\rho\sim\mathcal{O}(7~\text{g/cm}^3)$ is a typical ferromagnetic material's density, and we take $\alpha_c=10^3$~J/m$^3$~\cite{masciocchi2023optimizationpermalloypropertiesmagnetic} as the expected energy absorbed per magnetization flip. Assuming a levitated sensor at $T\sim 100$~K, this absorbed power is smaller than the emitted blackbody radiation (recalling we assume no mechanical coupling of ferromagnets to the external environment) of the sensor and so is not expected to lead to a significant contribution to the temperature.

As shown in Ref.~\cite{Hamaide:2025buy}, in the absence of heating from magnetization flips, a high aspect ratio sensor has a temperature that is mainly driven by absorption of the large optical beam necessary for levitation, where a the absorbed power $P_\text{abs}$ can be obtained as $P_\text{abs}\approx P_{L,g}\times \alpha \times 1~\mu$m, and where the absorption mainly occurs inside a ultrahigh reflectivity Bragg reflector on the sensor's surface whose absorption coefficient is given by $\alpha\approx 10^{-10}$~m$^{-1}$~\cite{Steinlechner:22,Steinlechner_2018}. We also use the levitation beam power $P_{L,g=}mg/2$ where $g$ is the gravitational acceleration on Earth, and $1\mu$m is the approximate thickness of the reflector. At thermal equilibrium this is balanced by the blackbody radiation of the sensor $P_\text{BB}$ given by $P_\text{BB}=\sigma_\text{BB} T^4 A$ where $A$ is the sensor's surface area and $\sigma_\text{BB}$ is the Stefan-Boltzman constant, such that the temperature\footnote{This is independent of the sensor mass or geometry since absorption from the reflective coating typically dominates over absorption from the bulk (substrate) of the sensor.} of the sensor is expected to be $T\sim 100$~K. Above $\sim 6$~kHz, the temperature of the sensor was found to increase due to the laser power of the trapping (horizontal) lasers dominating over the levitation beam power. The power of a pair of horizontal beams can be estimated when operating the sensor at the off-resonance SQL to be $P_L=m\omega_\text{SQL}^2/4\omega_L>P_{L,g}$~\cite{Hamaide:2025buy}. One then finds $T \propto \omega_\text{SQL}^{1/2}$ when matching the blackbody power to the absorbed power. This ultimately leads to a melting of the sensor when operated at sufficiently high SQL off-resonance frequencies.

Ferromagnetic materials have relatively high optical absorption coefficients $\alpha$ at typical infrared laser frequencies, often reaching $\alpha\gtrsim 1/\mu$m, meaning approximately all the light transmitted by the reflective coating on the sensor would be absorbed inside the sensor. To check whether this exceeds the emitted blackbody power that could be emitted by the ferromagnet's reflective coating at $T\sim 100$~K, we note the light absorbed by the ferromagnet first has to penetrate through the reflective coating discussed in Ref.~\cite{Hamaide:2025buy,Guccione_2013} which has reflectivity of $R\approx 99.998\%$, such that the transmitted (absorbed) light is given by $\mathcal{O}(1$~mW) up to $\omega_\text{SQL}\sim 6$~kHz, which is only an order or magnitude larger than the radiation absorbed by the coating and re-emitted at blackbody radiation. Therefore we expect the ferromagnetic bulk of the material to only heat up to $T\sim 200$~K for off-resonance SQL frequencies below 6~kHz. We note this is still expected to lead to a subdominant source of noise compared to quantum noises. Above this frequency, the material is expected to heat as $T\sim \omega_\text{SQL}^{1/2}$, until the material's Curie temperature of $T\gtrsim 500$~K is reached (which is below the melting temperature) around $\omega_a\gtrsim 240$~kHz. 

To improve the prospects for ferromagnets further, several directions are available. First, an extensive exploration of ferromagnetic materials with higher Curie temperatures, lower absorption and similar polarization per electron could lead to better performing sensors. Next, the coating's reflectivity can be improved. Indeed, as a Bragg reflector the transmission coefficient scales exponentially with the number of layers, such that increasing their number by $\mathcal{O}(10~\%)$ compared to recently achieved exampled discussed in Ref.~\cite{Guccione_2013} can improve its reflectivity by several orders of magnitude as required to bring the transmitted power to this threshold. Finally, improvements in the radiation pressure noise which dominates in these sensors could further decrease the impinging radiation by several orders of magnitude~\cite{Melo:2020prx}. In the rest of this work, we will consider the temperature of a ferromagnet used to probe the axion-electron coupling is maintained below the Curie temperature at all frequencies below $240$~kHz and leave a more detailed examination of possible improvements for future work.

%%%%%%%%%%%%%%%%%%%%%%%%%%%%%%%%%%%%%%%%%%%%%%
\paragraph{Active cooling -- paramagnetic materials\\}\label{paragraph:active_cooling}
%%%%%%%%%%%%%%%%%%%%%%%%%%%%%%%%%%%%%%%%%%%%%%
Although temperatures $T\sim\mathcal{O}(100$~K) suffice to operate ferromagnetic sensors without further loss of polarization and thus sensitivity (on Earth), when using a paramagnet to search for effects of the axion-fermion and axion-EDM coupling, the axion-induced signal strongly depends on the temperature since it directly affects the polarization. Indeed, spin polarizations in paramagnetic materials are well approximated by a thermal distribution $P=\kappa\tanh(\mu_fB/k_BT)\approx \kappa(\mu_nB/k_BT)$ where $\mu_f$ is the fermion's magnetic dipole moment, such that we always have $P\lesssim \kappa$ (as done \textit{e.g.}~in Ref.~\cite{JacksonKimball:2017elr}). Thus the polarization would be expected to reach $P\approx \kappa$ for magnetic fields $B\sim 10$~T and temperatures $T\sim 10$~mK, and constant for lower temperatures and/or higher magnetic fields. For larger temperatures $T\sim 100$~K for example, the polarization is expected to be suppressed by $\mathcal{O}(10^{-4})$ as it scales as $P\sim \kappa\,(10~\text{mK}/10~T)$ (and this decreases further for $\omega_\text{SQL}\gtrsim \omega_\text{SQL}^\text{min}$ where $P\sim \omega_\text{SQL}^{-1/2}$). Such temperatures are expected if the sole cooling mechanism available is blackbody radiation as is typically the case of optically levitated/trapped sensors discussed above.

In order to improve temperatures when using paramagnets, we may wish to use a sensor which is mechanically coupled to a cold environment while operated at the off-resonance SQL. We note that although the spring constant associated with a mechanical coupling typically leads to a fixed mechanical frequency, we assume in this work this mechanical spring constant can be made to be small, and that we add an optically-induced spring constant by focusing a laser beam as done in Ref.~\cite{Hamaide:2025buy} (rather than with the optical spring effect for convenience, although both methods lead to identical results). This optical field can then be tuned to reach the off-resonance SQL at a desired frequency, or (equivalently) to reach a desired mechanical frequency of the sensor at or above that induced by the mechanical coupling. Although this mechanical coupling of the sensor to the external environment can introduce noise which is larger than off-resonance SQL noise, for sufficiently low temperatures $T\sim 10$~mK this can be kept below the seismic and off-resonance SQL quantum noise floors at all frequencies, including near $\omega_\text{SQL}^\text{min}$ where it is lowest, see Ref.~\cite{Hamaide:2025buy} and references therein for further details. To reach these temperatures we will therefore consider additional cooling resources such as a dilution fridge. 

Thankfully, the mechanical coupling also allows stronger cooling power capacity due to the increased efficiency with which thermal energy is transmitted ({\it e.g.}~for a suspended or softly clamped system). Assuming $\sim 30~\mu$W of cooling power at $T\sim 10$~mK provided by a standard dilution fridge, our sensor can be kept near equilibrium with an environmental temperature $T\sim 10$~mK when operating the sensor at the off-resonance SQL for any $\omega_\text{SQL}= \omega_a\lesssim 60$~kHz. This is because the available cooling power is expected to be larger than absorbed power $P_\text{abs}$ inside the ultra high-reflectivity Bragg reflective coating~\cite{Hamaide:2025buy,Guccione_2013} as considered for the passive cooling case. Although reaching lower temperatures, \textit{e.g.}~down to $T\sim 1$~mK does not lead to significant improvement in the axion signals, as we will discuss shortly in Sec.~\ref{paragraph:magnetization_noise} this may be needed to maintain sufficiently low noise induced by fluctuations of the sensor's magnetization. For $\omega_\text{SQL}\lesssim 60$~kHz, the cooling power requirements drop at $P_L\propto\omega_\text{SQL}^2$, which matches the scaling of the cooling power of a dilution fridge with temperature, \textit{i.e.}~$P_\text{cool}\propto T^2$, such that the achievable temperature is expected to scale as $T\propto\omega_\text{SQL}$ until $T\sim 1$~mK is reached\footnote{We note that maintaining temperatures below $T\approx 1.75$~mK will likely require supplementing dilution fridge cooling with adiabatic nuclear depolarization or methods such as dynamic spin polarization~\cite{AAbragam_1978} to reach 1~mK or below, but leave a discussion of its implementation to follow-up work.} for $\omega_\text{SQL}\approx 6$~kHz.

Based on this estimate, for paramagnetic materials used to probe the axion-fermion and axion-EDM couplings, we assume a mechanically trapped sensor whose temperature is expected to remain at $T\sim 10$~mK for both the off-resonance SQL frequency scan regime $\omega_a\gtrsim\omega_\text{SQL}^\text{min}$ (up to 60~kHz) and in the heterodyning regime for $\omega_a\lesssim \omega_\text{SQL}^\text{min}$. These temperatures are therefore expected to maintain the fermion spin polarization (and in particular nuclear spin polarizations) to its maximum value $P=\kappa$ for these frequencies, above which the laser power is strong enough such that the absorbed power overcomes the cooling power. Above $60$~kHz, blackbody radiation is then expected to be the sole cooling mechanism once again, such that we expect a rapidly increasing temperature of $T\approx 300$~K in order to provide sufficient blackbody cooling. Above $60$~kHz, the temperature is then expected to scale as $\omega_a^{1/2}$ until it reaches melting temperatures at approximately $\omega_a \gtrsim 600$~kHz. Thus our use of mechanical cooling allows us to maintain strong polarization over up to large axion frequency range cutoff, such that we prioritize a sensor with a mechanical coupling for the search for axion-fermion couplings (except when using ferromagnets) and the axion-EDM coupling.

%%%%%%%%%%%%%%%%%%%%%%%%%%%%%%%%%%%%%%%%%%%%%%%%
\paragraph{Magnetization noise \\}\label{paragraph:magnetization_noise}
%%%%%%%%%%%%%%%%%%%%%%%%%%%%%%%%%%%%%%%%%%%%%%%%

Beyond the effect of blackbody radiation, vibrations and other sources of thermal noise, sensors mechanically coupled to the external environment experience fluctuations in their overall magnetization (rather than local fluctuations within the sensor only). These thermal fluctuations typically leave a small portion of excited spin states, which can be seen as free to move across the material at a characteristic rate given by the microscopic (\textit{e.g.}~dipole-dipole) interactions inside the material. This is typically enhanced for electrons whose spin magnetic dipole moment is nearly $\sim 10^3$ larger than those of nucleons, but can also be large if the magnetized (ground state) population is low, \textit{e.g.}~for nucleons. In particular, two types of magnetization dissipation may occur: longitudinal (relating to the amplitude of magnetization) with characteristic dissipation time denoted $T_1$, and transverse (for the angle of the magnetization) with characteristic dissipation time denoted $1/T_2$.

Following Ref.~\cite{Isaac_2017}, over timescales larger than $T_1$ and $T_2$, the standard deviation of the fluctuations of the nuclear magnetic dipole moment of the material is given by
\begin{align}
    \sigma_\mu =\mu_n\sqrt{PN(1-\tanh(\mu_nB/k_BT)^2)} \,\, .
\end{align}
Assuming two-point functions of these fluctuations which exponentially decay in time over timescales $T_1$ or $T_2$, \textit{i.e.}~$\langle \abs{\mu(t)}\abs{\mu(0)}\rangle=\sigma_\mu^2\, e^{-t/T_1}$ and $\langle \theta_\mu(t)\theta_\mu(0)\rangle=\sigma_\mu^2\, e^{-t/T_2}$, the power spectrum of fluctuations of the magnetization (in the small angle limit) can be obtained as Lorentzians of characteristic width $1/T_1$ and $1/T_2$ respectively~\cite{Budker:2013hfa,Braun_2007}
\begin{align}
    S_{\mu\mu}^{T_1}(\omega)=\frac{1}{\pi}\frac{\sigma_\mu^2 \,T_1}{1+\omega^2T_1^2} \,\,\,\,\,\, , \,\,\,\,\,\, S_{\mu\mu}^{T_2}(\omega)=\frac{1}{\pi}\frac{\sigma_\mu^2\, T_2}{1+(\omega-\Omega_L)^2T_2^2} \,\, ,
\end{align}
where $\Omega_L$ denotes the Larmor frequency which appears due to the harmonic oscillator-type potential of angular displacement of the magnetization.

Magnetization noise primarily affects the gradient force perceived in paramagnetic sensors stemming from the gradient of the potential $\nabla_x(\vec{\mu}\cdot \vec{B})\approx PN\mu_fB_0x$. Although this does not lead to a constant force on the sensor at $x=0$, the fluctuations of the forces along the $\mathbf{\hat{x}}$-axis at $x\neq 0$ due to the finite half-width $L$ of the sensor lead to a noise power spectrum of forces on the sensor given by
\begin{align}
    S_\text{FF}^\mu(\omega)\approx \frac{B_0L^2}{6}\left(S_{\mu\mu}^{T_1}(\omega) + S_{\mu\mu}^{T_2}(\omega)\right) \,\, .
\end{align}
where the $T_1$ and $T_2$ contributions are peaked at zero and $\Omega_L$ respectively. This can subsequently be compared to the noise power spectrum of the sensor when operated at the off-resonance SQL over the (approximate) range $\omega\in[\omega_\text{SQL}^\text{min},100$~kHz]. Assuming the sensor is centered at $x=0$ and $y=1~\mu$m, at temperatures at (or below) $T\sim 10$~mK the polarization probability reaches $\tanh(\mu_f B/k_BT)\sim 1$ such that $\sqrt{1-\tanh(\mu_f B/k_BT)^2}\approx 2e^{-\mu_fB/k_BT}$. Assuming that $T_1\approx T_2\approx 1000$~s for nuclear spins, the transverse nuclear magnetization noise of paramagnets is expected to be negligible compared to off-resonance SQL noise, primarily due to the large Larmor frequency assumed to be $\Omega_L\gtrsim 100$~MHz. However, the longitudinal noise is expected to dominate over off-resonance SQL noise at low frequencies in the region $\omega\in[\omega_\text{SQL}^\text{min},3~$kHz] unless polarization probability sufficiently near unity can be achieved (\textit{i.e.}~better than simply $\mathcal{O}(1)$). Indeed, at low frequencies, reaching $S_\text{FF}^\mu(\omega)\approx 2B_0\,y\,S_{\mu\mu}^{T_1}(\omega)\lesssim S_\text{FF}^\text{SQL}(\omega)$ requires reaching a magnetic field-to-temperature ratios $B/T\sim 20$~T/mK. As discussed in Sec.~\ref{paragraph:active_cooling}, in this work we assume sufficiently lower temperatures (down to $T\sim 1$~mK) can be reached while operating the sensor at the off-resonance SQL at frequencies up to $\omega_\text{SQL}\sim 6$~kHz. We therefore consider in the rest of this work that nuclear magnetization noise can be kept negligible. Meanwhile, although electronic relaxation times are typically lower (assumed to be tens of seconds), electronic magnetization noise is expected to be exponentially suppressed by the electron's larger (roughly $\times 10^3$) magnetic dipole moment, such that we will also ignore it in this work.

%We note that further magnetization noise can occur when heterodyning the force from a low frequency axion, particularly due to the external magnetic field oscillating in the sensor's frame when the sensor is coherently oscillating along the $\mathbf{\hat{y}}$-axis (at $x=0$). Although a careful estimate of the associated amplitude of the force power spectrum peaked at $\omega_\text{SQL}^\text{min}$ is left for future work, %can be probably be estimated to be $S_\text{FF}(\omega)\approx (PN\mu_fB_0y_0/\pi(\omega_\text{SQL}T_1))^2(T_1/(1+(\omega_\text{SQL}^\text{min}-\omega)^2T_1^2)) - tbc $ 
%the width of the peak is given by $1/T_1$, which is larger than the distance between the axion-induced peaks expected at $\Omega\pm\omega_a$ for axion masses $m_a\lesssim 2\times 10^{-19}$~eV. Assuming the $\vec{\mu}\cdot \vec{B}$-induced power spectrum amplitude is wider than twice the axion's mass, this could potentially prevent detection of the axion-induces peaks for these ultralight axions. However, this issue can be prevented by flipping the sign of the sextupole magnetic field instead of driving motion in along the $\mathcal{\hat{y}}$-axis, assuming the damping rate $\gamma_f$ of the flipping $\gamma_f\lesssim 1$~mHz can be achieved, similarly to our assumption for the ferromagnetic case where we took $\gamma\sim 10^{-6}$~Hz, and we leave a careful determination of this requirement for future work.

To maintain low noises induced by magnetic field fluctuations, we note that both precise control of the magnetic field in our proposed setup, as well as shielding from external magnetic fields (\textit{e.g.}~from Earth's core) are crucial. In particular, given the requirement to maintain small relative fluctuations of the magnetization (likely at $\mathcal{O}(10^{-9}))$ in order to reach the off-resonance SQL at all frequencies of interest implies that the required shielding of magnetic fields must reach $\mathcal{O}(10~\text{nT})$. This is routinely achieved with specialized magnetic shielding~\cite{Altarev:2014wka,Ahmed:2026cvd,nEDM:2022nqd}. Therefore, in this work we ignore external magnetic field noise. We finally note that potential quantum fluctuations of the (classical) sextupole magnetic fields is also left to follow work.

%%%%%%%%%%%%%%%%%%%%%%%%%%%%%%%%%%%%%%%%%%%%%%%%%
\subsection{Prospects for axion-induced forces and torques}\label{sec:prospects_forces_torques}
%%%%%%%%%%%%%%%%%%%%%%%%%%%%%%%%%%%%%%%%%%%%%%%%%

%%%%%%%%%%%%%%%%%%%%%%%%%%%%%%%%%%%%%%%%%%%%%%%%%
\subsubsection{Axion-fermion couplings \texorpdfstring{$g_f$}{gf} -- ferromagnetic sensor}\label{sec:prospects_ferro_derivative_coupling}
%%%%%%%%%%%%%%%%%%%%%%%%%%%%%%%%%%%%%%%%%%%%%%%%%

Having now discussed the optimization of the signal-to-noise ratio over the different ranges of axion masses, we will estimate sensitivities to the various axion couplings. For ferromagnetic materials (considered only for the $g_e$ coupling) with a polarization $P\sim 10^{-2}$, 
we use Eq.~\ref{eq:SNR_torque} with the signal torque from Eq.~\ref{eq:torquesignal} and off-resonance SQL torque noise in Eq.~\ref{eq:off-resonance-SQL-torque} to determine the axion-fermion coupling sensitivities
\begin{align}\label{eq:gf_ferro_force_sensitivity}
    g_e \gtrsim \begin{cases}
\dfrac{2\sqrt{2}\,L\,m_n\,\sqrt{\gamma}\,m_\text{SQL}^\text{min}}{\sqrt{3\pi m\rho_\text{DM}} v_a P N_T^{1/4}} \ , \quad &m_a\lesssim m_\text{SQL}^\text{min} \ , \\
    \dfrac{2\sqrt{2}\,L\,m_n\,m_a^{3/2}}{\sqrt{3\pi m\rho_\text{DM}} P N_T^{1/4}} \ , \quad &m_a\gtrsim m_\text{SQL}^\text{min} \ ,
    \end{cases}
\end{align}
where we used $m_\text{SQL}^\text{min}\sim10^{-14}~\text{eV}$, and the usual values for $a_0=\sqrt{2\rho_\text{DM}}/\omega_a$, $\sigma_x = 1/2$, and $N=m / 2m_n$. We also remind the number of coherence times is taken to be $N_T = 50$
at all axion masses at which the SQL quantum noise dominates over classical noises ({\it i.e.}~where $\omega_a\gtrsim \mathcal{O}(10)$~Hz). 
%NOTE: USING NT~m_av_a^2 may allow for better overall coverage of parameter space but does not optimize new parameter space since lower masses is where new space is easier to reach, which is opitmized for constant NT. We don't discuss this in the paper because it's too in depth, but worth keeping in mind in the future.
For axion masses below this value, we resort to heterodyning the signals of axions at lower frequencies with fixed mechanical damping rate $\gamma_m \sim \gamma$.

To determine the optimal geometry of the levitated sensor in the ferromagnetic case, we use that the sensitivity to $g_f$ scales linearly with $L$ and as the square root of the sensor mass $m^{1/2}$. Given a cuboid whose cross section is rectangular, scaling the mass $m$ by scaling the largest direction $L$ (which dominates the moment of inertia used in Eq.~\ref{eq:gf_ferro_force_sensitivity} leads to a $m\propto L$, such that reducing the longer length until the cuboid's cross section is square optimizing the sensitivity. Fixing the cross section to be square leads to $m\propto L^2$ such that the sensitivity is no longer dependent on the length. However a long slim cuboid whose (half)width $L$ in the $\mathbf{\hat{x}}$ and $\mathbf{\hat{y}}$ axes is $L \sim \mathcal{O}(1~\mu\text{m})$ maximizes the aspect ratio which maintains lower temperatures in the sensor. The minimum length of $L$ is then set by the requirement of a sufficiently thick reflective coating around the ferromagnetic core.
The length in the $\mathbf{\hat{z}}$-axis can then be maximize and we consider $10$~cm such that the mass $m\sim 3\times 10^{-6}$~g, which is the geometry used in our discussion of the sensor's temperature (in Sec.~\ref{sec:thermal_limits}).

%%%%%%%%%%%%%%%%%%%%%%%%%%%%%%%%%%%%%%%%%%%%%%%%%
\subsubsection{Axion-fermion coupling \texorpdfstring{$g_f$} -- paramagnetic sensor}\label{sec:prospects_derivative_coupling}
%%%%%%%%%%%%%%%%%%%%%%%%%%%%%%%%%%%%%%%%%%%%%%%%%
For paramagnetic materials, using Eq.~\ref{eq:force_paramagnetic} in Eq.~\ref{eq:SNR_force}, with the force noise Eq.~\ref{eq:SNoiseoff} we obtain
\begin{align}\label{eq:gf_force_sensitivity}
    g_f \gtrsim \begin{cases}
        \dfrac{ 4 \,m_n \sqrt{\gamma}\,m_\text{SQL}^\text{min}}{\sqrt{\pi m\rho_\text{DM}} v_a(\partial P/\partial x)N_T^{1/4}} \ , \quad &m_a\lesssim m_\text{SQL}^\text{min} \\
    \dfrac{4 \,m_n\,m_a^{3/2}}{\sqrt{\pi m\rho_\text{DM}}(\partial P/\partial x)N_T^{1/4}} \ , \quad &m_a\gtrsim m_\text{SQL}^\text{min} \, .
\end{cases}
\end{align}

As seen in Eq.~\ref{eq:gf_force_sensitivity}, the optimal magnetic field configuration maximizes the achievable magnetic field gradient, which crucially depends on the (inverse) spatial scale of the gradient. Indeed, assuming the maximum magnetic field at either end of the sensor (include the sensor's size $L$ and position standard deviation $\bar{\sigma}_x\lesssim L$) to be limited to $\mathcal{O}(\pm10$~T), the gradient is typically limited by $\partial B_x/\partial x\sim 10$~T/$L$. This further motivates assuming $L\sim\mathcal{O}(1~\mu$m) as the size of the paramagnetic sensor, \textit{i.e.}~identically to the ferromagnetic case.

Assuming sextupole magnetic field configuration given by 
\begin{align}
    \vec{B}(\vec{x})= B_0(xy,\frac{1}{2}(x^2-y^2),0) \,\, ,
\end{align}
where $B_0\sim 20~\text{T/$\mu$m}^2$ over a relatively small region where $x$ and $y$ are each $\lesssim\mathcal{O}(1~\mu$m), limits the dimensions of the sensor which can fit in this region, except in the $\mathbf{\hat{z}}$ direction. Wishing to maximize the mass and gradient simultaneously, we therefore assume a long slim cuboid geometry for the sensor, as in the ferromagnetic case. Importantly, the magnetic field along the line defined by $y\approx 1~\mu$m is $\mathcal{O}(1)$ constant in amplitude for small motion $\delta x\ll y\sim 1~\mu$m centered around $x=0$, while the field's angle $\theta_B\approx 2x/y$ varies linearly for motion along the $\mathbf{\hat{x}}$-axis, such that the polarization angle will rotate and the polarization projected along the $\mathbf{\hat{x}}$-axis $P(\vec{x})\propto x$. As discussed in Sec.~\ref{paragraph:EOM_study_first_term_derivative_coupling2}, this fulfills the conditions  for paramagnets for the sensor to perceive an axion-induced force (potential gradient) along the $\mathbf{\hat{x}}$-axis, which can then oscillate via the axion field. We note that for sufficiently large motion or variance of the position along in the $\mathbf{\hat{x}}$-axis such that $x^2\sim y^2$, the polarization along the $\mathbf{\hat{x}}$-axis saturates to become $\theta\sim \pi/2$ such that the sensor no longer feels an axion gradient force. However at off-resonance SQL the sensor's variance is expected to be $\sigma_x\ll 1~\mu$m, as we will show in Sec.~\ref{sec:optimization_freq_shift}.

This geometry also allows for detection in the heterodyne case, where we wish to design the magnetic field gradient such that it varies by $\pm\mathcal{O}(1)$ along the $\mathbf{\hat{y}}$-axis over the distance $\pm\bar{\sigma}_y\sim 1\mu$m corresponding to the amplitude of the (coherent) motion we create in the sensor, as discussed in Sec.~\ref{paragraph:EOM_study_first_term_derivative_coupling2}. The gradient of the sextupole magnetic field is expected to vary $\pm\mathcal{O}(1)$ over a distance $L\sim\sqrt{20~\text{T}/B_0}\sim 1~\mu$m, meaning the gradient term $\partial^2P(\vec{x})/\partial x\partial y\propto y P_\text{max}/L^2$ where $P_\text{max}$ is the polarization at the maximum magnetic field amplitude along the ~$\mathbf{\hat{x}}$-axis, \textit{i.e.}~$B_x=20$~T.

We approximate the magnetic field gradient as reaching up $(\partial B_x/\partial x)(y)=10$~T/$\mu$m~$(y/\mu\text{m})$, assuming a motion $\bar{\sigma}_y\sim 2L$, and a temperature $T=10$~mK can be maintained up to $60$~kHz using a dilution fridge (above which the temperature scales rapidly to $T\sim 100$~K and then as $T\approx 100~\text{K}(m_a/60~\text{kHz})^{1/2}$ until the melting is reached at $\omega_a=600$~kHz, as discussed in Sec.~\ref{sec:thermal_limits}). We thus expect to achieve a polarization gradient $\partial P/\partial x=2\times 10^{-2}$~eV for nucleons and $\partial P/\partial x=2\times 10^{-6}$~eV for electrons below $m_a\lesssim 4\times 10^{-11}$~eV ({\it i.e.}~$\omega_a=60$~kHz), and $\partial P/\partial x=2\times 10^{-6}~\text{eV}^{-1}\,(4\times 10^{-11}~\text{eV}/m_a)^{1/2}$ for nucleons and $\partial P/\partial x=2\times 10^{-10}$~eV for electrons at larger axion masses up to $m_a\sim 4\times 10^{-10}$~eV.

%%%%%%%%%%%%%%%%%%%%%%%%%%%%%%%%%%%%%%%%%%%%%%%%
\subsubsection{Axion-EDM coupling \texorpdfstring{$g_d$}{gd}}\label{sec:prospect_force_EDM}
%%%%%%%%%%%%%%%%%%%%%%%%%%%%%%%%%%%%%%%%%%%%%%%%
Similarly to the paramagnetic case in Sec.~\ref{sec:prospects_derivative_coupling} for the axion-fermion couplings, for the axion-EDM coupling we assume the same magnetic field configuration and an identically long slim cuboid, for both homodyne and heterodyne detection. Using the effective SNR condition Eq.~\ref{eq:SNR_force} with force given in Eq.~\ref{eq:EDM_force} and off-resonance SQL noise from Eq.~\ref{eq:SNoiseoff}, we obtain projected sensitivities
\begin{align}
    g_d\gtrsim \begin{cases}
        \dfrac{\sqrt{2\gamma}\,m_n\,m_a\,m_\text{SQL}^\text{min}}{\sqrt{m\rho_\text{DM}}(\partial P/\partial  x) E N_T^{1/4}} \,\,\,\,\,\,\, , \,\,\, m_a\lesssim m_\text{SQL}^\text{min} \\
        \dfrac{\sqrt{2}\,m_n\,m_a^{5/2}v_a}{\sqrt{m\rho_\text{DM}}(\partial P/\partial x) E N_T^{1/4}} \,\,\,\,\,\,\, , \,\,\, m_a\gtrsim m_\text{SQL}^\text{min}
    \end{cases}\label{eq:gd_sensitivity}
\end{align}
where again we used $a_0=\sqrt{2\rho_\text{DM}}/\omega_a$ and $N=m/m_n$ to simplify the expression. This force benefits from an additional $m_a$ scaling which was not present in the case of the axion-fermion coupling since it incurred an additional factor of $m_av_\text{DM}$. 

To optimize the sensitivity to the axion-EDM coupling we assume $\kappa\sim 10^{-3}$ portion of neutrons are polarizable in a ferroelectric material, such that the polarization gradient $\partial P/\partial x=2\times 10^{-4}$~eV below $m_a\lesssim 4\times 10^{-11}$~eV ({\it i.e.}~$\omega_a=60$~kHz), and $\partial P/\partial x=2\times 10^{-8}~\text{eV}^{-1}\,(4\times 10^{-11}~\text{eV}/m_a)^{1/2}$ at larger axion masses up to $m_a\sim 4\times 10^{-10}$~eV. Ferroelectric materials also present large internal electric fields reaching $E=10^5$~eV$^2$, which is further corrected by the Schiff suppression factor which we take to be $\epsilon\sim 10^{-2}$~\cite{JacksonKimball:2017elr} such that $E=10^3$~eV$^2$.

%%%%%%%%%%%%%%%%%%%%%%%%%%%%%%%%%%%%%%%%%%%%%%%%%
\subsection{Optimizing the search strategy for frequency shifts in optomechanical sensors}\label{sec:optimization_freq_shift}
%%%%%%%%%%%%%%%%%%%%%%%%%%%%%%%%%%%%%%%%%%%%%%%%%

As shown in Secs.~\ref{sec:axionpeakderivative}~\&~\ref{sec:peak_split_EDM}, both the axion-electron coupling and the axion-EDM coupling can lead to shift in the resonant frequency of mechanical sensors. Harmonic oscillators are very sensitive to changes in frequency given their sharply peaked power spectra, and this method is commonly used to detect Casimir forces or even fifth forces (see \textit{e.g.}~\cite{Ma:2026afn,Sedmik:2021iaw}). Frequency shifts are directly visible in the position power spectrum of harmonic oscillators $S_{xx}(\omega)$, which is typically the initial observable from which one obtains the force power spectrum $S_\text{FF}(\omega)$ via the (amplitude squared of the) mechanical susceptibility of the sensor 
\begin{align}
\abs{\chi_m(\omega)}^2=\dfrac{1}{\abs{m^2((\Omega^2-\omega^2)^2+\gamma^2\omega^2}^2} 
 ,
\end{align} 
and the relation 
\begin{align}
    S_{xx}(\omega) = \abs{\chi_m(\omega)}^2 S_\text{FF}(\omega) \ .
\end{align}
The mechanical susceptibility serves as the transfer function between forces and the position of the oscillator in frequency space, and is responsible for deviation of power spectra from white noise. As we will see below, it typically leads to a peak at the resonant frequency in the position power spectrum given an equally distributed ({\it i.e.}~white) spectrum of forces on the sensor. 
Assuming $\Omega$ is coupled to the axion (either via the fermion or EDM couplings), we can make the replacement $\Omega^2\rightarrow\Omega^2+\delta\Omega^2(t)$, such that the frequency shift is given by $\sqrt{\Omega^2+\delta\Omega^2(t)}-\Omega\approx \delta\Omega^2(t)/2\Omega$.  This 
shift in the mechanical frequency $\Omega$ mainly leads to a change of the mechanical susceptibility $\abs{\chi_m(\omega)}^2$, \textit{i.e.}~a change in the response of the sensor to external driving force (\textit{e.g.}~from noise).

\begin{figure}
    \centering
    \includegraphics[width=0.95\linewidth,page=2]{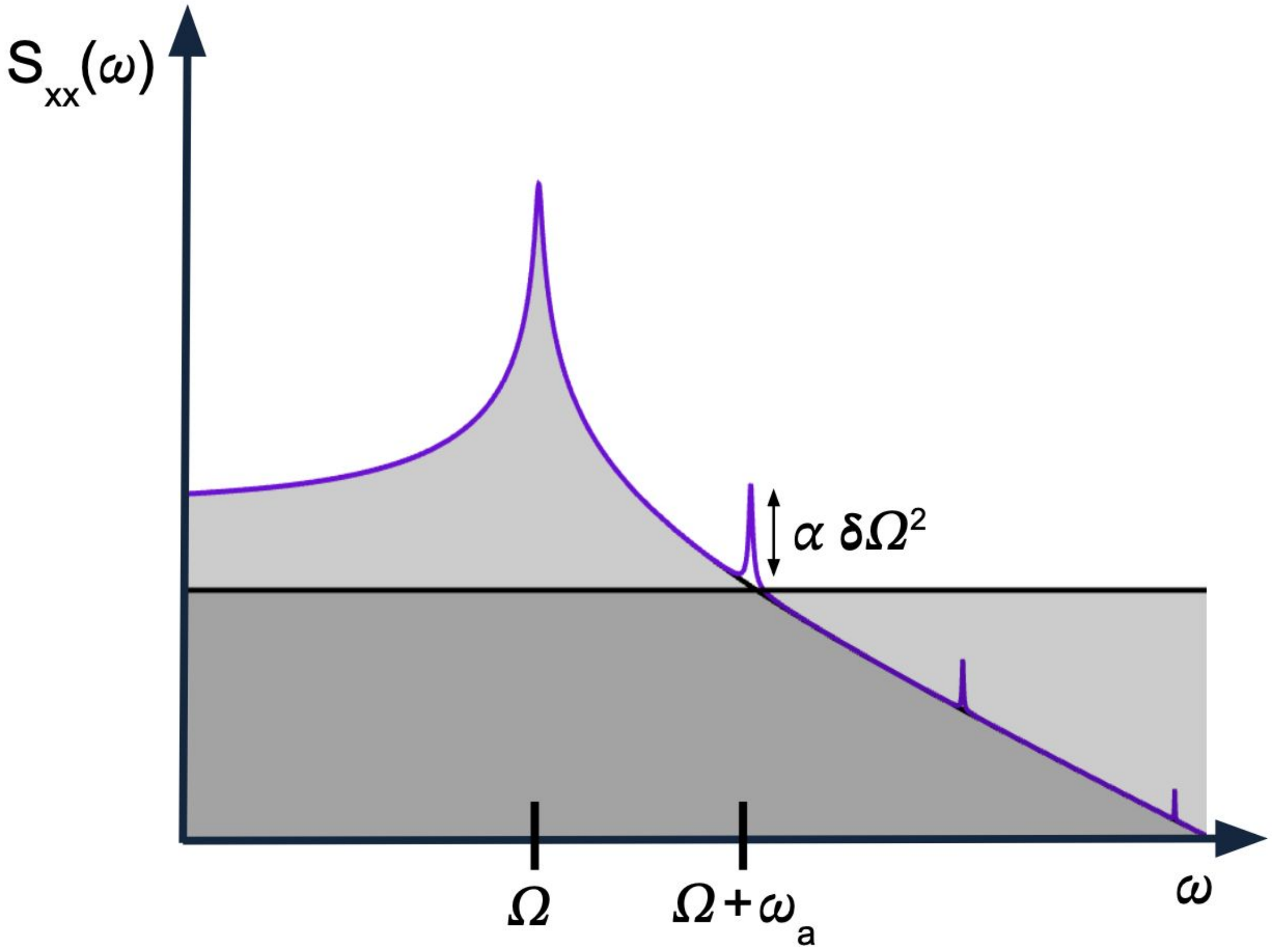}
    \caption{Illustration of the effect of frequency shifts oscillating at (axion Compton) frequency $\omega_a$ on the position power spectrum $S_{xx}(\omega)$ of a harmonic oscillator with mechanical frequency $\Omega$. For $\omega_a>\gamma$ the axion signal (dashed purple) is seen as adding a series of peaks to the power spectrum starting at $\Omega+\omega_a$. The expected (gray region) power spectrum is tuned to be at the SQL noise floor, where the shot noise (flat black line) is equal to the backaction (black peaked line which mostly coincides with the purple line) at $\Omega+\omega_a$.}
    \label{fig:freq_shift_effects}
\end{figure}

To detect a change in frequency, we require a  SNR~$\gtrsim\mathcal{O}(1)$ when comparing the observed power spectrum to the \textit{expected} noise power spectrum. In particular we search for a deviation in the shape of the power spectrum from the expected shape, as it is in practice difficult to predict exactly the value of $\Omega$ (which is often tuned if aiming for a particular frequency) in advance. The expected shape of the noise position power spectrum is flat in the region $\omega\lesssim\Omega-\gamma$, followed by a single Lorentzian peak centered at $\Omega$ with width $\gamma$ and ending with a $\omega^{-4}$ behavior for $\omega\gtrsim\Omega+\gamma$\footnote{Although in practice more complex harmonic oscillators may present several peaks {\it e.g.}~from phonon modes, optically trapped sensors do not present this difficulty such that in this work we assume a simple harmonic oscillator.}. The effective SNR discussed in Sec.~\ref{sec:force_noise_optimization} can be rewritten in terms of position power spectra $S_{xx}(\omega)$ by multiplying the the numerator and denominator by the susceptibility, which gives
\begin{align}\label{eq:SNR_freq_shift_Sxx}
    \text{SNR}_x^\text{eff}=\frac{\sqrt{\int_{\delta\omega} S_{xx}^a(\omega) d\omega}}{\sqrt{\sqrt{N_T}\int_{\delta\omega} S_{xx}^N(\omega) d\omega}} \,\,\,\, ,
\end{align}
where again we took the incoherent sum of noise variance over $N_T$, and the total noise power spectrum $S^N_{xx}$ is explicitly defined as the sum of the backaction noise (denoted BA), shot noise (denoted SN) and thermal noise (denoted T), {\it i.e.}
\begin{align}\label{eq:total_noise_PSD}
    S^N_{xx}(\omega)\equiv S_{xx}^{BA}(\omega)+S_{xx}^{\text{SN}}(\omega)+S_{xx}^{T}(\omega) \,\, .
\end{align} 
Given the peak of the mechanical susceptibility is translated when $\Omega$ is shifted, a common method to detect frequency shifts is thus simply to use that $S_{xx}^{BA}(\omega)=4\omega_LP_L\abs{\chi_m(\omega)}^{2}$ or $S_{xx}^T(\omega)=4m\gamma_T k_B T\abs{\chi_m(\omega)}^{2}$, which depend on the mechanical frequency $\Omega$ and dominate over the shot noise  $S_{xx}^{\text{SN}}=1/(4\omega_LP_L)$, where $\omega_L$ and $P_L$ are the optical frequency and laser power, and $T$ and $\gamma_T$ are the temperature and damping rate of the thermal bath which the sensor is in equilibrium respectively. In this case, the backaction and thermal noise lead to the expected peak-like structure of the observed $S_{xx}^N(\omega)$ while $S_{xx}^\text{SN}(\omega)$ is a flat background, as illustrated in Fig.~\ref{fig:freq_shift_effects}. To notice a shift in the peak we thus simply need the peak to be sufficiently larger than the (flat) shot noise background. This is readily obtained experimentally through the use of sufficient laser power $P_L$ and/or temperature (within experimental constraints such as melting, as discussed in Sec.~\ref{sec:thermal_limits}).

Beyond a simple (constant) frequency shift which would be unnoticeable given the relative ab initio unpredictability of $\Omega$, harmonic oscillators with time-dependent resonant frequencies shifts $\delta\Omega(t)^2/2\Omega=(\delta\Omega_0^2/2\Omega)\cos(\omega_a t)$ have power spectra with different shapes than the expected shape in the absence of an axion coupling, making the effect detectable. These effects can often studied as power spectra resulting from solutions to the Mathieu equation shown in the form
\begin{align}\label{eq:peak_split}
    m \ddot{x}+m\gamma\dot{x}+m(\Omega^2+\delta\Omega(t)^2) x=F_N(t) \;\; ,
\end{align}
when looking at translational frequency shifts, and  
\begin{align}
    I \ddot{\theta}+I\gamma\dot{\theta}+I(\Omega_\theta^2+\delta\Omega_\theta(t)^2) \theta=\tau_N(t) \;\; ,
\end{align}
when looking at angular frequency shifts. In what follows we will treat the translational case explicitly, and note the treatment of the angular Mathieu equation is identical up to $x\rightarrow\theta$, $\Omega\rightarrow\Omega_\theta$ and $m\rightarrow I$. The solutions to these EOMs are Mathieu functions of the first kind, namely series of cosines with frequencies $n\omega_a$ for all positive integers $n$. The behavior (and stability) of Mathieu functions and their associated power spectra are given by the relative hierarchy of the parameters in the EOM. Given $\omega_a>\gamma$ for all $\omega_a\gtrsim  10^{-20}$~eV where the peak at $n=1$ is expected to have the largest amplitude, and we ignore the case where $\omega_a<\gamma$. 

In this regime, considering the contribution from $\delta\Omega^2$ in the EOM is assumed to be small compared to the main forces in the EOM, we begin by writing the effect as an effective force on the oscillator. Indeed, taking the zeroth-order $x_0(t)$ motion of the sensor as the motion in the absence of the axion coupling, \textit{i.e.}~solely driven by force noise, to first order the last term of the left hand side of Eq.~\ref{eq:peak_split} can be written as $F_\text{eff}^a(t)=m\delta\Omega^2(t)x_0(t)$, which appears as a driving force on the oscillator's motion. The power spectrum of $F_\text{eff}^a$ is then given by $S_{\text{F}_\text{eff}^a\text{F}_\text{eff}^a}=m^2 S_{x * \delta\Omega^2}(\omega)$ where $S_{x * \delta\Omega^2}(\omega)$ is the power spectra of the convolution of $\delta\Omega^2(t)$, whose power spectrum is given by that of the axion, with $x_0(t)$ whose power spectrum is given by the unperturbed harmonic oscillator spectrum (gray shaded region in Fig.~\ref{fig:freq_shift_effects}). The resulting power spectrum presents a peak at $\omega=\Omega+\omega_a$ with an amplitude estimated by
\begin{align}
    S_{x * \delta\Omega^2}(\Omega+\omega_a)\approx\begin{cases}
        \,\,\delta\Omega_0^4S_{xx}(\Omega) \,\,\,\,\,\,\,\,\,\,\,\,\, , \,\,\,\,\, m_av_a^2<\gamma \\
        \,\, S_{\delta\Omega^2\delta\Omega^2}(\omega_a)\bar{\sigma}_x^2 \,\,\,\,\, , \,\,\,\,\, m_av_a^2>\gamma
    \end{cases}
\end{align}
To obtain the $m_av_a^2<\gamma$ and $m_av_a^2>\gamma$ cases, we approximated $S_{\delta\Omega^2\delta\Omega^2}(\omega)$ and $S_{xx}(\omega)$ respectively as a delta function while approximating the other as a Lorentzians of full width half maxima $m_av_a^2$ and $\gamma$ respectively, and used that the integral of the position power spectrum is given by $\bar{\sigma}_x^2 \equiv \int_{\omega>0} S_{xx}^N (\omega) d\omega = (4\pi\omega_LP_L)/(2m^2\Omega^2\gamma)$~\cite{Millen_2020}, and that the peak of the (Lorentzian) power spectrum of the axion-induced frequency shift is given by $S_{\delta\Omega^2\delta\Omega^2}(\omega_a)\approx \delta\Omega_0^4/(4\pi\,\Omega^2 m_av_a^2)$.
Assuming the damping rate $\gamma\sim 1.5\times 10^{-5}$~Hz, the axion's power spectrum width $\gamma<m_av_a^2$ for $m_a\gtrsim 10^{-14}$~eV (and vice versa at lower axion masses), such that we must consider both cases.

The amplitude of the axion-induced peak of the power spectrum (indigo peak in Fig.~\ref{fig:freq_shift_effects}) is thus given by
\begin{align}\label{eq:freq_shift_PSD}
    S_{xx}^a(\Omega+\omega_a)=m^2S_{x * \delta\Omega^2}(\Omega+\omega_a)\abs{\chi_m(\Omega+\omega_a)}^2 \,\, .
\end{align}
In our work, we assume that the effect of the axion effective force can be seen when it reaches a similar level to the external driving noise force power spectrum, which is the the point where the perturbative formalism breaks down. Thus, we may obtain a $\mathcal{O}(1)$ estimate of the sensitivity by plugging Eqs.~\ref{eq:freq_shift_PSD}~\&~\ref{eq:total_noise_PSD} in Eq.~\ref{eq:SNR_freq_shift_Sxx} at $\omega=\Omega+\omega_a$. The detectability condition then becomes
\begin{align}\label{eq:SNR_freq_shift}
    \text{SNR}^\text{eff}_x\gtrsim 1 \iff \begin{cases}
        \,\, \,\,\,\,\,\,\dfrac{2\delta\Omega^2}{\Omega\gamma N_T^{1/4}}\gtrsim 1 \,\,\,\,\,\,\,\,\,\,\, , \,\,\,\,\, m_av_a^2<\gamma \\ \,\, \dfrac{2\delta\Omega^2}{\Omega\sqrt{m_av_a^2\gamma}N_T^{1/4}}\gtrsim 1\,\,\,\,\, , \,\,\,\,\, m_av_a^2>\gamma 
    \end{cases}
\end{align}
where we used that the noise power spectrum is dominated by the backaction noise (rather than the shot noise) $S^N_{xx}(\Omega+\omega_a)\approx S_{xx}^\text{BA}(\Omega+\omega_a)=4\omega_LP_L\abs{\chi_m(\Omega+\omega_a)}^2$. We note that because $N_T=50$, the SNR$_\text{x}^\text{eff}$ is smaller than unity in any one bin, such that studying a perturbative change to the power spectrum in this regime remains valid\footnote{Although in the case $N_T\sim 1$ taking SNR$_x^\text{eff}\sim 1$ would lie outside of the perturbative regime, we may expect a monotonic dependence of the SNR$_x^\text{eff}$ on the axion couplings such that values larger than those which that which maximize the condition SNR~$\lesssim 1$ still lead to observable effects, albeit of unknown amplitude.}. We note that in the relatively small region of $\omega_a$ where on-resonance frequency shifts occur, \textit{i.e.}~$\abs{\omega_a-\Omega}<\gamma$, the Mathieu equation presents (an often unstable) parametric resonance which also lead to visible effect in the power spectrum, typically through a significantly increased motion excitation rate which affects the amplitude of the central peak at $\Omega$ and which often leads to the particle escaping the trap if non-linear damping is not applied. 
We see in Eq.~\ref{eq:SNR_freq_shift} that in the regime where $m_av_a^2<\gamma$, the frequency shift condition shows that these shifts can be seen at any $\omega_a$ (equivalently, $m_a$) as long as $\delta\Omega^2/\Omega\gtrsim\gamma$, assuming $\delta\Omega^2$ is independent of $m_a$. Meanwhile in the regime where $m_av_a^2>\gamma$, the frequency shift needs to scale as $\sqrt{\omega_a}$ because the bandwidth of the effect grows width $m_a$.

Similarly to the appearance of a noise power spectrum due to the magnetization noise of the sensor, we must also check the effect of magnetization fluctuations on the frequency of the sensor, and determine whether it limits the detection of (axion-induced) frequency shifts. We recall for frequency shifts we use a sextupole magnetic rotated with an additional constant magnetic field as shown in Eq.~\ref{eq:B_field_freq_shift}. The presence of the potential term 
\begin{align}
    \mu_xB_x=\mu_fN\tanh(\mu_fB_1/k_BT)\frac{B_0^2}{2B_1}x^4 \,\, ,
\end{align}
where we used $\mu_x=\mu_fP_xN$ and the polarization projection along the $\mathbf{\hat{x}}$-axis $P_x \approx (B_x/B_y) \tanh(\mu_f B_1 / k_B T)$ in the small angle limit, leads to an additional time-dependent frequency shift of the oscillator frequency (referred to as a `Duffing' oscillator in the presence of a quartic potential) given by $\Omega^2\rightarrow\Omega^2+\Omega_\mu^2$ where
\begin{align}
    \Omega_\mu^2=\mu_f\tanh(\mu_fB_1/k_BT)\frac{B_0^2}{4m_nB_1}\sigma_x^2\approx  10^{-4}~\text{Hz}\,\left(\frac{\omega_\text{SQL}}{30~\text{Hz}}\right)^2\left(\frac{\mu_f}{\mu_p}\right) \,\, ,
\end{align}
and where we used the identity $\sigma_x^2=\omega_\text{SQL}^2/(\Omega^2m\gamma)$~\cite{Hamaide:2025buy} and assumed a fixed mechanical frequency $\Omega\sim 15$~Hz (which we will justify shortly), and show the result with protons as a benchmark case. Thus the frequency shift is expected to be subdominant up to $\omega_\text{SQL}\approx 15$~kHz when assuming polarized nucleons and up to $\omega_\text{SQL}\approx 700$~Hz when assuming polarized electrons, above which the linear increase in the sensor position's variance with $\omega_\text{SQL}$ leads to a (approximately) linear increase in the oscillator frequency. 

Further, the longitudinal and transverse fluctuations of the magnetization with variance $\sigma_\mu^2$ (as described in Sec.~\ref{paragraph:magnetization_noise}) lead to fluctuations of the oscillator frequency around the total mechanical frequency given by $\sqrt{\Omega^2+\Omega_\mu^2}$. This can be modeled as an additional contribution to the damping rate $\gamma\rightarrow \gamma+\delta\Omega_\mu^2/\Omega$ where~\cite{Isaac_2017}
\begin{align}
    \delta\Omega_\mu^2=\mu_f\sqrt{N(1-\tanh(\mu_fB_1/k_BT)^2)}\frac{B_0^2}{2m B_1}\sigma_x^2= \Omega_\mu^2\frac{\sqrt{1-\tanh(\mu_fB_1/k_BT)^2}}{\sqrt{N}\tanh(\mu_fB_1/k_BT)} \,\, ,
\end{align}
such that $\delta\Omega_\mu^2/\Omega_\mu\propto\Omega_\mu\propto \omega_\text{SQL}$ is expected to be smaller than the oscillator's damping rate $\gamma\sim 1.5\times 10^{-5}$~Hz for all off-resonance SQL frequencies until $\omega_\text{SQL}$ reaches $\omega_\text{SQL}\approx 2$~kHz when assuming polarized nucleons or $\omega_\text{SQL}\approx 100$~Hz when assuming polarized electrons. Above this point $\delta\Omega_\mu^2/\Omega_\mu$ is expected to increase linearly %(\text{i.e.}~$\delta\Omega_\mu^2\sim\Omega_\mu^2(k_BT/\mu_fB_1\sqrt{N})$) 
until it reaches $\gamma\approx 4.5\times 10^{-4}$~Hz at $\omega_\text{SQL}\approx 60$~kHz. In summary, although accounting for magnetization fluctuations by replacing $\Omega\rightarrow\sqrt{\Omega^2+\Omega_\mu^2}$ and $\gamma\rightarrow \gamma+\delta\Omega_\mu^2/\Omega$ in the detectability conditions of the axion coupling Eq.~\ref{eq:SNR_freq_shift} leads to negligible corrections for $\omega_\text{SQL}\lesssim 2$~kHz or $100$~Hz for nucleons and electrons respectively, the detectability of the axion-induced frequency shift $\delta\Omega^2$ for higher axion Compton frequencies scales with an additional factor of $\omega_\text{SQL}$ until $\omega_\text{SQL}\approx 15$~kHz or $700$~Hz (for nucleons and electrons respectively), above which it scales with an additional factor of $\omega_\text{SQL}^2$. We note that this applies to both $g_f$ and $g_d$, and since $\delta\Omega^2\propto g_f$ or $g_d$, the sensitivity to these couplings has an additional $m_a$ and $m_a^2$ scaling above these frequencies.

At frequencies where magnetization noise can be neglected (\textit{i.e.}~where Eq.~\ref{eq:SNR_freq_shift} holds), we note that choosing a relatively low mechanical frequency for the sensor (which is nearly independent of the off-resonance SQL frequency for $\omega_a\gg\Omega$) directly improves the sensor sensitivity to both axion couplings at all frequencies. Therefore, we set the mechanical frequency $\Omega$ for frequency shift searches of axions to be $\Omega\approx 15~\text{Hz}\lesssim\omega_\text{SQL}^\text{min}\approx 30$~Hz. This assumes thermal noises (such as magnetization noise, blackbody radiation, or vibrations) do not dominate at this frequency for either levitated or suspended sensors.
Thankfully, although the analysis we performed for frequency shifts uses $S_{xx}$ instead of $S_\text{FF}$, since thermal noise was shown to always be subdominant to the off-resonance SQL noise floor in $S_\text{FF}$ it is necessarily subdominant when compared using $S_{xx}$ since both thermal and quantum noises in position or force units are related by an identical factor of $\abs{\chi_m(\omega)}^{\pm2}$.

In both cases, the sensitivity condition Eq.~\ref{eq:SNR_freq_shift} hinges on the need for sufficiently large laser power $P_L$ such that the backaction is larger than the shot noise at the first axion-induced peak located at $\Omega+\omega_a$. Although this can also be in principle achieved with large thermal noise instead of a large backaction noise, however this lowers the proton spin polarization of the sensor which scales as $P\sim 1/T$, so we do not consider this approach in this work. 
The required laser power for the shot noise to match the backaction noise is typically achievable and given by the SQL condition at $\omega_a$ given by $P_L=m(\Omega+\omega_a)^{2}/4\omega_L$~\cite{Hamaide:2025buy}. 
This may increase the temperature of the sensor (and thus decrease its polarization) due to larger optical absorption which is expected due to the larger required laser powers at sufficiently high frequencies at which we reach the off-resonance SQL. Crucially however, since the laser power required is identical to that for torque and force off-resonance SQL detection, assuming a long slim cuboid as in the force and torque sensing cases considered previously, we expect the same sensor temperatures and polarization at different axion frequencies $\omega_a$ as previously discussed in Sec.~\ref{sec:thermal_limits}. 

Remarkably, the detectability condition in Eq.~\ref{eq:SNR_freq_shift} also does not depends on $\omega_a$ beyond any dependence of $\delta\Omega^2$ on $\omega_a$ (ignoring the effect of magnetization noise at higher frequencies). Further, it does not depend on the off-resonance SQL frequency as long as the axion frequency of interest respects $\omega_a\lesssim\omega_\text{SQL}$, which is due to the backaction force noise $S_\text{FF}^\text{BA}(\omega)$ being spectrally flat below $\omega_\text{SQL}$ such that both the signal and noise scale as  $\omega_a^{-4}$ and $\omega^{-4}$ respectively. This however assumes laser absorption-induced heating does not affect polarization or heat the material above the Curie temperature or melting temperature when operating at this off-resonance SQL frequency. Frequency shifting is thus a naturally broadband search method for any axion signal with frequency in the region $\omega_a\lesssim \omega_\text{SQL}$. Once sufficiently high frequencies where absorption leads to significant heating of the material are reached, the search for the axion-induced frequency shift requires scanning at the off-resonance SQL to optimize the sensitivity at every frequency, analogously to force sensing which performs broadband searches at low frequency via heterodyning, while homodyne detection requires frequency scanning at higher frequencies until the melting temperature or Curie temperature is reached.

We also note that operating the sensor at the off-resonance SQL increases the total integrated motion power spectrum (called center of mass temperature~\cite{Aspelmeyer:2013lha,Millen_2020}) $\int_{\omega>0} S_{xx}(\omega)d\omega=\bar{\sigma}_x^2$ as a function of $\omega_\text{SQL}=\Omega+\omega_a$, %large excitation of the motion.
and thus increases the variance of the motion $\sigma_x^2$. At large enough variance, typically order $\sigma_x\sim\lambda_L /Q$~\cite{Gieseler:2013ael} (although this can be designed to be larger in principle) where $\lambda_L$ is the wavelength of the optical field, this leads to non-linearities in the oscillator's EOM, which would require an increased damping rate $\gamma$. We leave to future work the determination of the upper frequency at which non-linearities cannot be avoided, and assume in this work that thermal limitations dominate instead.

%%%%%%%%%%%%%%%%%%%%%%%%%%%%%%%%%%%%%%%%%%%%%%%%%
\subsection{Prospects for axion-induced frequency shifts}\label{sec:prospects_freq_shift}
%%%%%%%%%%%%%%%%%%%%%%%%%%%%%%%%%%%%%%%%%%%%%%%%%

%%%%%%%%%%%%%%%%%%%%%%%%%%%%%%%%%%%%%%%%%%%%%%%%%
\subsubsection{Axion-fermion couplings \texorpdfstring{$g_f$}{gf} -- ferromagnetic sensor}\label{sec:g_f_ferro_freq_shift_prospects}
%%%%%%%%%%%%%%%%%%%%%%%%%%%%%%%%%%%%%%%%%%%%%%%%%
We begin by discussing the detection of a frequency shift with ferromagnets in the small angle limit when setting the spin polarization is aligned with the axion wind, assumed in the $\mathbf{\hat{x}}$-axis, and focusing on the axion-electron coupling's first term. As discussed in Sec.~\ref{sec:axionpeakderivative_ferro}, this leads to a frequency shift $\delta\Omega_\theta^2\approx gm_av_aa(t)\,\sigma_xPN/I$. This frequency shift is independent of $m_a$ since $m_aa_0\approx\sqrt{2\rho_\text{DM}}$, assuming the polarization can be kept constant. It is also independent of the sensor mass $m$ since $I$ and the number of polarized spins and moment of inertia scale identically with mass. Using SNR$^\text{eff}_x\gtrsim 1$ condition Eq.~\ref{eq:SNR_freq_shift} with $\delta\Omega_\theta^2$ taken from Eq.~\ref{eq:gf_ferro_peak_split}, mechanical sensors are expected to be sensitive to the region
\begin{align}\label{eq:gf_ferro_freq_shift_sensitivity}
    g_e \gtrsim \frac{m_nL^2\Omega_\theta\sqrt{\gamma\max(\gamma,m_av_a^2)}}{6 \sqrt{2\rho_\text{DM}}\,v_aP N_T^{1/4}} \ ,
\end{align}
where we used $N=m/2m_n$, $I=mL^2/6$, $\sigma_x=1/2$ for a nucleon spin polarized in the $x$ direction, and that $\gamma\sim 1.5\times 10^{-5}$~Hz~$\sim 10^{-20}$~eV. We note the minimum coupling we are sensitive to using a ferromagnet does not require a large polarization gradient (simply a large overall electronic polarization) which does not constrain the geometry, and that the dependence on the mass has also disappeared compared to the force/torque sensing case. However the dependence on the half-length of the sensor (moment arm $L$) remains such that to minimize this we assume a microrod geometry with $L\sim\mathcal{O}(1~\mu\text{m})$ as used {\it e.g.}~in the force sensing case. This conveniently allows to have a single sensor design optimizing both frequency shift and torque sensing.

Using a mechanical sensor with a fixed mechanical frequency $\Omega_\theta\sim 15~\text{Hz}\lesssim\omega_\text{SQL}^\text{min}$, we also assume a constant integration time at all frequencies assuming a single power spectrum integration time with SQL frequency set to $\omega_\text{SQL}\approx 240$~kHz is necessary to detect any signal with frequencies $\omega_a\lesssim 240~\text{kHz}\approx 2\times 10^{-10}$~eV. In the region $m_a\lesssim 10^{-14}$~eV where the signal width is fixed, we may fix $N_T=50$ independently of the axion mass, assuming a year of continuous integration of the power spectrum. For larger axion masses $10^{-14}~\text{eV}\lesssim m_a\lesssim 2\times 10^{-10}$~eV, this integration time corresponds to $N_T=50\,(m_av_a^2/\gamma)$, which contributes to an overall scaling of the bounds as $m_a^{1/4}$.

%%%%%%%%%%%%%%%%%%%%%%%%%%%%%%%%%%%%%%%%%%%%%%
\subsubsection{Axion-fermion couplings \texorpdfstring{$g_f$}{gf} -- paramagnetic sensor}\label{sec:g_f_freq_shift_prospects}
%%%%%%%%%%%%%%%%%%%%%%%%%%%%%%%%%%%%%%%%%%%%%%
We now look at the case of paramagnetic materials, where the detectability condition using Eq.~\ref{eq:SNR_freq_shift} with Eq.~\ref{eq:gf_peak_split}, and including the magnetization noise effects discussed in Sec.~\ref{sec:optimization_freq_shift} (via substitutions of $\Omega$ and $\gamma$), can be written as
\begin{align}\label{eq:gf_freq_shift_sensitivity}
    g_f \gtrsim \frac{2m_n\sqrt{(\Omega^2+\Omega_\mu^2)}(\sqrt{\gamma\max(\gamma,m_av_a^2)}+\delta\Omega_\mu^2/\Omega)}{\sqrt{2\rho_\text{DM}} v_a(\partial^2 P/\partial x^2)N_T^{1/4}} \ ,
\end{align}
where again we used $N=m/2m_n$ and that $\sigma_x=1/2$. We may assume a similar mechanical sensor with a fixed mechanical frequency $\Omega\sim 15$~Hz.

We recall that we obtain the paramagnetic polarization of fermions using the sextupole magnetic field, with an additional $B_1$ homogeneous component, and which is rotated by $\theta=\pi/2$ in the $x-y$ plane (but otherwise identical to that used for force and torque sensing as displayed in Fig.~\ref{fig:setup}). In this setup, motion along the $\mathbf{\hat{x}}$-axis (for $\abs{x}\lesssim\mathcal{O}(1~\mu$m)) leaves the amplitude of the polarization approximately constant for small motion $\delta x$ in the $\mathbf{\hat{x}}$-axis around the coordinate $\vec{x}=(0,1\mu\text{m},z)$, while the polarization angle rotates linearly such that $\theta_P\propto x$. This allows the projection of the polarization along the $\mathbf{\hat{x}}$-axis to oscillate freely, such that we may estimate second gradient of the polarization as a function of the sensor position along the $\mathbf{\hat{x}}$-axis as $\partial^2 P/\partial x^2=(\kappa\mu_n/k_BT)\partial^2B_x/\partial x^2\approx 4 \times 10^{-3}$~eV$^{2}$ for protons and neutrons, and $\partial^2 P/\partial x^2=(\kappa\mu_n/k_BT)\partial^2B_x/\partial x^2\approx 4 \times 10^{-7}$~eV$^{2}$ for electrons, for axion mass $10^{-14}\text{eV}\lesssim m_a\lesssim 4\times 10^{-11}$~eV. Meanwhile, for axion masses $4\times 10^{-11}~\text{eV}\lesssim m_a \lesssim 4\times 10^{-10}~\text{eV}$ we can estimate $\partial^2 P/\partial x^2\approx 4 \times 10^{-7}$~eV$^{2}~(4\times 10^{-11}~\text{eV}/m_a)^{1/2}$ for protons and neutrons, while we obtain $\partial^2 P/\partial x^2\approx 4 \times 10^{-11}$~eV$^{2}~(4\times 10^{-11}~\text{eV}/m_a)^{1/2}$ for electrons given the larger temperature due to cooling limitations provided by the mechanical coupling. At these larger axion masses, to optimize the sensor's sensitivity we search the signal by tuning the sensor to operate at the off-resonance SQL at every axion mass. 

We note that in the region where the polarization gradient is constant and the effect of magnetization noise is negligible, we can assume values of $N_T=50(m_av^2/\gamma)$ for $m_a\gtrsim 10^{-14}$~eV and $N_T=50$ constant for $m_av^2\lesssim\gamma$ (\textit{i.e.}~$m_a\lesssim 10^{-14}$~eV). This leads to projected sensitivities to $g_f$ which scale as $m_a^{1/4}$ identically to the ferromagnetic case, and which are expected to be constant for $m_a\lesssim 10^{-14}$~eV. Meanwhile, in the region, $4\times 10^{-11}~\text{eV}\lesssim m_a \lesssim 4\times 10^{-10}~\text{eV}$ we may assume a constant $N_T\sim 10^5$ (\textit{i.e.}~$50(m_av^2/\gamma)$ where $m_a\sim 4\times 10^{-11}~\text{eV}$) at every frequency bin (similarly to how $N_T$ is set to be fixed across frequencies in Ref.~\cite{Hamaide:2025buy} when searching for DM forces by scanning frequencies at the off-resonance SQL). This leads to projected bounds on $g_f$ which are expected to scale as $m_a^3$ (including all effects). We note the sensitivity limit is now independent of the sensor's mass for all axion masses, such that a long slim cuboid whose longest direction is along the $\mathbf{\hat{z}}$-axis, as used for force and torque sensing (see Sec.~\ref{sec:prospects_derivative_coupling}), is also ideally suited to sense axion-induced frequency shifts.

%%%%%%%%%%%%%%%%%%%%%%%%%%%%%%%%%%%%%%%%%%%%%%%%%
\subsubsection{Axion-EDM coupling \texorpdfstring{$g_d$}{gd}} \label{sec:edm_freq_shift_prospects}
%%%%%%%%%%%%%%%%%%%%%%%%%%%%%%%%%%%%%%%%%%%%%%%%%
We now look at the frequency shift induced by the EDM-coupling's first term, as discussed in Sec.~\ref{sec:peak_split_EDM}. We again use the detectability (SNR) condition from Eq.~\ref{eq:SNR_freq_shift} with Eq.~\ref{eq:gd_peak_split}, and include the effect of magnetization noise as done with the axion-fermion coupling, to obtain 
\begin{align}\label{eq:gd_freq_shift_sensitivity}
    g_d\gtrsim \frac{2m_n\,m_a\sqrt{(\Omega^2+\Omega_\mu^2)}(\sqrt{\gamma\max(\gamma,m_av_a^2)}+\delta\Omega_\mu^2/\Omega)}{\sqrt{2\rho_\text{DM}}E (\partial^2 P/\partial x^2) N_T^{1/4}} \ ,
\end{align}
We assume a suspended mechanical sensor cooled via its mechanical coupling with the same mechanical frequency, and damping rate, as discussed in Sec.~\ref{sec:g_f_freq_shift_prospects}. We also assume same internal (ferroelectric) electric field as discussed in Sec.~\ref{sec:prospect_force_EDM} given by $E=10^3$~eV$^2$ (including the Schiff suppression factor). 

We consider the same second gradient of the polarization and $N_T$ (\textit{i.e.}~integration time) as in the paramagnetic case for $g_f$ in Sec.~\ref{sec:g_f_freq_shift_prospects} as well as the same frequency-dependent $N_T$ for paramagnetic materials, \textit{i.e.}~$N_T=50$ for $m_a\lesssim 10^{-14}$~eV, while $N_T=50\,(m_av_a^2/\gamma)$ for $10^{-14}~\text{eV}\lesssim m_a\lesssim 4\times 10^{-11}$~eV, and $N_T\sim 10^4$ for $m_a\gtrsim 4\times 10^{-11}$~eV where the polarization's second gradient scales as $\partial^2 P/\partial x^2\propto m_a^{-1/2}$. Therefore, up to an additional factor of $m_a$ (from Eq.~\ref{eq:gd_freq_shift_sensitivity}) we expect the projected sensitivity to scale identically with respect to axion mass as the sensitivity discussion for the paramagnetic case for protons and neutrons in Sec.~\ref{sec:g_f_freq_shift_prospects}.

%%%%%%%%%%%%%%%%%%%%%%%%%%%%%%%%%%%%%%%%%%%%%%%%%%
\subsection{Sensitivity projections and comparison to current bounds}\label{sec:sensitivity_projections}
%%%%%%%%%%%%%%%%%%%%%%%%%%%%%%%%%%%%%%%%%%%%%%%%%%

We now compare the sensitivities of the methods discussed in the previous sections, which we tabulated in Table~\ref{tab:summary}, to current bounds.

%%%%%%%%%%%%%%%%%%%%%%%%%%%%%%%%%%%%%%%%%%%%%%%%%%
\subsubsection{Axion-electron coupling \texorpdfstring{$g_e$}{ge}}\label{sec:bounds_g_e}
\begin{figure}[htb!]
    \centering
    \includegraphics[width=1.0\linewidth,page=3]{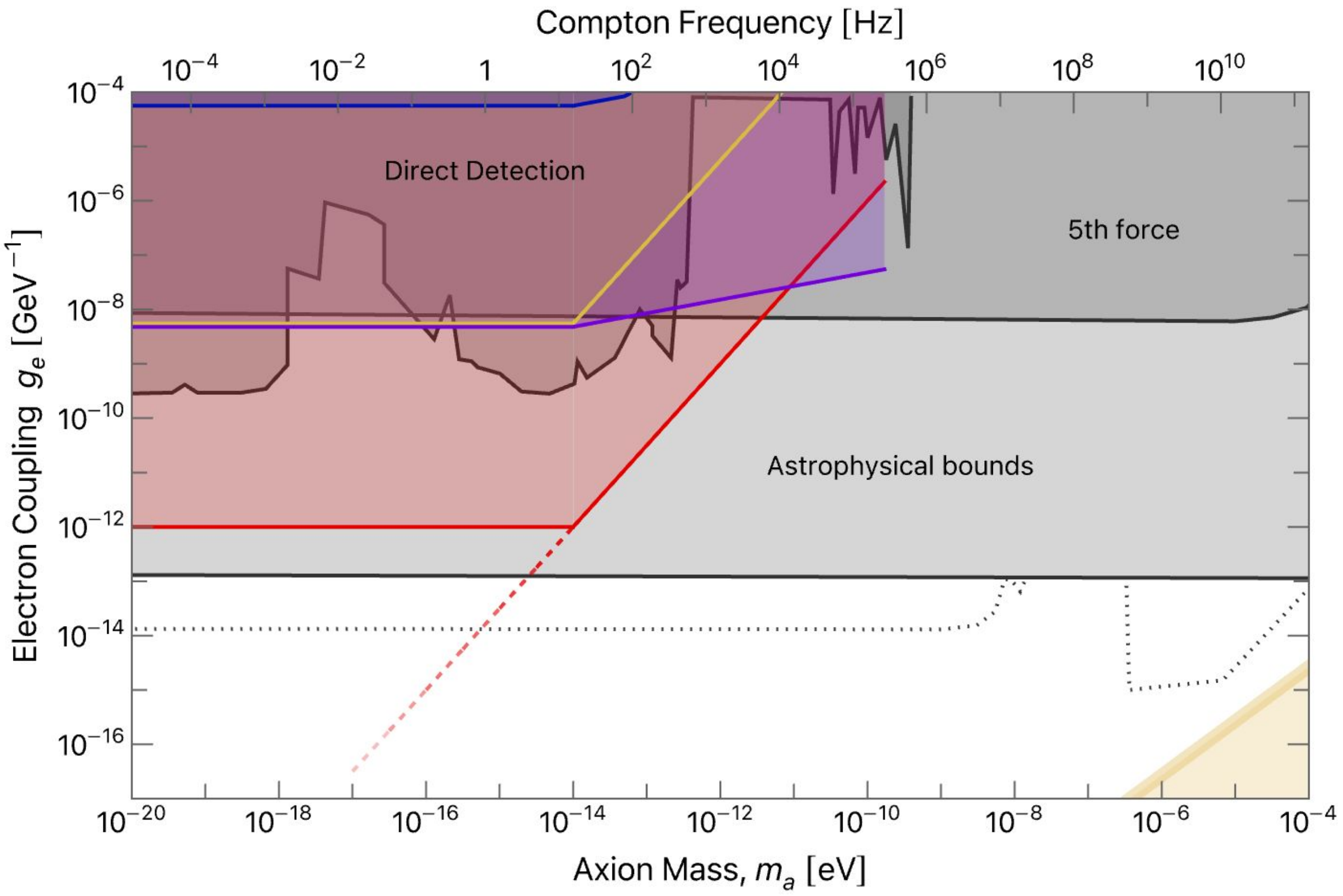}%{gae_constraints_plot.pdf}
    \caption{Projected sensitivity on the axion-electron coupling $g_e$ using a ferromagnetic mechanical quantum sensor, using both searches for axion-induced oscillating torque (red) and frequency shift searches (indigo). We also show the equivalent projected sensitivities for paramagnetic sensors for force sensing (yellow) and frequency shift searches (blue). We show competing bounds in gray, which we class into three categories: direct detection experiments (typically underground), astrophysical bounds, and fifth force searches. We also show future proposed experiments with black dashed lines, as well as parameter space of KFSZ and DFSZ QCD axion models in gold in the bottom right corner.}
    \label{fig:gae_sensitivity}
\end{figure}

%%%%%%%%%%%%%%%%%%%%%%%%%%%%%%%%%%%%%%%%%%%%%%%%%%
We first discuss the sensitivity to the axion-electron coupling $g_e$ using a long slim cuboid-like to searching for force and torques, by plotting  Eq.~\ref{eq:gf_ferro_force_sensitivity} as a function of axion mass in Fig.~\ref{fig:gae_sensitivity}. Assuming a ferromagnetic material (red line) is projected to reach $g_e\gtrsim 10^{-12}$~GeV$^{-1}$ for masses $m_a\lesssim 10^{-14}$~eV in the heterodyne regime. In the homodyne regime this scales as $m_a^{3/2}$ for $m_a\gtrsim 10^{-14}~\text{eV}$ since we scan frequencies by tuning the sensor be at off-resonance SQL. We see a ferromagnet is expected to be more sensitive than current direct detection bounds by over two orders of magnitude for all axion with mass $m_a\lesssim 3\times 10^{-11}$, and by six orders of magnitude near $m_a\sim 10^{-16}$~eV. Our sensor also significantly outperforms current fifth force sensors by four orders of magnitude for $m_a\lesssim 10^{-14}$~eV. 

In the short term, astrophysical bounds are expect to outperform our proposed setup by an order of magnitude, however we note that a larger sensor with $m\sim \mathcal{O}(0.1~$mg) would be expected to outperform these experiments. Given the small mass and resources necessary to create such a setup, scaling the number of sensors in an array (rather than increasingly long sensors in the $\mathbf{\hat{z}}$-axis) may be preferable. Next, using a ferromagnet to search for a frequency shifts (shown in indigo, obtained from Eq.~\ref{eq:gf_ferro_freq_shift_sensitivity}) is found to be complementary to torque searches, surpassing torque sensing sensitivity for axion masses $m_a\gtrsim 6\times 10^{-12}$~eV. This is because of the naturally weak dependence of the sensitivity to $m_a$ at higher masses, in addition to their naturally broadband behavior at low masses where frequency shifts can reach $g_e\gtrsim 4.8\times 10^{-9}$~GeV$^{-1}$. When comparing the detectability conditions Eqs.~\ref{eq:gf_ferro_force_sensitivity}~\&~\ref{eq:gf_ferro_freq_shift_sensitivity}, we see this is due to scaling differences with respect to $m_a$.

In the (electron) paramagnetic case, where we search for axion-induced forces in a long slim cuboid geometry coupled to a dilution fridge at axion masses below $m_a\lesssim 4\times 10^{-11}$~eV, plotting Eq.~\ref{eq:gf_force_sensitivity} leads to a sensitivity $g_e\gtrsim 5.5\times 10^{-9}$~GeV$^{-1}$ for all axion masses below $m_a\sim 10^{-14}~\text{eV}$, while $g_e\gtrsim 5.5\times 10^{-9}$~GeV$^{-1}(m_a/10^{-14}~\text{eV})^{3/2}$ for axion masses $10^{-14}~\text{eV}\lesssim m_a\lesssim 10^{-9}$~eV. These projections (plotted in yellow in Fig.~\ref{fig:gae_sensitivity}) are, as expected, less sensitive than ferromagnets due to their relatively lower electron polarization. As our conservative benchmark, we nevertheless find electron paramagnetic materials are more sensitive than current fifth force bounds and can surpass direct detect experiments by over two orders of magnitude. However, when conducting a frequency shift search using electron paramagnets (plotted in blue, obtained from Eq.~\ref{eq:gf_freq_shift_sensitivity}), we obtain $g_e\gtrsim 2.8\times 10^{-5}$~GeV$^{-1}$, which is not sensitivity to new parameter space in the near future.

%%%%%%%%%%%%%%%%%%%%%%%%%%%%%%%%%%%%%%%%%%%%%%%%%%
\subsubsection{Axion-proton coupling \texorpdfstring{$g_p$}{gp} and axion-neutron coupling \texorpdfstring{$g_n$}{gn}}\label{sec:bounds_g_p}
%%%%%%%%%%%%%%%%%%%%%%%%%%%%%%%%%%%%%%%%%%%%%%%%%%

\begin{figure}
    \centering
    \includegraphics[width=1.0\linewidth,page=4]{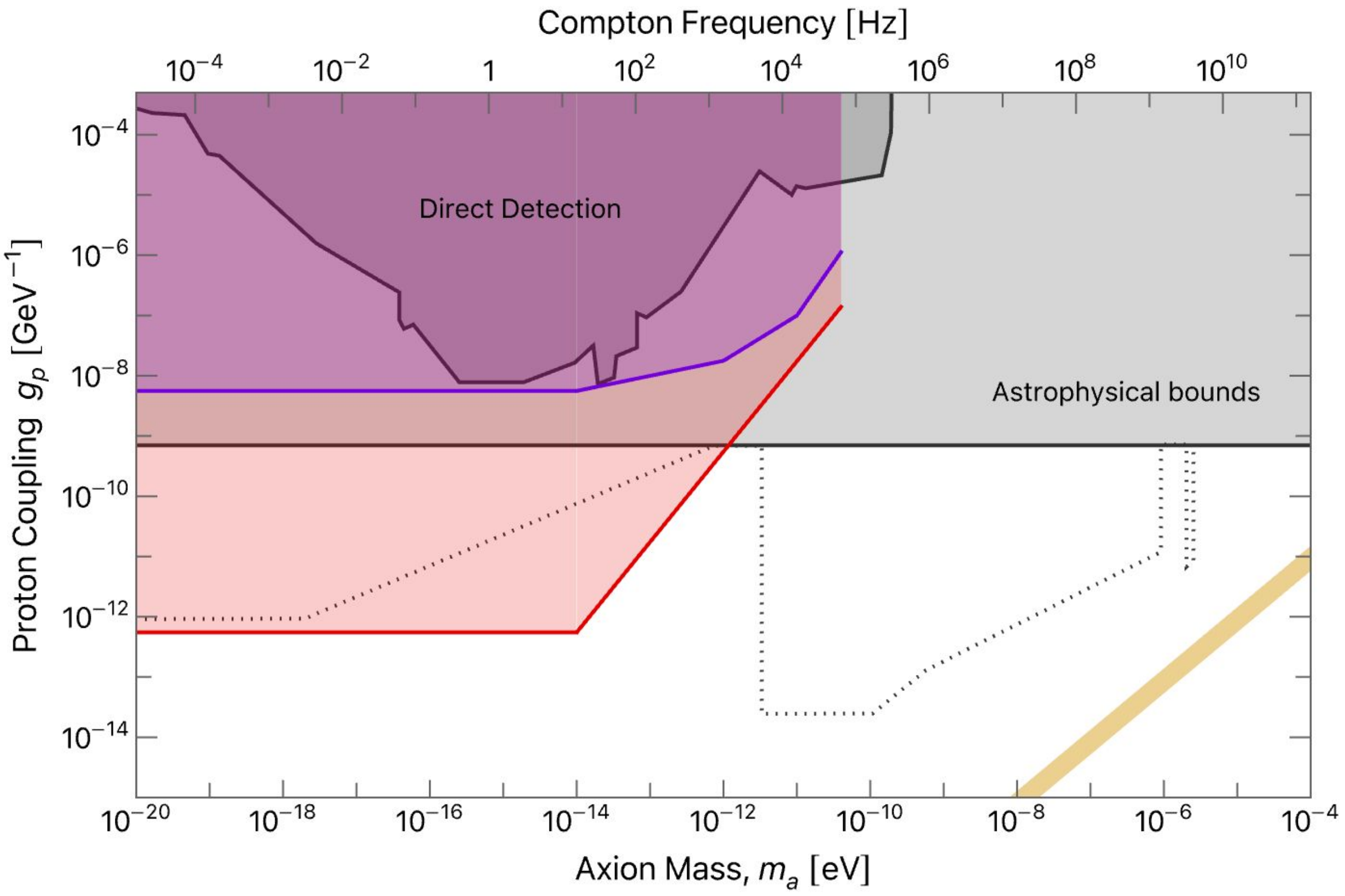}%{gap_constraints_plot.pdf}
    \caption{Projected sensitivity on the axion-proton coupling $g_p$ using mechanical quantum sensors, using both searches for axion-induced oscillating force (red) and frequency shift searches (indigo). We show competing bounds in gray, which we class into three categories: underground direct detection experiments, astrophysical bounds, and fifth force searches. We also show future proposed experiments with black dashed lines, as well as parameter space of KFSZ and DFSZ QCD axion models in gold in the bottom right corner.}
    \label{fig:gap_sensitivity}
\end{figure}

\begin{figure}
    \centering
    \includegraphics[width=1.0\linewidth,page=6]{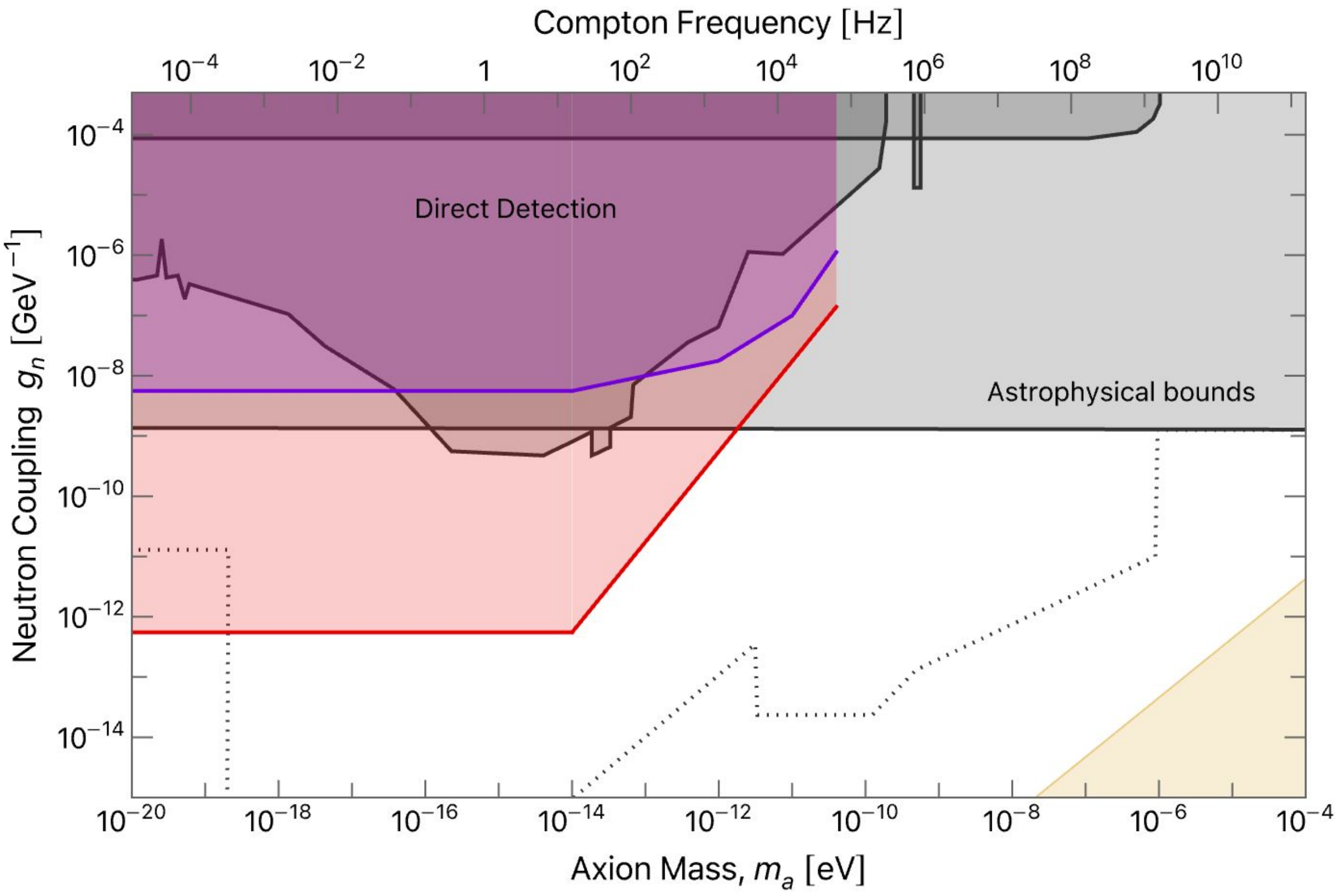}
    \caption{Projected sensitivity on the axion-neutron coupling $g_n$ using mechanical quantum sensors, using both searches for axion-induced oscillating force (red) and frequency shift searches (indigo). We show competing bounds in gray, which we class into three categories: underground direct detection experiments, astrophysical bounds, and fifth force searches. We also show future proposed experiments with black dashed lines, as well as parameter space of DFSZ QCD axion models in gold in the bottom right corner.}
    \label{fig:gan_sensitivity}
\end{figure}

We next show the projected sensitivity to the axion-proton and axion neutron coupling in paramagnetic materials in Figs.~\ref{fig:gap_sensitivity}~\&~\ref{fig:gan_sensitivity}, assuming a dilution fridge at axion masses below $m_a\lesssim 4\times 10^{-11}$~eV. Given we assume an identical polarizability for both (this may also be achieved using a single material such as lithium deuteride to simultaneously search for both couplings), when searching for oscillating forces, Eq.~\ref{eq:gf_force_sensitivity} leads to projected sensitivities for both couplings shown in red and given by $g_f\sim 5.5\times 10^{-13}$~GeV$^{-1}$ where $f=(p,n)$ for $m_a \lesssim m_\text{SQL}^\text{min}=10^{-14}~\text{eV}$. 
For axion masses $10^{-14}~\text{eV}\lesssim m_a\lesssim 4\times 10^{-11}~\text{eV}$ this further scales as $m_a^{3/2}$ where the additional scaling (as in the electron paramagnetic case) is due to the SQL noise scaling and the bandwidth of the axion signal. The sensitivity of a paramagnetic material to $g_f$ therefore surpasses current experimental bounds for all axion masses $m_a\lesssim 10^{-12}$~eV by over three orders of magnitude, and surpasses all currently proposed experiments (to our knowledge) for all $m_a\lesssim 10^{-12}$~eV for $g_p$ and all $m_a\lesssim 2\times10^{-19}$~eV for $g_n$. 
Meanwhile, above $m_a\gtrsim 4\times 10^{-11}$~eV, the polarization gradient rapidly decreases significantly due to blackbody cooling which dominates over the cryogenic cooling such that we obtain the projections in this mass range that lie just outside of the parameter space shown in Fig.~\ref{fig:gap_sensitivity}, and are therefore not sensitive to new parameter space. 

Meanwhile, the projected sensitivity to the axion-proton and axion-neutron couplings when searching for frequency shifts given by Eq.~\ref{eq:gf_freq_shift_sensitivity} is shown in indigo in Figs.~\ref{fig:gap_sensitivity}~\&~\ref{fig:gan_sensitivity} and covers the region $g_f\gtrsim 5\times 10^{-9}~\text{GeV}^{-1}$ where $f=(p,n)$ for axion masses $m_a\lesssim 10^{-14}~\text{eV}$, while above this an additional scaling as $m_a^{1/4}$ appears, until magnetization noise leads to more rapid scaling above $m_a\sim 10^{-12}$~eV and $m_a\sim 10^{-11}$~eV due to the increases of the mechanical frequency and damping rate respectively above these axion masses, and such that the sensitivity scales $g_f\propto m_a^{3/4}$ and $g_f\propto m_a^{7/4}$ respectively. Although sensitivity in this region is still roughly a factor of four above astrophysical observations, it surpasses direct detection bounds by up to two orders of magnitude at the lowest allowed axion masses, providing crucial complementary Earth-based bounds in this parameter space. Finally, for axion masses $ m_a\gtrsim 4\times 10^{-11}~\text{eV}$, projected sensitivities cover the region $g_p\gtrsim 5\times 10^{-4}~\text{GeV}^{-1}~(m_a/4\times 10^{-11}~\text{eV})^{5/2}$, which is just outside the parameter space shown in Fig.~\ref{fig:gap_sensitivity} and therefore not sensitive to new parameter space.

%%%%%%%%%%%%%%%%%%%%%%%%%%%%%%%%%%%%%%%%%%%%%%%%%%
\subsubsection{Axion-EDM coupling \texorpdfstring{$g_d$}{gd}}\label{sec:bounds_g_d}
\begin{figure}
    \centering
    \includegraphics[width=1.0\linewidth,page=5]{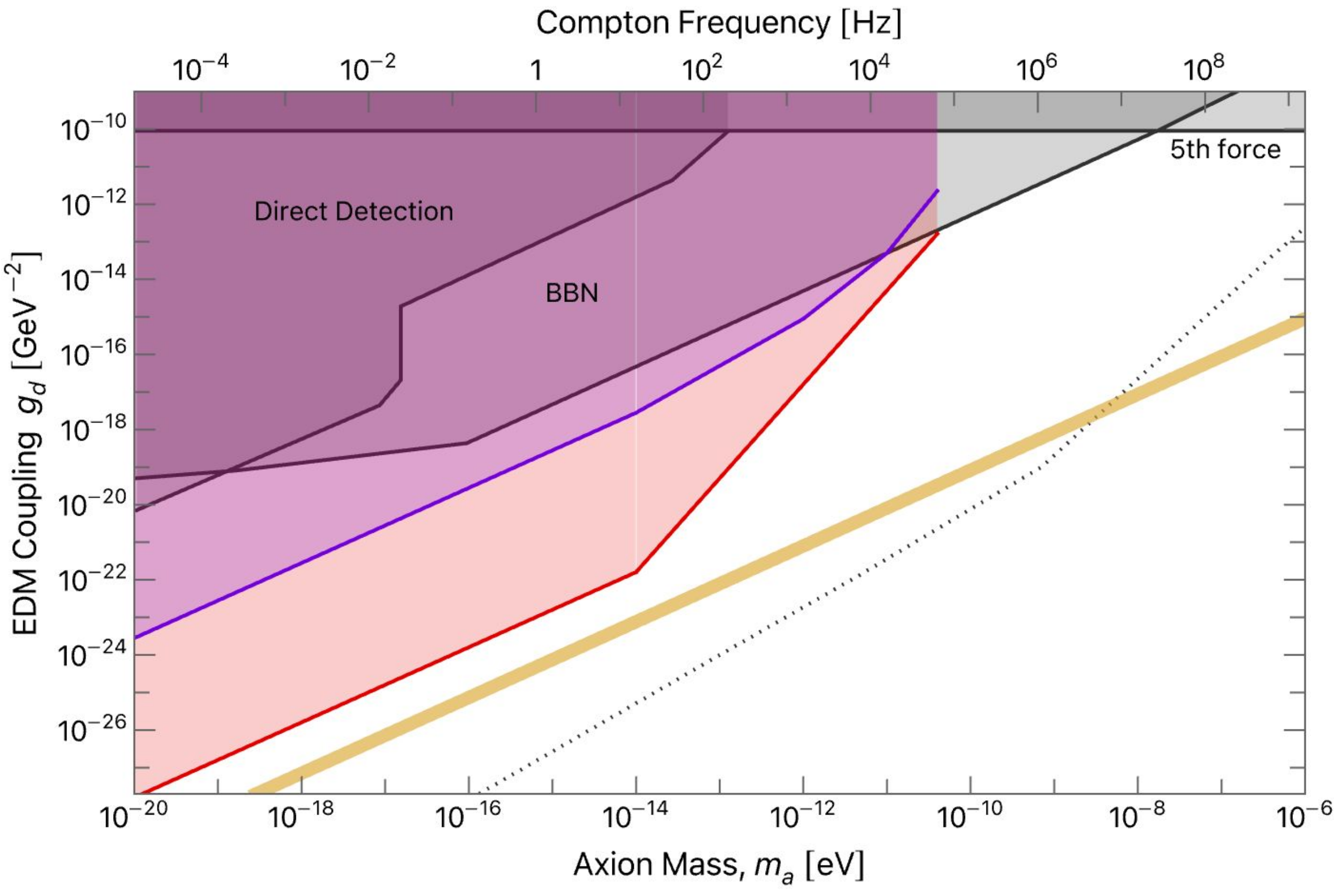}%{gd_constraints_plot.pdf}
    \caption{Projected sensitivity on the axion-EDM coupling $g_d$ using mechanical quantum sensors, using both searches for axion-induced oscillating force (red) and frequency shift searches (indigo). We show competing bounds in gray, which we class into three categories: underground direct detection experiments, big bang nucleosynthesis (BBN) bounds, and fifth force searches. We also show future proposed experiments with black dashed lines, and parameter space of KFSZ and DFSZ QCD axion models in gold.}
    \label{fig:EDM_sensitivity}
\end{figure}

%%%%%%%%%%%%%%%%%%%%%%%%%%%%%%%%%%%%%%%%%%%%%%%%%%
We now show the projected sensitivity to the axion-EDM coupling $g_d$ using a ferroelectric material in Fig.~\ref{fig:EDM_sensitivity}, assuming the same conditions as with searches for $g_f$ discussed in Sec.~\ref{sec:bounds_g_p}, with the exception of a different polarizability. When looking for force on the sensor, Eq.~\ref{eq:gd_sensitivity} leads to a lower bound on the axion-EDM coupling shown in red which is given by $g_d\sim 1.6\times 10^{-22}$~GeV$^{-2}~(m_a/10^{-14}~\text{eV})^{3/2}$ for axion masses $m_a\lesssim 10^{-14}~\text{eV}$, while for axion masses $10^{-14}~\text{eV}\lesssim m_a\lesssim 4\times 10^{-11}~\text{eV}$, we obtain $g_d\sim 1.6\times 10^{-22}$~GeV$^{-2}~(m_a/10^{-14}~\text{eV})^{5/2}$. Finally, for $m_a\gtrsim 4\times 10^{-11}$~eV, we obtain $g_d \sim  1.6\times 10^{-9}$~GeV$^{-2}~(m_a/4\times 10^{-11}~\text{eV})^3$, which is just outside of the range of the figure and therefore not sensitive to new parameter space.

Our force sensing projections are overall significantly more sensitive than current bounds, outperforming direct detection bounds by seven orders of magnitude and Big Bang nucleosynthesis by up to eight orders of magnitude at the lowest axion masses, as shown in Fig.~\ref{fig:EDM_sensitivity}. Although our setup is not expected to be as sensitive as long term phase III projections from CASPEr Electric~\cite{JacksonKimball:2017elr} shown with a black dotted line, our sensitivity is expected to outperform CASPEr phase I projections by several orders of magnitude and is expected to be similar to CASPEr phase II projections. Importantly, these projections are only an order of magnitude away from probing DFSZ and KFSZ QCD axion model parameter space, implying this may be attainable in the future with further development, notably by pushing for larger sensor masses, higher spin polarizability, higher effective electric fields, or reaching lower quantum noises.

Next, when looking for the axion-EDM coupling via a frequency shift, the projected sensitivity region of our setup as given by Eq.~\ref{eq:gd_freq_shift_sensitivity} is shown in indigo and is given by $g_d\gtrsim 2.8\times 10^{-18}$~GeV$^{-2}~(m_a/10^{-14}~\text{eV})$ for axion masses below $m_a\lesssim 10^{-14}$~eV, while for $10^{-14}~\text{eV}\lesssim m_a\lesssim 4\times 10^{-11}$~eV the sensitivity scales as $m_a^{5/4}$. This scaling is respected until magnetization noise leads to faster scaling with $m_a$ above $m_a\sim 10^{-12}$~eV and $m_a\sim 10^{-11}$~eV due to the linear scaling of the mechanical frequency and damping rate respectively above these masses, and such that the sensitivity scales $g_f\propto m_a^{7/4}$ and $g_f\propto m_a^{11/4}$ respectively. Finally, in the region $10^{-9}~\text{eV}\lesssim m_a\lesssim 10^{-7}$~eV we obtain $g_d\gtrsim 2.2\times 10^{-8}$~GeV$^{-2}(m_a/10^{-9}~\text{eV})^{7/4}$, where the scaling change is due to the scaling of $\partial^2 P/\partial x^2\propto m_a^{-1/2}$ in the blackbody cooling regime.

Looking at Fig.~\ref{fig:EDM_sensitivity}, we see that frequency shifts are not complementary to force searches, unlike in the case of ferromagnets used to search for the axion-electron coupling, and perform less well overall compared to force searches. Nevertheless they are expected to surpass current bounds for all axion masses below by over three orders of magnitude. Future development of methods allowing to increase the maximum axion frequency would allow to exploit the scaling of its sensitivity at higher axion masses. Nevertheless, compared to current experimental bounds on $g_d$, frequency shift search remains a powerful detection that is projected to surpass all current experimental bounds at all axion masses up to $m_a\sim 10^{-11}$~eV, and by up to three orders of magnitude at the lowest axion masses.

%%%%%%%%%%%%%%%%%%%%%%%%%%%%%%%%%%%%%%%%%%%%%%%%%
\subsection{Future improvements in thermal, seismic and quantum noises}
%%%%%%%%%%%%%%%%%%%%%%%%%%%%%%%%%%%%%%%%%%%%%%%%%
We now consider what near-future improvements could lead to sensors which are more sensitive to the fermion and EDM dark matter couplings in different frequency ranges, and beginning with low mass axions. At low frequencies $\omega_a\lesssim\mathcal{O}(10)$~Hz, the noise in our detection schemes (both force and frequency shift) is primarily limited by the seismic noise, which is experienced by all Earth-based sensors despite state-of-the-art mitigation efforts aimed at noise mitigation~\cite{galaxies10010020}. The achievable sensitivity expected in a space mission is therefore expected to be much larger at these frequency, in the absence of seismic noise. Although work to determine this noise is still required, we note that if the residual noise at low frequencies is white (\textit{e.g.}~thermal), the off-resonance SQL noise at low frequencies may also be used instead of heterodyning given the spectrally flat force noise power spectrum below $\omega_\text{SQL}$. Further, encouragingly such an experiment is modest in scale and resources as a setup similar to the one discussed here (but slightly shorter in the $\mathbf{\hat{z}}$-axis) would fit inside a commercial cubesat, such as the proof-of-concept LOMS~\cite{Homans:2025eks} experiment which was recently launched.

Assuming sufficiently low gas pressures and temperatures are achievable in space, at low frequencies, noises are expected to be dominated by the off-resonance SQL noise floor. However this is tempered by the integration time, which (assuming a space mission with around 3 years of data), would evolve such that $N_T\propto 1/m_a$ until that $N_T=1$ is reached at $m_a\sim10^{-16}$~Hz, and such that the sensitivity would be expected to scale as $m_a^{5/4}$ (as opposed to $m_a^{3/2}$ \textit{e.g.}~as in Eqs.~\ref{eq:gf_ferro_force_sensitivity}~\&~\ref{eq:gf_force_sensitivity}) in the region $10^{-16}~\text{eV}\lesssim m_a\lesssim10^{-14}$~eV. Assuming noises can be further suppressed, below this point we assume the bandwidth $1/T_\text{int}$ would be constant such that the sensitivity would scale linearly with $m_a$ (corresponding to the scaling at ultralow frequencies when $N_T<1$). At the lowest point in the noise power spectrum,  In Fig.~\ref{fig:gae_sensitivity} we thus included the sensitivity which a space mission would be expected to achieve assuming an ideal environment and control of other systematic as a long term goal. If made shorter in the $\mathbf{\hat{z}}$-axis, a slim cuboid sensor such as the geometry considered in this work could fit in a cubesat and would require laser powers significantly lower than 1~Watt to operate at the off-resonance SQL at these frequencies, which is well within the capabilities of commercial cubsats. However this also requires more careful study of the noises ({\it e.g.}~thermal and vibrational) in space, which lies beyond of the scope of this work.

Next, we discuss a possible experimental improvement for axion masses $m_a\gtrsim 10^{-14}$~eV. Although optically trapped sensors have the advantage of having highly tunable mechanical frequencies, currently this tunability comes at the cost of larger laser powers such that the SQL can only ever be reached off-resonance. Assuming a more advanced optical configuration (\textit{e.g.}~inspired by recent work by Ref.~\cite{Melo:2020prx}) which can trap the sensor at a point of relatively low laser intensity, on-resonance SQL noise may be achievable in optically trapped systems in the near future. The potential improvement on the SQL noise compared to the off-resonance SQL is well-approximated by the inverse quality factor $Q^{-1}=\gamma/\Omega$, which is typically much smaller than one in optomechanical systems. However, this also quickly reduces the bandwidth (given by $\gamma$) of a single frequency bin and the available integration time over each frequency bin, such that assuming the integration time remains constant, the potential improvement only weakly scales with the quality factor of the achievable on-resonance SQL. Thus, although on-resonance SQL operation is expected to improve sensitivity to axions above $m_a\gtrsim 10^{-14}$~eV, we postpone a precise determination of the maximum expected improvement from on-resonance to future work.

\section{Conclusion}
%%%%%%%%%%%%%%%%%%%%%%%%%%%%%%%%%%%%%%%%%%%%%%%%%%

The search for axion dark matter through several couplings to the SM has been the object of increasingly diverse effort in recent years, with many new proposed experiments and detection strategies expected to lead to improved bounds. In particular, while astrophysical bounds on axion-fermion and axion-EDM couplings in the ultralight mass range have been historically dominated by astrophysical bounds, novel Earth-based experiments are expected to improve on these in the near future and reach new ALP and even QCD axion parameter space. Despite this, detection of axions through their fermion couplings, especially to protons and neutrons, is still understudied when compared to the relatively large range of novel quantum detectors. By using mechanical quantum sensors designed to probe these coupling, this can be improved upon. Such sensors are developed to sense forces, torques and frequency shift and can be made using a variety of materials and trapping mechanisms (see \textit{e.g.}~Ref.~\cite{Aspelmeyer:2013lha} for a review). These parameters can be carefully chosen to require large spin polarizations and couplings to the axion field, which requires controlled experimental conditions.

As we have shown, well chosen mechanical quantum sensors can be highly sensitive to these axion couplings which lead to high quality oscillating forces and torques. In particular, both levitated and mechanically coupled and uncoupled optically trapped sensors have complementary roles to play given the temperature requirements of different types of materials to obtain strong axion signals. Further, a variety of detection schemes developed in the field of force and torque sensing can be used to drastically broaden the reach of these sensors. First, heterodyning the axion signal can lead to naturally broadband sensitivity to the axion signal at low frequencies. Broadband searches are useful for searches of both a single axion of unknown mass or multiple axions over a range of masses (as in `axiverse' theories, see {\it e.g.}~\cite{Stott:2017hvl,Arvanitaki:2009fg}) signals. Next, high quality frequency shift detection allows for broadening the reach of these sensors, in particular to frequencies above the resonant frequency of the sensor, due to the advantageous scaling of the signal of frequency oscillations at higher frequencies. Finally, carefully engineered magnetic fields with large gradients can be used to lead to oscillatory translational forces, notably making the detection of the axion-electron couplings possible with materials that so far had not been considered for axion detection. 

Among these materials, paramagnetic materials with advantageous properties for optomechanical experiments can be made to be sensitive to axion-proton and axion-neutron couplings at levels beyond all current bounds. Materials such as ferroelectrics, can also be used in novel ways (as mechanical sensors), with sensitivity to the axion-EDM coupling projected to far surpass current experiments and reaching parameter space less than two orders of magnitude away from the QCD axion band. More generally, once magnetically polarized, an optically trapped system as shown in Sec.~\ref{sec:Noises} can be used set significantly improved bounds on the axion-proton, axion-neutron and axion-EDM couplings in the near future with off-the-shelf technologies, and significantly surpass current Earth-based bounds on the axion-electron coupling. Although this depends on the benchmark polarization one considers, preliminary estimate of the temperature and thermal noise of highly polarized materials shows they can be operated as optomechanical sensors. Nevertheless a more detailed discussion of the experimental parameters, thermal dynamics, and fabrication feasibility remain important questions which we plan to address in future work comprising a detailed experimental proposal. 

In general, the methods studied here demonstrate how mechanical quantum sensors can be highly versatile sensors when their three primary parameters $(m,\gamma,\Omega)$ (in any given direction) can be optimized to obtain the best results for different couplings. We also note that the desired axion-induced effects used for these methods/regimes can remarkably be achieved with a single sextupole (or even parallel wire) magnetic field configuration, {\it i.e.}~for paramagnets and ferromagnets, although a more extensive study of various magnetic field configurations may lead to further improvements to the signal. Although in this study we assume optically trapped sensors, this presents unique challenges for paramagnetic and ferromagnetic materials, whose higher optical absorption emphasizes the need for ultrahigh reflectivity mirrors. Expanding the possible magnetic field configuration to involve a large homogeneous static external magnetic field used to trap ferromagnets, such that the average magnetic field felt by the sensor can no longer be made to be zero (or nearly zero), leads to quantum fluctuations of the magnetic field which may be advantageous (over optical fluctuations) but is left for future work. However, preliminary studies~\cite{Higgins:2023gwq} of the quantum noises ({\it i.e.}~both induced by fluctuations of these fields) of both optical and magnetic field noises suggests the scaling of the noise in both types of mechanical sensor respect universal scaling with the size of the external trapping and readout fields, such that the expected sensitivities are expected to be similar for similar sized systems.

We also note several directions to improve further on these projected sensitivities. In particular, surpassing the current (astrophysical) bounds on the axion-electron coupling is also possible for the axion-electron coupling if either one of several different improvements can be made. First, a larger mass, as already discussed for the long term prospects of both magnetically and optically levitated sensors~\cite{Hamaide:2025buy,Kalia:2024eml,Amaral:2024rbj}, would be expected to lead to improved sensitivity projection which scale as $m^{1/2}$. Although in the short term optically levitated sensors up to the $\mathcal{O}(1$~mg) scale are expected to be achievable which increases further to $\mathcal{O}(1$~g) in the long term, studies of magnetically trapped sensors with potential masses up to $\mathcal{O}(100$~g) as discuss in Refs.~\cite{Kalia:2024eml,Amaral:2024tjg} have been considered which may ultimately be more desirable for probing the axion-electron coupling. 

Next, simultaneous optimization of quantum sensors to reach the SQL on-resonance while maintaining tunability of the mechanical frequency is an open critical question in the field of mechanical quantum sensors, which has notably motivated the recent study and variety of proposed levitated mechanical sensors in recent years. 
The rapid development of the field, notably with the use of so-called `dark foci' traps~\cite{Melo:2020prx} fuels optimism that reaching SQL on-resonance over a range of frequencies will be possible in the near future. Finally, optomechanical sensors are expected to perform better in space, primarily due to naturally lower thermal cooling requirements and the expected absence of seismic noise. Although a study of these noises is outside the scope of this work, a quiet environment in space is expected to lead to lowering of noise by several orders of magnitude below the current minimum off-resonance SQL frequency and which could be used to detect axions via a heterodyning strategy. We believe these promising directions justify current excitement in the field and an in-depth investigation of these questions.

%%%%%%%%%%%%%%%%%%%%%%%%%%%%%%%%%%%%%%%%%%%%%%%%
\section*{Acknowledgements}

The authors would like to thank Hannah Banks and Saarik Kalia for their helpful comments on an early draft of this paper, as well as Asher Berlin and Dmitry Budker for useful discussions. The authors would like to thank the IPPP Durham, Fermilab Theory group and King's College London for their hospitality during the final stages of this work. This work is supported by the Cluster of Excellence {\em Precision Physics, Fundamental Interactions and Structure of Matter\/} (PRISMA${}^{++}$ -- EXC~2118/2) within the German Excellence Strategy (project ID 390831469). LH is supported by the Alexander von Humboldt Foundation. The authors acknowledge the use of LLMs, namely Copilot, for literature searches and cross-checking analytic expressions.

\appendix
%%%%%%%%%%%%%%%%%%%%%%%%%%%%%%%%%%%%%%%%%%%%%%%%%
\section{Precession in equations of motion}\label{app:EOM_precession}
%%%%%%%%%%%%%%%%%%%%%%%%%%%%%%%%%%%%%%%%%%%%%%%%%
The coupled angular EOMs in $\phi$ and $\theta$ which we reiterate for convenience:
\begin{align}\label{eq:EOM_angular_app}
    \begin{cases}
        I\ddot{\phi} + I \gamma_\phi \dot{\phi} + I \Omega^2_\phi \phi + \dot{\theta}PN\sigma_y(t) = \tau_N^\phi(t) \\
        I\ddot{\theta} + I \gamma_\theta \dot{\theta} + I \Omega^2_\theta \theta - \dot{\phi}PN\sigma_x(t) = \tau_N^\theta(t)- \tau_a(t) \,\, ,
    \end{cases}
\end{align}
may be expected to lead to a procession in $\phi$ when the axion coupling is sufficiently large to excite significant motion in $\theta$. Depending on the amplitude of this precession, it can then partly (or mostly) slow the motion in $\theta$ by canceling $\tau_a$. To understand these effects, we begin by assuming the sensor is at rest in $\phi$ and is excited in $\theta$ by an axion-induced torque such that $\tau_a>\tau_N$ in the frequency range $\omega\in[m_a(1-v^2_a/2),m_a(1+v_a^2)]$, such that the axion effect dominates. Assuming the sensor is operated at the off-resonance SQL at  a frequency $\omega_a=m_a$ for both $\theta$ and $\phi$\footnote{This is expected for a long slim cuboid geometry since $\Omega_\phi$ is estimated by $\Omega_\phi=\sqrt{\tau_\phi(\phi)/I\phi}\approx\sqrt{P_L L_x/I_\phi}<\Omega_\theta$.}, the resulting angular power spectrum of the sensor at $\omega=\omega_a$ is given by
\begin{align}
    S_{\theta\theta}(m_a)=S_{\tau\tau}(m_a)\abs{\chi_m(m_a)}^{2}=\frac{\tau_a^2}{I_\theta^2 m_a^4\max(m_av_a^2,\gamma)} \,\, ,
\end{align}
where we used that $\abs{\chi_m(m_a)}\approx(I m_a^2)^{-1}$ at frequencies such that $\omega_a-\Omega>\gamma$~\cite{Hamaide:2025buy}. From this we obtain the angular power spectrum near around $\omega=\omega_a$ of $\phi$ induced by torque from motion in $\theta$ in the usual way identifying the torque from Eq.~\ref{eq:EOM_angular_app} to obtain
\begin{align}\label{eq:phi_precession}
    S_{\phi\phi}(m_a)
    %= \frac{m_a^2 S_{\theta\theta}(m_a)(m_av_a^2)(PN\sigma_y(t))^2}{I_\phi^2 m_a^4(m_av_a^2)}
    =\frac{S_{\theta\theta}(m_a)(PN\sigma_y(t))^2}{I_\phi^2m_a^2}  \,\, .
\end{align}
%where the $m_a^2$ in the numerator comes from the time derivative of $\theta$ and . 
Since the torque off-resonance SQL condition is given by~\cite{Hamaide:2025buy}
\begin{align}\label{eq:angular_off_res_SQL}
    S_{\tau\tau}^\text{SQL}(\omega)=I\omega^2\sim S^\text{BA}_\text{FF}(\omega)L^2=\omega_LP_LL^2 \,\, ,
\end{align}
the requirement on the laser power is independent of the sensor moment arm length $L$. Thus, assuming we operate the sensor in $\theta$ at the off-resonance SQL, this implies the $\phi$ degree of freedom is also operated at the off-resonance SQL. Therefore, comparing Eq.~\ref{eq:phi_precession} to the noise power spectrum of $\phi$ in Eq.~\ref{eq:angular_off_res_SQL} taken at $\omega=\omega_a$, we see that for all $m_a\gtrsim m_\text{SQL}^\text{min}\sim 10^{-14}$~eV and for axion couplings at (or somewhat larger than) the sensitivity projections in Figs.~\ref{fig:gae_sensitivity},~\ref{fig:gap_sensitivity}~\&~\ref{fig:EDM_sensitivity}, the power spectrum in Eq.~\ref{eq:phi_precession} is smaller, such that the induced precession is not detectable. We note this is partly due to the relatively large value of $I_\phi$ (assuming a long slim cuboid geometry as in Sec.~\ref{sec:Noises}) and that a sensor with smaller anisotropy would likely lead to detectable gyroscopic precession as expected. We also note that for sufficiently large axion couplings this may no longer be the case. Although these couplings would be large, one may also check that regardless of the value of the coupling the direct effect on the EOM $\theta$ is the largest driving force on the motion, such that observing axion-induced motion in $\phi$ may contribute a subleading detectable observable of the dark matter effect in this case.

Finally, we may also estimate the effect of the motion in $\phi$ when operating at the off-resonance SQL on the motion in $\theta$, to check the effect of this motion is smaller than the motion in $\theta$ expected at the off-resonance SQL. Similarly to  Eq.~\ref{eq:phi_precession}, we estimate the contribution of this motion on the power spectrum of $\theta$ as
\begin{align}
    S_{\theta\theta}(m_a)%=\frac{m_a^2 S_{\phi\phi}^\text{SQL}(m_a)m_av_a^2(PN\sigma_y(t))^2\abs{\chi_{m,\theta}}^2}{m_av_a^2}
    =\frac{m_a S_{\phi\phi}^\text{SQL}(PN\sigma_y(t))^2}{I_\theta^2 m_a^2}=S_{\theta\theta}^\text{SQL}\frac{(PN\sigma_y(t))^2}{I_\theta I_\phi m_a^2}
\end{align}
which is much smaller than the off-resonance SQL noise in $\theta$ denoted $S_{\theta\theta}^\text{SQL}$ for all $m_a\lesssim 10^{-14}$~eV mainly due to the relatively large moment of inertia $I_\phi$. Thus we may neglect the $\dot{\phi}$-term in Eq.~\ref{eq:EOM_angular_app} such that the EOM of $\theta$ may be treated as decoupled from the EOM of $\phi$ in this work.

%%%%%%%%%%%%%%%%%%%%%%%%%%%%%%%%%%%%%%%%%%%%%%%%%

\bibliographystyle{JHEP}
\bibliography{references}

@article{Arias:2012az,
    author = "Arias, Paola and Cadamuro, Davide and Goodsell, Mark and Jaeckel, Joerg and Redondo, Javier and Ringwald, Andreas",
    title = "{WISPy Cold Dark Matter}",
    eprint = "1201.5902",
    archivePrefix = "arXiv",
    primaryClass = "hep-ph",
    reportNumber = "DESY-11-226, MPP-2011-140, CERN-PH-TH-2011-323, IPPP-11-80, DCPT-11-160",
    doi = "10.1088/1475-7516/2012/06/013",
    journal = "JCAP",
    volume = "06",
    pages = "013",
    year = "2012"
}

@article{Co:2021rhi,
    author = "Co, Raymond T. and Harigaya, Keisuke and Pierce, Aaron",
    title = "{Gravitational waves and dark photon dark matter from axion rotations}",
    eprint = "2104.02077",
    archivePrefix = "arXiv",
    primaryClass = "hep-ph",
    reportNumber = "LCTP-21-08, UMN-TH-4010/21, FTPI-MINN-21-03",
    doi = "10.1007/JHEP12(2021)099",
    journal = "JHEP",
    volume = "12",
    pages = "099",
    year = "2021"
}

@article{Graham:2015rva,
    author = "Graham, Peter W. and Mardon, Jeremy and Rajendran, Surjeet",
    title = "{Vector Dark Matter from Inflationary Fluctuations}",
    eprint = "1504.02102",
    archivePrefix = "arXiv",
    primaryClass = "hep-ph",
    doi = "10.1103/PhysRevD.93.103520",
    journal = "Phys. Rev. D",
    volume = "93",
    number = "10",
    pages = "103520",
    year = "2016"
}

@article{Fayet:2025brx,
    author = "Fayet, Pierre",
    title = "{Yukawa potential of a nonhomogeneous sphere with new limits on an ultralight boson}",
    eprint = "2507.02723",
    archivePrefix = "arXiv",
    primaryClass = "hep-ph",
    doi = "10.1103/17wc-xxjm",
    journal = "Phys. Rev. D",
    volume = "112",
    number = "9",
    pages = "095018",
    year = "2025"
}

@article{Cirelli:2024ssz,
    author = "Cirelli, Marco and Strumia, Alessandro and Zupan, Jure",
    title = "{Dark Matter}",
    eprint = "2406.01705",
    archivePrefix = "arXiv",
    primaryClass = "hep-ph",
    month = "6",
    volume=" ",
journal="",
    year = "2024"
}

@article{Arvanitaki:2009fg,
    author = "Arvanitaki, Asimina and Dimopoulos, Savas and Dubovsky, Sergei and Kaloper, Nemanja and March-Russell, John",
    title = "{String Axiverse}",
    eprint = "0905.4720",
    archivePrefix = "arXiv",
    primaryClass = "hep-th",
    doi = "10.1103/PhysRevD.81.123530",
    journal = "Phys. Rev. D",
    volume = "81",
    pages = "123530",
    year = "2010"
}

@article{Svrcek:2006yi,
    author = "Svrcek, Peter and Witten, Edward",
    title = "{Axions In String Theory}",
    eprint = "hep-th/0605206",
    archivePrefix = "arXiv",
    reportNumber = "SLAC-PUB-11894",
    doi = "10.1088/1126-6708/2006/06/051",
    journal = "JHEP",
    volume = "06",
    pages = "051",
    year = "2006"
}

@article{Dine:1982ah,
    author = "Dine, Michael and Fischler, Willy",
    editor = "Srednicki, M. A.",
    title = "{The Not So Harmless Axion}",
    reportNumber = "UPR-0201T",
    doi = "10.1016/0370-2693(83)90639-1",
    journal = "Phys. Lett. B",
    volume = "120",
    pages = "137--141",
    year = "1983"
}

@article{Abbott:1982af,
    author = "Abbott, L. F. and Sikivie, P.",
    editor = "Srednicki, M. A.",
    title = "{A Cosmological Bound on the Invisible Axion}",
    reportNumber = "PRINT-82-0695 (BRANDEIS)",
    doi = "10.1016/0370-2693(83)90638-X",
    journal = "Phys. Lett. B",
    volume = "120",
    pages = "133--136",
    year = "1983"
}

@article{Preskill:1982cy,
    author = "Preskill, John and Wise, Mark B. and Wilczek, Frank",
    editor = "Srednicki, M. A.",
    title = "{Cosmology of the Invisible Axion}",
    reportNumber = "HUTP-82-A048, NSF-ITP-82-103",
    doi = "10.1016/0370-2693(83)90637-8",
    journal = "Phys. Lett. B",
    volume = "120",
    pages = "127--132",
    year = "1983"
}

@article{DiLuzio:2021pxd,
    author = "Di Luzio, Luca and Gavela, Belen and Quilez, Pablo and Ringwald, Andreas",
    title = "{An even lighter QCD axion}",
    eprint = "2102.00012",
    archivePrefix = "arXiv",
    primaryClass = "hep-ph",
    reportNumber = "DESY-21-010, DESY 21-010, IFT-UAM/CSIC-20-143, FTUAM-20-21",
    doi = "10.1007/JHEP05(2021)184",
    journal = "JHEP",
    volume = "05",
    pages = "184",
    year = "2021"
}

@article{Eberhardt:2025caq,
    author = "Eberhardt, Andrew and Ferreira, Elisa G. M.",
    title = "{Ultralight fuzzy dark matter review}",
    eprint = "2507.00705",
    archivePrefix = "arXiv",
    primaryClass = "astro-ph.CO",
    month = "7",
    year = "2025"
}

@article{Zhitnitsky:1980tq,
    author = "Zhitnitsky, A. R.",
    title = "{On Possible Suppression of the Axion Hadron Interactions. (In Russian)}",
    journal = "Sov. J. Nucl. Phys.",
    volume = "31",
    pages = "260",
    year = "1980"
}

@article{Dine:1981rt,
    author = "Dine, Michael and Fischler, Willy and Srednicki, Mark",
    title = "{A Simple Solution to the Strong CP Problem with a Harmless Axion}",
    reportNumber = "Print-81-0320 (IAS,PRINCETON)",
    doi = "10.1016/0370-2693(81)90590-6",
    journal = "Phys. Lett. B",
    volume = "104",
    pages = "199--202",
    year = "1981"
}

@article{Kim:1979if,
    author = "Kim, Jihn E.",
    title = "{Weak Interaction Singlet and Strong CP Invariance}",
    reportNumber = "UPR-0120T",
    doi = "10.1103/PhysRevLett.43.103",
    journal = "Phys. Rev. Lett.",
    volume = "43",
    pages = "103",
    year = "1979"
}

@article{Shifman:1979if,
    author = "Shifman, Mikhail A. and Vainshtein, A. I. and Zakharov, Valentin I.",
    title = "{Can Confinement Ensure Natural CP Invariance of Strong Interactions?}",
    reportNumber = "ITEP-64-1979",
    doi = "10.1016/0550-3213(80)90209-6",
    journal = "Nucl. Phys. B",
    volume = "166",
    pages = "493--506",
    year = "1980"
}

@article{Wilczek:1977pj,
    author = "Wilczek, Frank",
    title = "{Problem of Strong  $P$  and  $T$  Invariance in the Presence of Instantons}",
    reportNumber = "Print-77-0939 (COLUMBIA)",
    doi = "10.1103/PhysRevLett.40.279",
    journal = "Phys. Rev. Lett.",
    volume = "40",
    pages = "279--282",
    year = "1978"
}

@article{Weinberg:1977ma,
    author = "Weinberg, Steven",
    title = "{A New Light Boson?}",
    reportNumber = "HUTP-77/A074",
    doi = "10.1103/PhysRevLett.40.223",
    journal = "Phys. Rev. Lett.",
    volume = "40",
    pages = "223--226",
    year = "1978"
}

@article{Peccei:1977ur,
    author = "Peccei, R. D. and Quinn, Helen R.",
    title = "{Constraints Imposed by CP Conservation in the Presence of Instantons}",
    reportNumber = "ITP-572-STANFORD",
    doi = "10.1103/PhysRevD.16.1791",
    journal = "Phys. Rev. D",
    volume = "16",
    pages = "1791--1797",
    year = "1977"
}

@article{Cyncynates:2024yxm,
    author = "Cyncynates, David and Weiner, Zachary J.",
    title = "{Experimental targets for dark photon dark matter}",
    eprint = "2410.14774",
    archivePrefix = "arXiv",
    primaryClass = "hep-ph",
    doi = "10.1103/PhysRevD.111.103535",
    journal = "Phys. Rev. D",
    volume = "111",
    number = "10",
    pages = "103535",
    year = "2025"
}

@article{Cyncynates:2023zwj,
    author = "Cyncynates, David and Weiner, Zachary J.",
    title = "{Detectable and Defect-Free Dark Photon Dark Matter}",
    eprint = "2310.18397",
    archivePrefix = "arXiv",
    primaryClass = "hep-ph",
    doi = "10.1103/PhysRevLett.134.211002",
    journal = "Phys. Rev. Lett.",
    volume = "134",
    number = "21",
    pages = "211002",
    year = "2025"
}

@article{Adshead:2023qiw,
    author = "Adshead, Peter and Lozanov, Kaloian D. and Weiner, Zachary J.",
    title = "{Dark photon dark matter from an oscillating dilaton}",
    eprint = "2301.07718",
    archivePrefix = "arXiv",
    primaryClass = "hep-ph",
    doi = "10.1103/PhysRevD.107.083519",
    journal = "Phys. Rev. D",
    volume = "107",
    number = "8",
    pages = "083519",
    year = "2023"
}

@article{Caputo:2021eaa,
    author = "Caputo, Andrea and Millar, Alexander J. and O'Hare, Ciaran A. J. and Vitagliano, Edoardo",
    title = "{Dark photon limits: A handbook}",
    eprint = "2105.04565",
    archivePrefix = "arXiv",
    primaryClass = "hep-ph",
    reportNumber = "NORDITA-2021-036",
    doi = "10.1103/PhysRevD.104.095029",
    journal = "Phys. Rev. D",
    volume = "104",
    number = "9",
    pages = "095029",
    year = "2021"
}

@article{Dror:2018pdh,
    author = "Dror, Jeff A. and Harigaya, Keisuke and Narayan, Vijay",
    title = "{Parametric Resonance Production of Ultralight Vector Dark Matter}",
    eprint = "1810.07195",
    archivePrefix = "arXiv",
    primaryClass = "hep-ph",
    doi = "10.1103/PhysRevD.99.035036",
    journal = "Phys. Rev. D",
    volume = "99",
    number = "3",
    pages = "035036",
    year = "2019"
}

@article{Bastero-Gil:2018uel,
    author = "Bastero-Gil, Mar and Santiago, Jose and Ubaldi, Lorenzo and Vega-Morales, Roberto",
    title = "{Vector dark matter production at the end of inflation}",
    eprint = "1810.07208",
    archivePrefix = "arXiv",
    primaryClass = "hep-ph",
    reportNumber = "UG-FT 328-18, CAFPE 198-18, SISSA 44/2018/FISI",
    doi = "10.1088/1475-7516/2019/04/015",
    journal = "JCAP",
    volume = "04",
    pages = "015",
    year = "2019"
}

@article{Agrawal:2018vin,
    author = "Agrawal, Prateek and Kitajima, Naoya and Reece, Matthew and Sekiguchi, Toyokazu and Takahashi, Fuminobu",
    title = "{Relic Abundance of Dark Photon Dark Matter}",
    eprint = "1810.07188",
    archivePrefix = "arXiv",
    primaryClass = "hep-ph",
    reportNumber = "TU-1074,IPMU18-015,RESCEU-13/18,MIT-CTP/5066, TU-1074, IPMU18-015, RESCEU-13/18, MIT-CTP/5066",
    doi = "10.1016/j.physletb.2019.135136",
    journal = "Phys. Lett. B",
    volume = "801",
    pages = "135136",
    year = "2020"
}

@article{Goodsell:2009xc,
    author = "Goodsell, Mark and Jaeckel, Joerg and Redondo, Javier and Ringwald, Andreas",
    title = "{Naturally Light Hidden Photons in LARGE Volume String Compactifications}",
    eprint = "0909.0515",
    archivePrefix = "arXiv",
    primaryClass = "hep-ph",
    reportNumber = "IPPP-09-63, DCPT-09-126, DESY-09-123",
    doi = "10.1088/1126-6708/2009/11/027",
    journal = "JHEP",
    volume = "11",
    pages = "027",
    year = "2009"
}

@article{Nelson:2011sf,
    author = "Nelson, Ann E. and Scholtz, Jakub",
    title = "{Dark Light, Dark Matter and the Misalignment Mechanism}",
    eprint = "1105.2812",
    archivePrefix = "arXiv",
    primaryClass = "hep-ph",
    doi = "10.1103/PhysRevD.84.103501",
    journal = "Phys. Rev. D",
    volume = "84",
    pages = "103501",
    year = "2011"
}

@article{Abel:2008ai,
    author = "Abel, S. A. and Goodsell, M. D. and Jaeckel, J. and Khoze, V. V. and Ringwald, A.",
    title = "{Kinetic Mixing of the Photon with Hidden U(1)s in String Phenomenology}",
    eprint = "0803.1449",
    archivePrefix = "arXiv",
    primaryClass = "hep-ph",
    reportNumber = "IPPP-08-14, DESY-08-026",
    doi = "10.1088/1126-6708/2008/07/124",
    journal = "JHEP",
    volume = "07",
    pages = "124",
    year = "2008"
}

@article{Fabbrichesi:2020wbt,
    author = "Fabbrichesi, Marco and Gabrielli, Emidio and Lanfranchi, Gaia",
    title = "{The Dark Photon}",
    eprint = "2005.01515",
    journal = "SpringerBriefs in Physics 2020",
    archivePrefix = "arXiv",
    primaryClass = "hep-ph",
    doi = "10.1007/978-3-030-62519-1",
    month = "5",
    year = "2020"
}

@article{Banerjee:2018xmn,
    author = "Banerjee, Abhishek and Kim, Hyungjin and Perez, Gilad",
    title = "{Coherent relaxion dark matter}",
    eprint = "1810.01889",
    archivePrefix = "arXiv",
    primaryClass = "hep-ph",
    doi = "10.1103/PhysRevD.100.115026",
    journal = "Phys. Rev. D",
    volume = "100",
    number = "11",
    pages = "115026",
    year = "2019"
}

@article{Piazza:2010ye,
    author = "Piazza, Federico and Pospelov, Maxim",
    title = "{Sub-eV scalar dark matter through the super-renormalizable Higgs portal}",
    eprint = "1003.2313",
    archivePrefix = "arXiv",
    primaryClass = "hep-ph",
    doi = "10.1103/PhysRevD.82.043533",
    journal = "Phys. Rev. D",
    volume = "82",
    pages = "043533",
    year = "2010"
}

@article{TAYLOR1988450,
title = {Dilaton couplings at large distances},
journal = {Physics Letters B},
volume = {213},
number = {4},
pages = {450-454},
year = {1988},
issn = {0370-2693},
doi = {https://doi.org/10.1016/0370-2693(88)91290-7},
url = {https://www.sciencedirect.com/science/article/pii/0370269388912907},
author = {T.R. Taylor and G. Veneziano}
}

@article{Cicoli:2011yy,
    author = "Cicoli, M. and Burgess, C. P. and Quevedo, F.",
    title = "{Anisotropic Modulus Stabilisation: Strings at LHC Scales with Micron-sized Extra Dimensions}",
    eprint = "1105.2107",
    archivePrefix = "arXiv",
    primaryClass = "hep-th",
    reportNumber = "DESY-11-073",
    doi = "10.1007/JHEP10(2011)119",
    journal = "JHEP",
    volume = "10",
    pages = "119",
    year = "2011"
}

@article{Burgess:2010sy,
    author = "Burgess, C. P. and Maharana, Anshuman and Quevedo, F.",
    title = "{Uber-naturalness: unexpectedly light scalars from supersymmetric extra dimensions}",
    eprint = "1005.1199",
    archivePrefix = "arXiv",
    primaryClass = "hep-th",
    reportNumber = "IC-2010-013, DAMTP-2010-31",
    doi = "10.1007/JHEP05(2011)010",
    journal = "JHEP",
    volume = "05",
    pages = "010",
    year = "2011"
}

@article{Arkani-Hamed:1999lsd,
    author = "Arkani-Hamed, Nima and Hall, Lawrence J. and Tucker-Smith, David and Weiner, Neal",
    title = "{Solving the hierarchy problem with exponentially large dimensions}",
    eprint = "hep-ph/9912453",
    archivePrefix = "arXiv",
    reportNumber = "LBNL-44717, UCB-PTH-99-55, LBL-44717",
    doi = "10.1103/PhysRevD.62.105002",
    journal = "Phys. Rev. D",
    volume = "62",
    pages = "105002",
    year = "2000"
}

@article{Dimopoulos:1996kp,
    author = "Dimopoulos, S. and Giudice, G. F.",
    title = "{Macroscopic forces from supersymmetry}",
    eprint = "hep-ph/9602350",
    archivePrefix = "arXiv",
    reportNumber = "CERN-TH-96-47",
    doi = "10.1016/0370-2693(96)00390-5",
    journal = "Phys. Lett. B",
    volume = "379",
    pages = "105--114",
    year = "1996"
}

@article{FAYET1990743,
title = {Extra U(1)'s and new forces},
journal = {Nuclear Physics B},
volume = {347},
number = {3},
pages = {743-768},
year = {1990},
issn = {0550-3213},
doi = {https://doi.org/10.1016/0550-3213(90)90381-M},
url = {https://www.sciencedirect.com/science/article/pii/055032139090381M},
author = {Pierre Fayet}
}

@article{HOLDOM1986196,
title = {Two U(1)'s and epsilon charge shifts},
journal = {Physics Letters B},
volume = {166},
number = {2},
pages = {196-198},
year = {1986},
issn = {0370-2693},
doi = {https://doi.org/10.1016/0370-2693(86)91377-8},
url = {https://www.sciencedirect.com/science/article/pii/0370269386913778},
author = {Bob Holdom}
}

@article{DiLuzio:2020wdo,
    author = "Di Luzio, Luca and Giannotti, Maurizio and Nardi, Enrico and Visinelli, Luca",
    title = "{The landscape of QCD axion models}",
    eprint = "2003.01100",
    archivePrefix = "arXiv",
    primaryClass = "hep-ph",
    reportNumber = "DESY 20-036, DESY-20-036",
    doi = "10.1016/j.physrep.2020.06.002",
    journal = "Phys. Rept.",
    volume = "870",
    pages = "1--117",
    year = "2020"
}

@article{PhysRevLett.38.1440,
  title = {$\mathrm{CP}$ Conservation in the Presence of Pseudoparticles},
  author = {Peccei, R. D. and Quinn, Helen R.},
  journal = {Phys. Rev. Lett.},
  volume = {38},
  issue = {25},
  pages = {1440--1443},
  numpages = {0},
  year = {1977},
  month = {Jun},
  publisher = {American Physical Society},
  doi = {10.1103/PhysRevLett.38.1440},
  url = {https://link.aps.org/doi/10.1103/PhysRevLett.38.1440}
}

@article{SHIFMAN1980493,
title = {Can confinement ensure natural CP invariance of strong interactions?},
journal = {Nuclear Physics B},
volume = {166},
number = {3},
pages = {493-506},
year = {1980},
issn = {0550-3213},
doi = {https://doi.org/10.1016/0550-3213(80)90209-6},
url = {https://www.sciencedirect.com/science/article/pii/0550321380902096},
author = {M.A. Shifman and A.I. Vainshtein and V.I. Zakharov}
}

@inproceedings{Battaglieri:2017aum,
    author = "Battaglieri, Marco and others",
    title = "{US Cosmic Visions: New Ideas in Dark Matter 2017: Community Report}",
    booktitle = "{U.S. Cosmic Visions: New Ideas in Dark Matter}",
    eprint = "1707.04591",
    archivePrefix = "arXiv",
    primaryClass = "hep-ph",
    reportNumber = "FERMILAB-CONF-17-282-AE-PPD-T",
    month = "7",
    year = "2017"
}

@article{Jaeckel:2010ni,
    author = "Jaeckel, Joerg and Ringwald, Andreas",
    title = "{The Low-Energy Frontier of Particle Physics}",
    eprint = "1002.0329",
    archivePrefix = "arXiv",
    primaryClass = "hep-ph",
    reportNumber = "CPT-10-18, DESY-10-016, IPPP-10-09",
    doi = "10.1146/annurev.nucl.012809.104433",
    journal = "Ann. Rev. Nucl. Part. Sci.",
    volume = "60",
    pages = "405--437",
    year = "2010"
}

@article{Antypas:2022asj,
    author = "Antypas, D. and others",
    title = "{New Horizons: Scalar and Vector Ultralight Dark Matter}",
    eprint = "2203.14915",
    archivePrefix = "arXiv",
    primaryClass = "hep-ex",
    reportNumber = "FERMILAB-PUB-22-262-AD-PPD-T",
    month = "3",
journal="",
    year = "2022"
}

@article{PDG,
    author = {Particle Data Group and Workman, R L and Burkert, V D and Crede, V and Klempt, E and Thoma, U and Tiator, L and others },
    title = {Review of Particle Physics},
    journal = {Progress of Theoretical and Experimental Physics},
    volume = {2022},
    number = {8},
    pages = {083C01},
    year = {2022},
    month = {08},
    issn = {2050-3911},
    doi = {10.1093/ptep/ptac097},
    url = {https://doi.org/10.1093/ptep/ptac097},
    eprint = {https://academic.oup.com/ptep/article-pdf/2022/8/083C01/49175539/ptac097.pdf},
}

@article{Jaeckel:2022kwg,
    author = "Jaeckel, Joerg and Rybka, Gray and Winslow, Lindley",
    title = "{Report of the Topical Group on Wave Dark Matter for Snowmass 2021}",
    eprint = "2209.08125",
    journal = "Snowmass 2021",
    archivePrefix = "arXiv",
    primaryClass = "hep-ph",
    month = "9",
    year = "2022"
}

@article{Hui:2021tkt,
    author = "Hui, Lam",
    title = "{Wave Dark Matter}",
    eprint = "2101.11735",
    archivePrefix = "arXiv",
    primaryClass = "astro-ph.CO",
    doi = "10.1146/annurev-astro-120920-010024",
    journal = "Ann. Rev. Astron. Astrophys.",
    volume = "59",
    pages = "247--289",
    year = "2021"
}

@article{Choi:2020rgn,
    author = "Choi, Kiwoon and Im, Sang Hui and Sub Shin, Chang",
    title = "{Recent Progress in the Physics of Axions and Axion-Like Particles}",
    eprint = "2012.05029",
    archivePrefix = "arXiv",
    primaryClass = "hep-ph",
    reportNumber = "CTPU-PTC-20-28",
    doi = "10.1146/annurev-nucl-120720-031147",
    journal = "Ann. Rev. Nucl. Part. Sci.",
    volume = "71",
    pages = "225--252",
    year = "2021"
}

@article{Fayet:1980ad,
    author = "Fayet, Pierre",
    title = "{Effects of the Spin 1 Partner of the Goldstino (Gravitino) on Neutral Current Phenomenology}",
    reportNumber = "CERN-TH-2895",
    doi = "10.1016/0370-2693(80)90488-8",
    journal = "Phys. Lett. B",
    volume = "95",
    pages = "285--289",
    year = "1980"
}

@article{Fayet:1980rr,
    author = "Fayet, Pierre",
    title = "{On the Search for a New Spin 1 Boson}",
    reportNumber = "CERN-TH-2972",
    doi = "10.1016/0550-3213(81)90122-X",
    journal = "Nucl. Phys. B",
    volume = "187",
    pages = "184--204",
    year = "1981"
}

@article{Arvanitaki:2014faa,
    author = "Arvanitaki, Asimina and Huang, Junwu and Van Tilburg, Ken",
    title = "{Searching for dilaton dark matter with atomic clocks}",
    eprint = "1405.2925",
    archivePrefix = "arXiv",
    primaryClass = "hep-ph",
    doi = "10.1103/PhysRevD.91.015015",
    journal = "Phys. Rev. D",
    volume = "91",
    number = "1",
    pages = "015015",
    year = "2015"
}

@article{Amaral:2024rbj,
    author = "Amaral, Dorian W. P. and Uitenbroek, Dennis G. and Oosterkamp, Tjerk H. and Tunnell, Christopher D.",
    title = "{First Search for Ultralight Dark Matter Using a Magnetically Levitated Particle}",
    eprint = "2409.03814",
    archivePrefix = "arXiv",
    primaryClass = "hep-ph",
    doi = "10.1103/PhysRevLett.134.251001",
    journal = "Phys. Rev. Lett.",
    volume = "134",
    number = "25",
    pages = "251001",
    year = "2025"
}

@article{Kawasaki:2014sqa,
    author = "Kawasaki, Masahiro and Saikawa, Ken'ichi and Sekiguchi, Toyokazu",
    title = "{Axion dark matter from topological defects}",
    eprint = "1412.0789",
    archivePrefix = "arXiv",
    primaryClass = "hep-ph",
    reportNumber = "ICRR-REPORT-696-2014-22, IPMU14-0348",
    doi = "10.1103/PhysRevD.91.065014",
    journal = "Phys. Rev. D",
    volume = "91",
    number = "6",
    pages = "065014",
    year = "2015"
}

@article{Hiramatsu:2012gg,
    author = "Hiramatsu, Takashi and Kawasaki, Masahiro and Saikawa, Ken'ichi and Sekiguchi, Toyokazu",
    title = "{Production of dark matter axions from collapse of string-wall systems}",
    eprint = "1202.5851",
    archivePrefix = "arXiv",
    primaryClass = "hep-ph",
    reportNumber = "ICRR-REPORT-608-2011-25, IPMU12-0025, YITP-12-9",
    doi = "10.1103/PhysRevD.85.105020",
    journal = "Phys. Rev. D",
    volume = "85",
    pages = "105020",
    year = "2012",
    note = "[Erratum: Phys.Rev.D 86, 089902 (2012)]"
}

@article{Battye:1994au,
    author = "Battye, R. A. and Shellard, E. P. S.",
    title = "{Axion string constraints}",
    eprint = "astro-ph/9403018",
    archivePrefix = "arXiv",
    reportNumber = "DAMTP-R-94-8",
    doi = "10.1103/PhysRevLett.73.2954",
    journal = "Phys. Rev. Lett.",
    volume = "73",
    pages = "2954--2957",
    year = "1994",
    note = "[Erratum: Phys.Rev.Lett. 76, 2203--2204 (1996)]"
}

@article{Kalia:2024eml,
    author = "Kalia, Saarik and Budker, Dmitry and Kimball, Derek F. Jackson and Ji, Wei and Liu, Zhen and Sushkov, Alexander O. and Timberlake, Chris and Ulbricht, Hendrik and Vinante, Andrea and Wang, Tao",
    title = "{Ultralight dark matter detection with levitated ferromagnets}",
    eprint = "2408.15330",
    archivePrefix = "arXiv",
    primaryClass = "hep-ph",
    reportNumber = "UMN-TH-4329/24, FERMILAB-PUB-24-0545-SQMS",
    doi = "10.1103/PhysRevD.110.115029",
    journal = "Phys. Rev. D",
    volume = "110",
    number = "11",
    pages = "115029",
    year = "2024"
}

@article{Stroberg:2019mxo,
    author = "Stroberg, S. Ragnar and Bogner, Scott K. and Hergert, Heiko and Holt, Jason D.",
    title = "{Nonempirical Interactions for the Nuclear Shell Model: An Update}",
    eprint = "1902.06154",
    archivePrefix = "arXiv",
    primaryClass = "nucl-th",
    doi = "10.1146/annurev-nucl-101917-021120",
    journal = "Ann. Rev. Nucl. Part. Sci.",
    volume = "69",
    pages = "307--362",
    year = "2019"
}

@article{Walter:2025ldb,
    author = "Walter, Julian and others",
    title = "{Search for axionlike dark matter using liquid-state nuclear magnetic resonance}",
    eprint = "2504.16044",
    archivePrefix = "arXiv",
    primaryClass = "hep-ex",
    doi = "10.1103/39nc-vr9m",
    journal = "Phys. Rev. D",
    volume = "112",
    number = "5",
    pages = "052008",
    year = "2025"
}

@article{BALL2003101,
title = {First results of the large COMPASS 6LiD polarized target},
journal = {Nuclear Instruments and Methods in Physics Research Section A: Accelerators, Spectrometers, Detectors and Associated Equipment},
volume = {498},
number = {1},
pages = {101-111},
year = {2003},
issn = {0168-9002},
doi = {https://doi.org/10.1016/S0168-9002(02)02116-2},
url = {https://www.sciencedirect.com/science/article/pii/S0168900202021162},
author = {J Ball and G Baum and P Berglund and I Daito and N Doshita and F Gautheron and St Goertz and J Harmsen and T Hasegawa and J Heckmann and N Horikawa and T Iwata and Yu Kisselev and J Koivuniemi and K Kondo and J.M {Le Goff} and A Magnon and A Meier and W Meyer and E Radtke and G Reicherz and N Takabayashi}
}

@article{AAbragam_1978,
doi = {10.1088/0034-4885/41/3/002},
url = {https://doi.org/10.1088/0034-4885/41/3/002},
year = {1978},
month = {mar},
publisher = {},
volume = {41},
number = {3},
pages = {395},
author = {A Abragam and M Goldman},
title = {Principles of dynamic nuclear polarisation},
journal = {Reports on Progress in Physics}
}

@book{abragam1982nuclear,
  author    = {Abragam, Anatole and Goldman, Maurice},
  title     = {Nuclear Magnetism: Order and Disorder},
  series    = {International Series of Monographs on Physics},
  volume    = {32},
  publisher = {Clarendon Press},
  address   = {Oxford},
  year      = {1982},
  isbn      = {0198512945}
}

@article{Jarmola_2012,
   title={Temperature- and Magnetic-Field-Dependent Longitudinal Spin Relaxation in Nitrogen-Vacancy Ensembles in Diamond},
   volume={108},
   ISSN={1079-7114},
   url={http://dx.doi.org/10.1103/PhysRevLett.108.197601},
   DOI={10.1103/physrevlett.108.197601},
   number={19},
   journal={Physical Review Letters},
   publisher={American Physical Society (APS)},
   author={Jarmola, A. and Acosta, V. M. and Jensen, K. and Chemerisov, S. and Budker, D.},
   year={2012},
   month=May }

@article{Isaac_2017,
   title={Correction: Dynamic nuclear polarization in a magnetic resonance force microscope experiment},
   volume={19},
   ISSN={1463-9084},
   url={http://dx.doi.org/10.1039/C7CP90126G},
   DOI={10.1039/c7cp90126g},
   number={24},
   journal={Physical Chemistry Chemical Physics},
   publisher={Royal Society of Chemistry (RSC)},
   author={Isaac, Corinne E. and Gleave, Christine M. and Nasr, Paméla T. and Nguyen, Hoang L. and Curley, Elizabeth A. and Yoder, Jonilyn L. and Moore, Eric W. and Chen, Lei and Marohn, John A.},
   year={2017},
   pages={16282–16282} }

@article{Braun_2007,
   title={Faraday-rotation fluctuation spectroscopy with static and oscillating magnetic fields},
   volume={75},
   ISSN={1550-235X},
   url={http://dx.doi.org/10.1103/PhysRevB.75.085310},
   DOI={10.1103/physrevb.75.085310},
   number={8},
   journal={Physical Review B},
   publisher={American Physical Society (APS)},
   author={Braun, Matthias and König, Jürgen},
   year={2007},
   month=Feb }

@article{Altarev:2014wka,
    author = "Altarev, I. and others",
    title = "{A magnetically shielded room with ultra low residual field and gradient}",
    eprint = "1403.6467",
    archivePrefix = "arXiv",
    primaryClass = "physics.ins-det",
    doi = "10.1063/1.4886146",
    journal = "Rev. Sci. Instrum.",
    volume = "85",
    pages = "075106",
    year = "2014"
}

@article{Ahmed:2026cvd,
    author = "Ahmed, S. and others",
    title = "{Performance of the TUCAN Magnetically Shielded Room}",
    eprint = "2604.15565",
    archivePrefix = "arXiv",
    primaryClass = "physics.ins-det",
    month = "4",
    year = "2026"
}

@article{nEDM:2022nqd,
    author = "Ayres, N. J. and others",
    collaboration = "nEDM",
    title = "{The very large n2EDM magnetically shielded room with an exceptional performance for fundamental physics measurements}",
    eprint = "2206.10714",
    archivePrefix = "arXiv",
    primaryClass = "physics.ins-det",
    doi = "10.1063/5.0101391",
    journal = "Rev. Sci. Instrum.",
    volume = "93",
    number = "9",
    pages = "095105",
    year = "2022",
    note = "[Erratum: Rev.Sci.Instrum. 93, 119902 (2022)]"
}

@article{Homans:2025eks,
    author = "Homans, Jack and others",
    title = "{An experimental platform for levitated mechanics in space}",
    eprint = "2502.17108",
    archivePrefix = "arXiv",
    primaryClass = "physics.ins-det",
    doi = "10.1088/2058-9565/ade624",
    journal = "Quantum Sci. Technol.",
    volume = "10",
    number = "3",
    pages = "035053",
    year = "2025"
}

@Article{nano15141062,
AUTHOR = {Ai, Wencheng and Wu, Huaning and Zhao, Lin and Xie, Hui},
TITLE = {High-Power-Density Miniaturized VLF Antenna with Nanocrystalline Core for Enhanced Field Strength},
JOURNAL = {Nanomaterials},
VOLUME = {15},
YEAR = {2025},
NUMBER = {14},
ARTICLE-NUMBER = {1062},
URL = {https://www.mdpi.com/2079-4991/15/14/1062},
PubMedID = {40711182},
ISSN = {2079-4991},
DOI = {10.3390/nano15141062}
}

@inproceedings{Steinlechner:22,
author = {Jessica Steinlechner and Peter G. Murray and Massimo Granata and Riccardo Bassiri},
booktitle = {Optical Interference Coatings Conference (OIC) 2022},
journal = {Optical Interference Coatings Conference (OIC) 2022},
pages = {WA.4},
publisher = {Optica Publishing Group},
title = {Coatings for Gravitational Wave Detectors},
year = {2022},
url = {https://opg.optica.org/abstract.cfm?URI=OIC-2022-WA.4},
doi = {10.1364/OIC.2022.WA.4},
}

@article{Steinlechner_2018,
   title={Development of coatings for gravitational wave detectors},
   volume={376},
   ISSN={1471-2962},
   url={https://doi.org/10.1098/rsta.2017.0282},
   DOI={10.1098/rsta.2017.0282},
   issue={2120},
   journal={Philosophical Transactions of the Royal Society A},
   publisher={IOP Publishing},
   author={Steinlechner, J.},
   year={2018},
   month=apr }

@misc{masciocchi2023optimizationpermalloypropertiesmagnetic,
      title={Optimization of Permalloy properties for magnetic field sensors using He$^+$ irradiation}, 
      author={Giovanni Masciocchi and Johannes Wilhelmus van der Jagt and Maria-Andromachi Syskaki and Jürgen Langer and Gerhard Jakob and Jeffrey McCord and Benjamin Borie and Andreas Kehlberger and Dafine Ravelosona and Mathias Kläui},
      year={2023},
      eprint={2302.14455},
      archivePrefix={arXiv},
      primaryClass={cond-mat.mtrl-sci},
      url={https://arxiv.org/abs/2302.14455}, 
}

@article{Carney:2019cio,
    author = "Carney, Daniel and Hook, Anson and Liu, Zhen and Taylor, Jacob M. and Zhao, Yue",
    title = "{Ultralight dark matter detection with mechanical quantum sensors}",
    eprint = "1908.04797",
    archivePrefix = "arXiv",
    primaryClass = "hep-ph",
    reportNumber = "FERMILAB-PUB-19-364-T",
    doi = "10.1088/1367-2630/abd9e7",
    journal = "New J. Phys.",
    volume = "23",
    number = "2",
    pages = "023041",
    year = "2021"
}

@article{Higgins:2023gwq,
    author = "Higgins, Gerard and Kalia, Saarik and Liu, Zhen",
    title = "{Maglev for dark matter: Dark-photon and axion dark matter sensing with levitated superconductors}",
    eprint = "2310.18398",
    archivePrefix = "arXiv",
    primaryClass = "hep-ph",
    reportNumber = "FERMILAB-PUB-23-624-SQMS-V",
    doi = "10.1103/PhysRevD.109.055024",
    journal = "Phys. Rev. D",
    volume = "109",
    number = "5",
    pages = "055024",
    year = "2024"
}

@article{Manley:2020mjq,
    author = "Manley, Jack and Chowdhury, Mitul Dey and Grin, Daniel and Singh, Swati and Wilson, Dalziel J.",
    title = "{Searching for vector dark matter with an optomechanical accelerometer}",
    eprint = "2007.04899",
    archivePrefix = "arXiv",
    primaryClass = "quant-ph",
    doi = "10.1103/PhysRevLett.126.061301",
    journal = "Phys. Rev. Lett.",
    volume = "126",
    number = "6",
    pages = "061301",
    year = "2021"
}

@article{Mitridate:2022tnv,
    author = "Mitridate, Andrea and Trickle, Tanner and Zhang, Zhengkang and Zurek, Kathryn M.",
    title = "{Snowmass white paper: Light dark matter direct detection at the interface with condensed matter physics}",
    eprint = "2203.07492",
    archivePrefix = "arXiv",
    primaryClass = "hep-ph",
    reportNumber = "FERMILAB-PUB-23-551-T",
    doi = "10.1016/j.dark.2023.101221",
    journal = "Phys. Dark Univ.",
    volume = "40",
    pages = "101221",
    year = "2023"
}

@Article{galaxies10010020,
AUTHOR = {Trozzo, Lucia and Badaracco, Francesca},
TITLE = {Seismic and Newtonian Noise in the GW Detectors},
JOURNAL = {Galaxies},
VOLUME = {10},
YEAR = {2022},
NUMBER = {1},
ARTICLE-NUMBER = {20},
URL = {https://www.mdpi.com/2075-4434/10/1/20},
ISSN = {2075-4434},
DOI = {10.3390/galaxies10010020}
}

@article{Manley:2019vxy,
    author = "Manley, Jack and Stump, Russell and Wilson, Dalziel and Grin, Daniel and Singh, Swati",
    title = "{Searching for scalar dark matter with compact mechanical resonators}",
    eprint = "1910.07574",
    archivePrefix = "arXiv",
    primaryClass = "astro-ph.IM",
    doi = "10.1103/PhysRevLett.124.151301",
    journal = "Phys. Rev. Lett.",
    volume = "124",
    number = "15",
    pages = "151301",
    year = "2020"
}

@article{Winstone:2023whl,
    author = "Winstone, George and Grinin, Alexey and Bhattacharya, Mishkat and Geraci, Andrew A. and Li, Tongcang and Pauzauskie, Peter J. and Vamivakas, Nick",
    title = "{Optomechanics of optically-levitated particles: A tutorial and perspective}",
    eprint = "2307.11858",
    archivePrefix = "arXiv",
    primaryClass = "quant-ph",
    month = "7",
    year = "2023"
}

@article{Gieseler:2013ael,
    author = "Gieseler, Jan and Novotny, Lukas and Quidant, Romain",
    title = "{Thermal nonlinearities in a nanomechanical oscillator}",
    eprint = "1307.4684",
    archivePrefix = "arXiv",
    primaryClass = "cond-mat.mes-hall",
    doi = "10.1038/nphys2798",
    month = "7",
    year = "2013"
}

@article{Aspelmeyer:2013lha,
    author = "Aspelmeyer, Markus and Kippenberg, Tobias J. and Marquardt, Florian",
    title = "{Cavity Optomechanics}",
    eprint = "1303.0733",
    archivePrefix = "arXiv",
    primaryClass = "cond-mat.mes-hall",
    doi = "10.1103/RevModPhys.86.1391",
    journal = "Rev. Mod. Phys.",
    volume = "86",
    pages = "1391",
    year = "2014"
}

@article{Capozzi:2020cbu,
    author = "Capozzi, Francesco and Raffelt, Georg",
    title = "{Axion and neutrino bounds improved with new calibrations of the tip of the red-giant branch using geometric distance determinations}",
    eprint = "2007.03694",
    archivePrefix = "arXiv",
    primaryClass = "astro-ph.SR",
    reportNumber = "MPP-2020-106",
    doi = "10.1103/PhysRevD.102.083007",
    journal = "Phys. Rev. D",
    volume = "102",
    number = "8",
    pages = "083007",
    year = "2020"
}

@article{Buschmann:2021juv,
    author = "Buschmann, Malte and Dessert, Christopher and Foster, Joshua W. and Long, Andrew J. and Safdi, Benjamin R.",
    title = "{Upper Limit on the QCD Axion Mass from Isolated Neutron Star Cooling}",
    eprint = "2111.09892",
    archivePrefix = "arXiv",
    primaryClass = "hep-ph",
    doi = "10.1103/PhysRevLett.128.091102",
    journal = "Phys. Rev. Lett.",
    volume = "128",
    number = "9",
    pages = "091102",
    year = "2022"
}

@article{Gomez-Banon:2024oux,
    author = "G{\'o}mez-Ba{\~n}{\'o}n, Antonio and Bartnick, Kai and Springmann, Konstantin and Pons, Jos{\'e} A.",
    title = "{Constraining Light QCD Axions with Isolated Neutron Star Cooling}",
    eprint = "2408.07740",
    archivePrefix = "arXiv",
    primaryClass = "hep-ph",
    doi = "10.1103/PhysRevLett.133.251002",
    journal = "Phys. Rev. Lett.",
    volume = "133",
    number = "25",
    pages = "251002",
    year = "2024"
}

@article{Kumamoto:2024wjd,
    author = "Kumamoto, Mia and Huang, Junwu and Drischler, Christian and Baryakhtar, Masha and Reddy, Sanjay",
    title = "{Neutron stars with exceptionally light QCD axions}",
    eprint = "2410.21590",
    archivePrefix = "arXiv",
    primaryClass = "hep-ph",
    reportNumber = "INT-PUB-24-055",
    doi = "10.1103/5zgm-z4v9",
    journal = "Phys. Rev. D",
    volume = "112",
    number = "4",
    pages = "043008",
    year = "2025"
}

@article{Balkin:2022qer,
    author = "Balkin, Reuven and Serra, Javi and Springmann, Konstantin and Stelzl, Stefan and Weiler, Andreas",
    title = "{White dwarfs as a probe of exceptionally light QCD axions}",
    eprint = "2211.02661",
    archivePrefix = "arXiv",
    primaryClass = "hep-ph",
    reportNumber = "IFT-UAM/CSIC-22-136",
    doi = "10.1103/PhysRevD.109.095032",
    journal = "Phys. Rev. D",
    volume = "109",
    number = "9",
    pages = "095032",
    year = "2024"
}

@article{Hook:2017psm,
    author = "Hook, Anson and Huang, Junwu",
    title = "{Probing axions with neutron star inspirals and other stellar processes}",
    eprint = "1708.08464",
    archivePrefix = "arXiv",
    primaryClass = "hep-ph",
    doi = "10.1007/JHEP06(2018)036",
    journal = "JHEP",
    volume = "06",
    pages = "036",
    year = "2018"
}

@article{Blum:2014vsa,
    author = "Blum, Kfir and D'Agnolo, Raffaele Tito and Lisanti, Mariangela and Safdi, Benjamin R.",
    title = "{Constraining Axion Dark Matter with Big Bang Nucleosynthesis}",
    eprint = "1401.6460",
    archivePrefix = "arXiv",
    primaryClass = "hep-ph",
    doi = "10.1016/j.physletb.2014.07.059",
    journal = "Phys. Lett. B",
    volume = "737",
    pages = "30--33",
    year = "2014"
}

@article{Madge:2024aot,
    author = "Madge, Eric and Perez, Gilad and Meir, Ziv",
    title = "{Prospects of nuclear-coupled-dark-matter detection via correlation spectroscopy of I2+ and Ca+}",
    eprint = "2404.00616",
    archivePrefix = "arXiv",
    primaryClass = "physics.atom-ph",
    doi = "10.1103/PhysRevD.110.015008",
    journal = "Phys. Rev. D",
    volume = "110",
    number = "1",
    pages = "015008",
    year = "2024"
}

@article{Abel:2017rtm,
    author = "Abel, C. and others",
    title = "{Search for Axionlike Dark Matter through Nuclear Spin Precession in Electric and Magnetic Fields}",
    eprint = "1708.06367",
    archivePrefix = "arXiv",
    primaryClass = "hep-ph",
    doi = "10.1103/PhysRevX.7.041034",
    journal = "Phys. Rev. X",
    volume = "7",
    number = "4",
    pages = "041034",
    year = "2017"
}

@article{Schulthess:2022pbp,
    author = "Schulthess, Ivo and others",
    title = "{New Limit on Axionlike Dark Matter Using Cold Neutrons}",
    eprint = "2204.01454",
    archivePrefix = "arXiv",
    primaryClass = "hep-ex",
    doi = "10.1103/PhysRevLett.129.191801",
    journal = "Phys. Rev. Lett.",
    volume = "129",
    number = "19",
    pages = "191801",
    year = "2022"
}

@article{Roussy:2020ily,
    author = "Roussy, Tanya S. and others",
    title = "{Experimental Constraint on Axionlike Particles over Seven Orders of Magnitude in Mass}",
    eprint = "2006.15787",
    archivePrefix = "arXiv",
    primaryClass = "hep-ph",
    doi = "10.1103/PhysRevLett.126.171301",
    journal = "Phys. Rev. Lett.",
    volume = "126",
    number = "17",
    pages = "171301",
    year = "2021"
}

@article{JEDI:2022hxa,
    author = "Karanth, S. and others",
    collaboration = "JEDI",
    title = "{First Search for Axionlike Particles in a Storage Ring Using a Polarized Deuteron Beam}",
    eprint = "2208.07293",
    archivePrefix = "arXiv",
    primaryClass = "hep-ex",
    doi = "10.1103/PhysRevX.13.031004",
    journal = "Phys. Rev. X",
    volume = "13",
    number = "3",
    pages = "031004",
    year = "2023"
}

@article{Gue:2025nxq,
    author = "Gu{\'e}, Jordan and Wolf, Peter and Hees, Aur{\'e}lien",
    title = "{Search for QCD Coupled Axion Dark Matter with Data from the MICROSCOPE Space Experiment}",
    eprint = "2504.00720",
    archivePrefix = "arXiv",
    primaryClass = "hep-ph",
    doi = "10.1103/fwzv-g1jg",
    journal = "Phys. Rev. Lett.",
    volume = "135",
    number = "17",
    pages = "171001",
    year = "2025"
}

@article{Fan:2024pxs,
    author = "Fan, Mingyu and Nima, Bassam and Radak, Aleksandar and Alonso-{\'A}lvarez, Gonzalo and Vutha, Amar",
    title = "{Wideband Search for Axionlike Dark Matter Using Octupolar Nuclei in a Crystal}",
    eprint = "2410.02218",
    archivePrefix = "arXiv",
    primaryClass = "physics.atom-ph",
    doi = "10.1103/dm9j-9pry",
    journal = "Phys. Rev. Lett.",
    volume = "136",
    number = "12",
    pages = "121802",
    year = "2026"
}

@article{Zhang:2022ewz,
    author = "Zhang, Xue and Banerjee, Abhishek and Leyser, Mahapan and Perez, Gilad and Schiller, Stephan and Budker, Dmitry and Antypas, Dionysios",
    title = "{Search for Ultralight Dark Matter with Spectroscopy of Radio-Frequency Atomic Transitions}",
    eprint = "2212.04413",
    archivePrefix = "arXiv",
    primaryClass = "physics.atom-ph",
    doi = "10.1103/PhysRevLett.130.251002",
    journal = "Phys. Rev. Lett.",
    volume = "130",
    number = "25",
    pages = "251002",
    year = "2023"
}

@article{Zhang:2023lem,
    author = "Zhang, Xin and Houston, Nick and Li, Tianjun",
    title = "{Nuclear decay anomalies as a signature of axion dark matter}",
    eprint = "2303.09865",
    archivePrefix = "arXiv",
    primaryClass = "hep-ph",
    doi = "10.1103/PhysRevD.108.L071101",
    journal = "Phys. Rev. D",
    volume = "108",
    number = "7",
    pages = "L071101",
    year = "2023"
}

@article{Alda:2024xxa,
    author = "Alda, Jorge and Broggini, Carlo and Di Carlo, Giuseppe and Di Luzio, Luca and Piatti, Denise and Rigolin, Stefano and Toni, Claudio",
    title = "{Weak nuclear decays deep-underground as a probe of axion dark matter}",
    eprint = "2412.20932",
    archivePrefix = "arXiv",
    primaryClass = "hep-ph",
    doi = "10.1103/PhysRevD.111.035022",
    journal = "Phys. Rev. D",
    volume = "111",
    number = "3",
    pages = "035022",
    year = "2025"
}

@article{Banerjee:2023bjc,
    author = "Banerjee, Abhishek and Budker, Dmitry and Filzinger, Melina and Huntemann, Nils and Paz, Gil and Perez, Gilad and Porsev, Sergey and Safronova, Marianna",
    title = "{Oscillating Nuclear Charge Radii as Sensors for Ultralight Dark Matter}",
    eprint = "2301.10784",
    archivePrefix = "arXiv",
    primaryClass = "hep-ph",
    reportNumber = "WSU-HEP-2301",
    doi = "10.1103/37vw-gc1r",
    journal = "Phys. Rev. Lett.",
    volume = "135",
    number = "22",
    pages = "22",
    year = "2025"
}

@article{Arvanitaki:2021wjk,
    author = "Arvanitaki, Asimina and Madden, Amalia and Van Tilburg, Ken",
    title = "{Piezoaxionic effect}",
    eprint = "2112.11466",
    archivePrefix = "arXiv",
    primaryClass = "hep-ph",
    doi = "10.1103/PhysRevD.109.072009",
    journal = "Phys. Rev. D",
    volume = "109",
    number = "7",
    pages = "072009",
    year = "2024"
}

@article{Marsh:2015xka,
    author = "Marsh, David J. E.",
    title = "{Axion Cosmology}",
    eprint = "1510.07633",
    archivePrefix = "arXiv",
    primaryClass = "astro-ph.CO",
    reportNumber = "KCL-PH-TH-2015-50",
    doi = "10.1016/j.physrep.2016.06.005",
    journal = "Phys. Rept.",
    volume = "643",
    pages = "1--79",
    year = "2016"
}

@article{Chang:2017ruk,
    author = "Chang, Seung Pyo and Haciomeroglu, Selcuk and Kim, On and Lee, Soohyung and Park, Seongtae and Semertzidis, Yannis K.",
    title = "{Axionlike dark matter search using the storage ring EDM method}",
    eprint = "1710.05271",
    archivePrefix = "arXiv",
    primaryClass = "hep-ex",
    doi = "10.1103/PhysRevD.99.083002",
    journal = "Phys. Rev. D",
    volume = "99",
    number = "8",
    pages = "083002",
    year = "2019"
}

@article{Graham:2013gfa,
    author = "Graham, Peter W. and Rajendran, Surjeet",
    title = "{New Observables for Direct Detection of Axion Dark Matter}",
    eprint = "1306.6088",
    archivePrefix = "arXiv",
    primaryClass = "hep-ph",
    doi = "10.1103/PhysRevD.88.035023",
    journal = "Phys. Rev. D",
    volume = "88",
    pages = "035023",
    year = "2013"
}

@article{JacksonKimball:2017elr,
    author = "Jackson Kimball, Derek F. and others",
    editor = "Carosi, Gianpaolo and Rybka, Gray",
    title = "{Overview of the Cosmic Axion Spin Precession Experiment (CASPEr)}",
    eprint = "1711.08999",
    archivePrefix = "arXiv",
    primaryClass = "physics.ins-det",
    doi = "10.1007/978-3-030-43761-9_13",
    journal = "Springer Proc. Phys.",
    volume = "245",
    pages = "105--121",
    year = "2020"
}

@article{Garcon:2019inh,
    author = "Garcon, Antoine and others",
    title = "{Constraints on bosonic dark matter from ultralow-field nuclear magnetic resonance}",
    eprint = "1902.04644",
    archivePrefix = "arXiv",
    primaryClass = "hep-ex",
    doi = "10.1126/sciadv.aax4539",
    journal = "Sci. Adv.",
    volume = "5",
    number = "10",
    pages = "eaax4539",
    year = "2019"
}

@article{Garcon:2017ixh,
    author = "Garcon, Antoine and others",
    title = "{The cosmic axion spin precession experiment (CASPEr): a dark-matter search with nuclear magnetic resonance}",
    eprint = "1707.05312",
    archivePrefix = "arXiv",
    primaryClass = "physics.ins-det",
    doi = "10.1088/2058-9565/aa9861",
    journal = "Quantum Sci. Technol.",
    volume = "3",
    number = "1",
    pages = "014008",
    year = "2017"
}

@article{Budker:2013hfa,
    author = "Budker, Dmitry and Graham, Peter W. and Ledbetter, Micah and Rajendran, Surjeet and Sushkov, Alex",
    title = "{Proposal for a Cosmic Axion Spin Precession Experiment (CASPEr)}",
    eprint = "1306.6089",
    archivePrefix = "arXiv",
    primaryClass = "hep-ph",
    doi = "10.1103/PhysRevX.4.021030",
    journal = "Phys. Rev. X",
    volume = "4",
    number = "2",
    pages = "021030",
    year = "2014"
}

@article{Stott:2017hvl,
    author = "Stott, Matthew J. and Marsh, David J. E. and Pongkitivanichkul, Chakrit and Price, Layne C. and Acharya, Bobby S.",
    title = "{Spectrum of the axion dark sector}",
    eprint = "1706.03236",
    archivePrefix = "arXiv",
    primaryClass = "astro-ph.CO",
    doi = "10.1103/PhysRevD.96.083510",
    journal = "Phys. Rev. D",
    volume = "96",
    number = "8",
    pages = "083510",
    year = "2017"
}

@article{Terrano:2019clh,
    author = "Terrano, William A. and Adelberger, Eric G. and Hagedorn, Charles A. and Heckel, Blayne R.",
    title = "{Constraints on axionlike dark matter with masses down to $10^{-23}$ eV/c$^2$}",
    eprint = "1902.04246",
    archivePrefix = "arXiv",
    primaryClass = "astro-ph.CO",
    doi = "10.1103/PhysRevLett.122.231301",
    journal = "Phys. Rev. Lett.",
    volume = "122",
    number = "23",
    pages = "231301",
    year = "2019"
}

@article{Graham:2017ivz,
    author = "Graham, Peter W. and Kaplan, David E. and Mardon, Jeremy and Rajendran, Surjeet and Terrano, William A. and Trahms, Lutz and Wilkason, Thomas",
    title = "{Spin Precession Experiments for Light Axionic Dark Matter}",
    eprint = "1709.07852",
    archivePrefix = "arXiv",
    primaryClass = "hep-ph",
    doi = "10.1103/PhysRevD.97.055006",
    journal = "Phys. Rev. D",
    volume = "97",
    number = "5",
    pages = "055006",
    year = "2018"
}

@article{Terrano:2015sna,
    author = "Terrano, W. A. and Adelberger, E. G. and Lee, J. G. and Heckel, B. R.",
    title = "{Short-range spin-dependent interactions of electrons: a probe for exotic pseudo-Goldstone bosons}",
    eprint = "1508.02463",
    archivePrefix = "arXiv",
    primaryClass = "hep-ex",
    doi = "10.1103/PhysRevLett.115.201801",
    journal = "Phys. Rev. Lett.",
    volume = "115",
    number = "20",
    pages = "201801",
    year = "2015"
}

@article{Guccione_2013,
   title={Scattering-Free Optical Levitation of a Cavity Mirror},
   volume={111},
   ISSN={1079-7114},
   url={http://dx.doi.org/10.1103/PhysRevLett.111.183001},
   DOI={10.1103/physrevlett.111.183001},
   number={18},
   journal={Physical Review Letters},
   publisher={American Physical Society (APS)},
   author={Guccione, G. and Hosseini, M. and Adlong, S. and Johnsson, M. T. and Hope, J. and Buchler, B. C. and Lam, P. K.},
   year={2013},
   month=oct }

@article{Beckey:2023shi,
    author = "Beckey, Jacob and Carney, Daniel and Marocco, Giacomo",
    title = "{Quantum measurements in fundamental physics: a user's manual}",
    journal = "SciPost Phys. Rev.",
    eprint = "2311.07270",
    archivePrefix = "arXiv",
    primaryClass = "hep-ph",
    doi = "10.21468/SciPostPhysRev.2",
    month = "11",
    year = "2023"
}

@article{Millen_2020,
   title={Optomechanics with levitated particles},
   volume={83},
   ISSN={1361-6633},
   url={http://dx.doi.org/10.1088/1361-6633/ab6100},
   DOI={10.1088/1361-6633/ab6100},
   number={2},
   journal={Reports on Progress in Physics},
   publisher={IOP Publishing},
   author={Millen, James and Monteiro, Tania S and Pettit, Robert and Vamivakas, A Nick},
   year={2020},
   month=Jan, pages={026401} }

@article{PhysRevLett.127.061804,
  title = {Trapped Electrons and Ions as Particle Detectors},
  author = {Carney, Daniel and H{\"a}ffner, Hartmut and Moore, David C. and Taylor, Jacob M.},
  journal = {Phys. Rev. Lett.},
  volume = {127},
  issue = {6},
  pages = {061804},
  numpages = {7},
  year = {2021},
  month = {Aug},
  publisher = {American Physical Society},
  doi = {10.1103/PhysRevLett.127.061804},
  url = {https://link.aps.org/doi/10.1103/PhysRevLett.127.061804}
}

@article{Iacoponi:2024mzu,
    author = "Iacoponi, J. H. and Rademacher, M. and Monteiro, T. S.",
    title = "{Nanoparticle arrays levitated in a cavity for quantum sensing}",
    eprint = "2406.00366",
    archivePrefix = "arXiv",
    primaryClass = "physics.optics",
    doi = "10.1103/wrd3-t5cf",
    journal = "Phys. Rev. Res.",
    volume = "8",
    number = "3",
    pages = "033286",
    year = "2026"
}

@article{Amaral:2024tjg,
    author = "Amaral, Dorian W. P. and Jain, Mudit and Amin, Mustafa A. and Tunnell, Christopher",
    title = "{Vector wave dark matter and terrestrial quantum sensors}",
    eprint = "2403.02381",
    archivePrefix = "arXiv",
    primaryClass = "hep-ph",
    reportNumber = "KCL-PH-TH/2024-09",
    doi = "10.1088/1475-7516/2024/06/050",
    journal = "JCAP",
    volume = "06",
    pages = "050",
    year = "2024"
}

@article{Manley:2024svv,
    author = "Manley, J. and Condos, C. A. and Schlamminger, S. and Pratt, J. R. and Wilson, D. J. and Terrano, W. A.",
    title = "{Microscale torsion resonators for short-range gravity experiments}",
    eprint = "2406.13020",
    archivePrefix = "arXiv",
    primaryClass = "gr-qc",
    doi = "10.1103/PhysRevD.110.122005",
    journal = "Phys. Rev. D",
    volume = "110",
    number = "12",
    pages = "122005",
    year = "2024"
}

@article{Chowdhury:2025avw,
    author = "Chowdhury, M. Dey and Manley, J. P. and Condos, C. A. and Agrawal, A. R. and Wilson, D. J.",
    title = "{Optomechanical accelerometer search for ultralight dark matter}",
    eprint = "2509.12175",
    archivePrefix = "arXiv",
    primaryClass = "hep-ex",
    doi = "10.1103/16yd-qj5v",
    journal = "Phys. Rev. D",
    volume = "113",
    number = "12",
    pages = "L121303",
    year = "2026"
}

@article{Geraci:2018fax,
    author = "Geraci, Andrew A. and Bradley, Colin and Gao, Dongfeng and Weinstein, Jonathan and Derevianko, Andrei",
    title = "{Searching for Ultralight Dark Matter with Optical Cavities}",
    eprint = "1808.00540",
    archivePrefix = "arXiv",
    primaryClass = "astro-ph.IM",
    doi = "10.1103/PhysRevLett.123.031304",
    journal = "Phys. Rev. Lett.",
    volume = "123",
    number = "3",
    pages = "031304",
    year = "2019"
}

@article{Mukhamedjanov:2004ki,
    author = "Mukhamedjanov, T. N. and Sushkov, O. P.",
    title = "{A Suggested search for Pb-207 nuclear Schiff moment in PbTiO(3) ferroelectric}",
    eprint = "physics/0411226",
    archivePrefix = "arXiv",
    doi = "10.1103/PhysRevA.72.034501",
    journal = "Phys. Rev. A",
    volume = "72",
    pages = "034501",
    year = "2005"
}

@article{Gov:1999fa,
    author = "Gov, S. and Shtrikman, S. and Thomas, H.",
    title = "{Magnetic trapping of neutral particles: classical and quantum-mechanical study of a ioffe-pritchard type trap}",
    eprint = "quant-ph/9907044",
    archivePrefix = "arXiv",
    doi = "10.1063/1.372444",
    journal = "J. Appl. Phys.",
    volume = "87",
    pages = "3989",
    year = "2000"
}

@article{Gomberoff:2007byo,
    author = "Gomberoff, K. and Fajans, J. and Friedman, A. and Grote, D. and Vay, J. -L. and Wurtele, J. S.",
    title = "{Simulations of plasma confinement in an antihydrogen trap}",
    doi = "10.1063/1.2778420",
    journal = "Phys. Plasmas",
    volume = "14",
    number = "10",
    pages = "102111",
    year = "2007"
}

@techreport{wolski2011maxwell,
  author      = {Wolski, Andrzej},
  title       = {Maxwell's Equations for Magnets},
  institution = {CERN},
  year        = {2011},
  type        = {CERN Accelerator School Report},
  number      = {CERN-2010-004},
  pages       = {1--38},
  url         = {https://cds.cern.ch/record/1333874/files/1.pdf}
}

@inproceedings{marks1994conventional,
  author       = {Marks, N.},
  title        = {Conventional magnets - I},
  booktitle    = {CAS - CERN Accelerator School : 5th General Accelerator Physics Course},
  year         = {1994},
  pages        = {867--890},
  publisher    = {CERN},
  series       = {CERN Report},
  number       = {CERN-1994-001},
  doi          = {10.5170/CERN-1994-001.867},
  url          = {https://cds.cern.ch/record/398441/files/p867.pdf}
}

@inproceedings{Prawanta:2019jwv,
    author = "Prawanta, Supachai and Boonwanna, Bombay and Klysubun, Prapong and Kwankasem, Apichai and Pongampai, Surachai and Sooksrimuang, Visitchai and Sudmuang, Porntip and Sunwong, Prapaiwan",
    title = "{Design and Construction of Sextupole Magnet Prototype for  Siam Photon Source II Project}",
    booktitle = "{10th International Particle Accelerator Conference}",
    doi = "10.18429/JACoW-IPAC2019-THPTS076",
    pages = "THPTS076",
    year = "2019"
}

@article{Crewther:1979pi,
    author = "Crewther, R. J. and Di Vecchia, P. and Veneziano, G. and Witten, Edward",
    title = "{Chiral Estimate of the Electric Dipole Moment of the Neutron in Quantum Chromodynamics}",
    reportNumber = "CERN-TH-2735",
    doi = "10.1016/0370-2693(79)90128-X",
    journal = "Phys. Lett. B",
    volume = "88",
    pages = "123",
    year = "1979",
    note = "[Erratum: Phys.Lett.B 91, 487 (1980)]"
}

@article{Magrini:2020agy,
    author = "Magrini, Lorenzo and Rosenzweig, Philipp and Bach, Constanze and Deutschmann-Olek, Andreas and Hofer, Sebastian G. and Hong, Sungkun and Kiesel, Nikolai and Kugi, Andreas and Aspelmeyer, Markus",
    title = "{Real-time optimal quantum control of mechanical motion at room temperature}",
    eprint = "2012.15188",
    archivePrefix = "arXiv",
    primaryClass = "quant-ph",
    doi = "10.1038/s41586-021-03602-3",
    month = "12",
    year = "2020"
}

@misc{Hamaide_2026,
   title={In preparation},
   author={Hamaide, L. and Aquino-Ahmed, A.},
   year={2026}}

@article{Aggarwal:2020umq,
    author = "Aggarwal, Nancy and Winstone, George P. and Teo, Mae and Baryakhtar, Masha and Larson, Shane L. and Kalogera, Vicky and Geraci, Andrew A.",
    title = "{Searching for New Physics with a Levitated-Sensor-Based Gravitational-Wave Detector}",
    eprint = "2010.13157",
    archivePrefix = "arXiv",
    primaryClass = "gr-qc",
    doi = "10.1103/PhysRevLett.128.111101",
    journal = "Phys. Rev. Lett.",
    volume = "128",
    number = "11",
    pages = "111101",
    year = "2022"
}

@article{Aggarwal:2025noe,
    author = "Aggarwal, Nancy and others",
    title = "{Challenges and opportunities of gravitational-wave searches above 10 kHz}",
    eprint = "2501.11723",
    archivePrefix = "arXiv",
    primaryClass = "gr-qc",
    reportNumber = "CERN-TH-2025-014, DESY-25-007",
    doi = "10.1007/s41114-025-00060-5",
    journal = "Living Rev. Rel.",
    volume = "28",
    number = "1",
    pages = "10",
    year = "2025"
}

@article{Michimura:2016urt,
    author = "Michimura, Yuta and Kuwahara, Yuya and Ushiba, Takafumi and Matsumoto, Nobuyuki and Ando, Masaki",
    title = "{Optical levitation of a mirror for reaching the standard quantum limit}",
    eprint = "1612.07127",
    archivePrefix = "arXiv",
    primaryClass = "quant-ph",
    doi = "10.1364/OE.25.013799",
    journal = "Opt. Express",
    volume = "25",
    number = "12",
    pages = "13799--13806",
    year = "2017"
}

@article{Melo:2020prx,
    author = "Melo, Bruno and Brand{\~a}o, Igor and da Silva, B. Pinheiro and Rodrigues, R. B. and Khoury, A. Z. and Guerreiro, Thiago",
    title = "{Optical Trapping in a Dark Focus}",
    eprint = "2006.15455",
    archivePrefix = "arXiv",
    primaryClass = "physics.optics",
    doi = "10.1103/PhysRevApplied.14.034069",
    journal = "Phys. Rev. Applied",
    volume = "14",
    pages = "034069",
    year = "2020"
}

@inbook{Saunders_2018,
   title={Realizing quantum materials with Helium: Helium films at ultralow temperatures, from strongly correlated atomically layered films to topological superfluidity},
   ISBN={9789813271340},
   url={http://dx.doi.org/10.1142/9789813271340_0012},
   DOI={10.1142/9789813271340_0012},
   booktitle={Topological Phase Transitions and New Developments},
   publisher={WORLD SCIENTIFIC},
   author={Saunders, J.},
   year={2018},
   month=Aug, pages={165–196} }

@article{Alesini:2019ajt,
    author = "Alesini, D. and others",
    title = "{Galactic axions search with a superconducting resonant cavity}",
    eprint = "1903.06547",
    archivePrefix = "arXiv",
    primaryClass = "physics.ins-det",
    doi = "10.1103/PhysRevD.99.101101",
    journal = "Phys. Rev. D",
    volume = "99",
    number = "10",
    pages = "101101",
    year = "2019"
}

@article{Beurthey:2020yuq,
    author = "Beurthey, S. and others",
    title = "{MADMAX Status Report}",
    eprint = "2003.10894",
    archivePrefix = "arXiv",
    primaryClass = "physics.ins-det",
    month = "3",
    year = "2020"
}

@article{HAYSTAC:2024jch,
    author = "Bai, Xiran and others",
    collaboration = "HAYSTAC",
    title = "{Dark Matter Axion Search with HAYSTAC Phase II}",
    eprint = "2409.08998",
    archivePrefix = "arXiv",
    primaryClass = "hep-ex",
    doi = "10.1103/PhysRevLett.134.151006",
    journal = "Phys. Rev. Lett.",
    volume = "134",
    number = "15",
    pages = "151006",
    year = "2025"
}

@article{BREAD:2021tpx,
    author = "Liu, Jesse and others",
    collaboration = "BREAD",
    title = "{Broadband Solenoidal Haloscope for Terahertz Axion Detection}",
    eprint = "2111.12103",
    archivePrefix = "arXiv",
    primaryClass = "physics.ins-det",
    reportNumber = "FERMILAB-PUB-21-694-AD-PPD-TD",
    doi = "10.1103/PhysRevLett.128.131801",
    journal = "Phys. Rev. Lett.",
    volume = "128",
    number = "13",
    pages = "131801",
    year = "2022"
}

@article{ADMX:2018ogs,
    author = "Boutan, C. and others",
    collaboration = "ADMX",
    title = "{Piezoelectrically Tuned Multimode Cavity Search for Axion Dark Matter}",
    eprint = "1901.00920",
    archivePrefix = "arXiv",
    primaryClass = "hep-ex",
    reportNumber = "FERMILAB-PUB-18-702-AD-AE",
    doi = "10.1103/PhysRevLett.121.261302",
    journal = "Phys. Rev. Lett.",
    volume = "121",
    number = "26",
    pages = "261302",
    year = "2018"
}

@article{ADMX:2025vom,
    author = "Carosi, G. and others",
    collaboration = "ADMX",
    title = "{Search for Axion Dark Matter from 1.1 to 1.3~GHz with ADMX}",
    eprint = "2504.07279",
    archivePrefix = "arXiv",
    primaryClass = "hep-ex",
    reportNumber = "FERMILAB-PUB-25-0298-PPD",
    doi = "10.1103/d7mg-6sqq",
    journal = "Phys. Rev. Lett.",
    volume = "135",
    number = "19",
    pages = "191001",
    year = "2025"
}

@article{Berlin:2020vrk,
    author = "Berlin, Asher and D'Agnolo, Raffaele Tito and Ellis, Sebastian A. R. and Zhou, Kevin",
    title = "{Heterodyne broadband detection of axion dark matter}",
    eprint = "2007.15656",
    archivePrefix = "arXiv",
    primaryClass = "hep-ph",
    doi = "10.1103/PhysRevD.104.L111701",
    journal = "Phys. Rev. D",
    volume = "104",
    number = "11",
    pages = "L111701",
    year = "2021"
}

@article{Li:2025pyi,
    author = "Li, Zenghai and Zhou, Kevin and Oriunno, Marco and Berlin, Asher and Calatroni, Sergio and Tito D'Agnolo, Raffaele and Ellis, Sebastian A. R. and Schuster, Philip and Tantawi, Sami G. and Toro, Natalia",
    title = "{A prototype hybrid mode cavity for heterodyne axion detection}",
    eprint = "2507.07173",
    archivePrefix = "arXiv",
    primaryClass = "physics.ins-det",
    reportNumber = "FERMILAB-PUB-25-0465-PPD",
    doi = "10.1016/j.nima.2026.171474",
    journal = "Nucl. Instrum. Meth. A",
    volume = "1088",
    pages = "171474",
    year = "2026"
}

@article{DMRadio:2022pkf,
    author = "Brouwer, L. and others",
    collaboration = "DMRadio",
    title = "{Projected sensitivity of DMRadio-m3: A search for the QCD axion below 1{\,}{\,}{\ensuremath{\mu}}eV}",
    eprint = "2204.13781",
    archivePrefix = "arXiv",
    primaryClass = "hep-ex",
    doi = "10.1103/PhysRevD.106.103008",
    journal = "Phys. Rev. D",
    volume = "106",
    number = "10",
    pages = "103008",
    year = "2022"
}

@article{Silva-Feaver:2016qhh,
    author = "Silva-Feaver, Maximiliano and others",
    title = "{Design Overview of DM Radio Pathfinder Experiment}",
    eprint = "1610.09344",
    archivePrefix = "arXiv",
    primaryClass = "astro-ph.IM",
    doi = "10.1109/TASC.2016.2631425",
    journal = "IEEE Trans. Appl. Supercond.",
    volume = "27",
    number = "4",
    pages = "1400204",
    year = "2017"
}

@article{Chaudhuri:2014dla,
    author = "Chaudhuri, Saptarshi and Graham, Peter W. and Irwin, Kent and Mardon, Jeremy and Rajendran, Surjeet and Zhao, Yue",
    title = "{Radio for hidden-photon dark matter detection}",
    eprint = "1411.7382",
    archivePrefix = "arXiv",
    primaryClass = "hep-ph",
    doi = "10.1103/PhysRevD.92.075012",
    journal = "Phys. Rev. D",
    volume = "92",
    number = "7",
    pages = "075012",
    year = "2015"
}

@article{Ma:2026afn,
    author = "Ma, Xiaolin and Takhistov, Volodymyr and Iizuka, Hideo",
    title = "{Casimir Geometry as a Probe of Short Range Forces}",
    eprint = "2603.22413",
    archivePrefix = "arXiv",
    primaryClass = "hep-ph",
    reportNumber = "KEK-QUP-2026-0005, KEK-TH-2819",
    month = "3",
    year = "2026"
}

@article{Sedmik:2021iaw,
    author = "Sedmik, Ren{\'e} I. P. and Pitschmann, Mario",
    title = "{Next Generation Design and Prospects for Cannex}",
    eprint = "2107.07645",
    archivePrefix = "arXiv",
    primaryClass = "physics.ins-det",
    doi = "10.3390/universe7070234",
    journal = "Universe",
    volume = "7",
    number = "7",
    pages = "234",
    year = "2021"
}

@book{BowenMilburn2016,
  author    = {Bowen, Warwick P. and Milburn, Gerard J.},
  title     = {Quantum Optomechanics},
  publisher = {CRC Press},
  year      = {2016},
  isbn      = {978-1482259155}
}

@incollection{10.1088/978-0-7503-2179-2ch5,
author = {Heyde, Kris and Wood, John L},
title = {The harmonic oscillator},
booktitle = {Quantum Mechanics for Nuclear Structure, Volume 1},
publisher = {IOP Publishing},
year = {2019},
series = {2053-2563},
type = {Book Chapter},
pages = {5-1 to 5-22},
url = {https://doi.org/10.1088/978-0-7503-2179-2ch5},
doi = {10.1088/978-0-7503-2179-2ch5},
isbn = {978-0-7503-2179-2}
}

@article{ARIADNE:2017tdd,
    author = "Geraci, A. A. and others",
    editor = "Carosi, Gianpaolo and Rybka, Gray and van Bibber, Karl",
    collaboration = "ARIADNE",
    title = "{Progress on the ARIADNE axion experiment}",
    eprint = "1710.05413",
    archivePrefix = "arXiv",
    primaryClass = "astro-ph.IM",
    doi = "10.1007/978-3-319-92726-8_18",
    journal = "Springer Proc. Phys.",
    volume = "211",
    pages = "151--161",
    year = "2018"
}

@article{Bloch:2019lcy,
    author = "Bloch, Itay M. and Hochberg, Yonit and Kuflik, Eric and Volansky, Tomer",
    title = "{Axion-like Relics: New Constraints from Old Comagnetometer Data}",
    eprint = "1907.03767",
    archivePrefix = "arXiv",
    primaryClass = "hep-ph",
    doi = "10.1007/JHEP01(2020)167",
    journal = "JHEP",
    volume = "01",
    pages = "167",
    year = "2020"
}

@article{deSalas:2020hbh,
    author = "de Salas, Pablo F. and Widmark, Axel",
    title = "{Dark matter local density determination: recent observations and future prospects}",
    eprint = "2012.11477",
    archivePrefix = "arXiv",
    primaryClass = "astro-ph.GA",
    doi = "10.1088/1361-6633/ac24e7",
    journal = "Rept. Prog. Phys.",
    volume = "84",
    number = "10",
    pages = "104901",
    year = "2021"
}

@article{Damour_2010,
   title={Equivalence principle violations and couplings of a light dilaton},
   volume={82},
   ISSN={1550-2368},
   url={http://dx.doi.org/10.1103/PhysRevD.82.084033},
   DOI={10.1103/physrevd.82.084033},
   number={8},
   journal={Physical Review D},
   publisher={American Physical Society (APS)},
   author={Damour, Thibault and Donoghue, John F.},
   year={2010},
   month=oct }

@article{Damour_1994,
   title={The string dilation and a least coupling principle},
   volume={423},
   ISSN={0550-3213},
   url={http://dx.doi.org/10.1016/0550-3213(94)90143-0},
   DOI={10.1016/0550-3213(94)90143-0},
   number={2–3},
   journal={Nuclear Physics B},
   publisher={Elsevier BV},
   author={Damour, T. and Polyakov, A.M.},
   year={1994},
   month=jul, pages={532–558} }

@article{Kivel:2022rzr,
    author = "Kivel, Alexey and Laux, Julien and Yu, Felix",
    title = "{Axion couplings in gauged U(1)' extensions of the Standard Model}",
    eprint = "2211.12155",
    archivePrefix = "arXiv",
    primaryClass = "hep-ph",
    doi = "10.1007/JHEP03(2023)078",
    journal = "JHEP",
    volume = "03",
    pages = "078",
    year = "2023"
}

@article{Ai:2024cnp,
    author = "Ai, Wen-Yuan and Garbrecht, Bjorn and Tamarit, Carlos",
    title = "{CP Conservation in the Strong Interactions}",
    eprint = "2404.16026",
    archivePrefix = "arXiv",
    primaryClass = "hep-ph",
    reportNumber = "KCL-PH-TH/2023-51, MITP-24-043, TUM-HEP-1508/24",
    doi = "10.3390/universe10050189",
    journal = "Universe",
    volume = "10",
    number = "5",
    pages = "189",
    year = "2024"
}

@article{Hamaide:2025buy,
    author = "Hamaide, Louis and Banks, Hannah and Barker, Peter and Geraci, Andrew A.",
    title = "{Searching for Ultralight Dark Matter with MOLeQuTE: a Massive Optically Levitated Quantum Tabletop Experiment}",
    eprint = "2512.00166",
    archivePrefix = "arXiv",
    primaryClass = "hep-ph",
    month = "11",
    year = "2025"
}

@article{Berlin:2023ubt,
    author = "Berlin, Asher and Millar, Alexander J. and Trickle, Tanner and Zhou, Kevin",
    title = "{Physical signatures of fermion-coupled axion dark matter}",
    eprint = "2312.11601",
    archivePrefix = "arXiv",
    primaryClass = "hep-ph",
    reportNumber = "FERMILAB-PUB-23-779-SQMS-T, SLAC-PUB-17758",
    doi = "10.1007/JHEP05(2024)314",
    journal = "JHEP",
    volume = "05",
    pages = "314",
    year = "2024"
}

@article{Yu:2023gdq,
    author = "Yu, Felix",
    title = "{Primer on Axion Physics}",
    eprint = "2308.08612",
    archivePrefix = "arXiv",
    primaryClass = "hep-ph",
    reportNumber = "MITP-23-038",
    doi = "10.1002/andp.202300106",
    journal = "Annalen Phys.",
    volume = "536",
    number = "1",
    pages = "2300106",
    year = "2024"
}

@article{Agrawal:2023lmw,
    author = "Agrawal, Prateek and Hutzler, Nicholas R. and Kaplan, David E. and Rajendran, Surjeet and Reig, Mario",
    title = "{Searching for axion forces with spin precession in atoms and molecules}",
    eprint = "2309.10023",
    archivePrefix = "arXiv",
    primaryClass = "hep-ph",
    reportNumber = "FERMILAB-PUB-23-651-SQMS-V",
    doi = "10.1007/JHEP07(2024)133",
    journal = "JHEP",
    volume = "07",
    pages = "133",
    year = "2024"
}

@article{Smith:2023htu,
    author = "Smith, Christopher",
    title = "{On the fermionic couplings of axionic dark matter}",
    eprint = "2302.01142",
    archivePrefix = "arXiv",
    primaryClass = "hep-ph",
    doi = "10.1140/epjc/s10052-023-12354-5",
    journal = "Eur. Phys. J. C",
    volume = "84",
    pages = "12",
    year = "2024"
}

@article{Foldy:1949wa,
    author = "Foldy, Leslie L. and Wouthuysen, Siegfried A.",
    title = "{On the Dirac theory of spin 1/2 particle and its nonrelativistic limit}",
    doi = "10.1103/PhysRev.78.29",
    journal = "Phys. Rev.",
    volume = "78",
    pages = "29--36",
    year = "1950"
}

@article{Pospelov:2008jk,
    author = "Pospelov, Maxim and Ritz, Adam and Voloshin, Mikhail B.",
    title = "{Bosonic super-WIMPs as keV-scale dark matter}",
    eprint = "0807.3279",
    archivePrefix = "arXiv",
    primaryClass = "hep-ph",
    reportNumber = "FTPI-MINN-08-29, UMN-TH-2709-08",
    doi = "10.1103/PhysRevD.78.115012",
    journal = "Phys. Rev. D",
    volume = "78",
    pages = "115012",
    year = "2008"
}

@article{Kivel:2022emq,
    author = "Kivel, Alexey and Laux, Julien and Yu, Felix",
    title = "{Supersizing axions with small size instantons}",
    eprint = "2207.08740",
    archivePrefix = "arXiv",
    primaryClass = "hep-ph",
    reportNumber = "MITP-22-058",
    doi = "10.1007/JHEP11(2022)088",
    journal = "JHEP",
    volume = "11",
    pages = "088",
    year = "2022"
}

@article{Elahi:2023vhu,
    author = "Elahi, Fatemeh and Elor, Gilly and Kivel, Alexey and Laux, Julien and Najjari, Saereh and Yu, Felix",
    title = "{Lighter QCD axion from anarchy}",
    eprint = "2301.08760",
    archivePrefix = "arXiv",
    primaryClass = "hep-ph",
    reportNumber = "MITP-23-002",
    doi = "10.1103/PhysRevD.108.L031701",
    journal = "Phys. Rev. D",
    volume = "108",
    number = "3",
    pages = "L031701",
    year = "2023"
}

@article{Ai:2020ptm,
    author = {Ai, Wen-Yuan and Cruz, Juan S. and Garbrecht, Bj{\"o}rn and Tamarit, Carlos},
    title = "{Consequences of the order of the limit of infinite spacetime volume and the sum over topological sectors for CP violation in the strong interactions}",
    eprint = "2001.07152",
    archivePrefix = "arXiv",
    primaryClass = "hep-th",
    reportNumber = "TUM-HEP-1249/20, CP3-20-02",
    doi = "10.1016/j.physletb.2021.136616",
    journal = "Phys. Lett. B",
    volume = "822",
    pages = "136616",
    year = "2021"
}

@article{Ai:2024vfa,
    author = "Ai, Wen-Yuan and Garbrecht, Bjorn and Tamarit, Carlos",
    title = "{The QCD theta-parameter in canonical quantization}",
    eprint = "2403.00747",
    archivePrefix = "arXiv",
    primaryClass = "hep-th",
    reportNumber = "KCL-PH-TH/2024-13, TUM-HEP-1499/24, MITP-24-031",
    month = "3",
    year = "2024"
}

@article{Abel:2020pzs,
    author = "Abel, C. and others",
    title = "{Measurement of the Permanent Electric Dipole Moment of the Neutron}",
    eprint = "2001.11966",
    archivePrefix = "arXiv",
    primaryClass = "hep-ex",
    doi = "10.1103/PhysRevLett.124.081803",
    journal = "Phys. Rev. Lett.",
    volume = "124",
    number = "8",
    pages = "081803",
    year = "2020"
}

@article{CAPP:2020utb,
    author = "Kwon, Ohjoon and others",
    collaboration = "CAPP",
    title = "{First Results from an Axion Haloscope at CAPP around 10.7  $\mu$eV}",
    eprint = "2012.10764",
    archivePrefix = "arXiv",
    primaryClass = "hep-ex",
    doi = "10.1103/PhysRevLett.126.191802",
    journal = "Phys. Rev. Lett.",
    volume = "126",
    number = "19",
    pages = "191802",
    year = "2021"
}

@article{QUAX:2025wtb,
    author = "Infirri, Giosu{\`e} Sardo and others",
    collaboration = "QUAX",
    title = "{Search for Postinflationary QCD Axions with a Quantum-Limited Tunable Microwave Receiver}",
    eprint = "2506.11589",
    archivePrefix = "arXiv",
    primaryClass = "hep-ex",
    reportNumber = "FERMILAB-PUB-25-1020-SQMS-V",
    doi = "10.1103/4dv9-72t5",
    journal = "Phys. Rev. Lett.",
    volume = "135",
    number = "21",
    pages = "211002",
    year = "2025"
}

\end{document}